\documentclass[ejs]{imsart}
\usepackage{latexsym}
\usepackage{amsthm}
\usepackage{amsmath,amssymb,amsfonts,bm,amsbsy,bbm, enumitem}
\usepackage{epsfig}
\usepackage{makecell}
\usepackage{graphicx,rotating,lscape}

\usepackage{booktabs}
\usepackage{verbatim}
\usepackage{natbib}
\usepackage{multirow,color}
\usepackage{subfigure}
\usepackage{float}

\DeclareFontFamily{U}{mathx}{}
\DeclareFontShape{U}{mathx}{m}{n}{<-> mathx10}{}
\DeclareSymbolFont{mathx}{U}{mathx}{m}{n}
\DeclareMathAccent{\widecheck}{0}{mathx}{"71}
\renewcommand{\theequation}{\arabic{equation}}
\newtheorem{Lem}{Lemma}
\newtheorem{Th}{Theorem}
\newtheorem{Cor}{Corollary}
\newtheorem{Pro}{Proposition}

\newtheorem{Rem}{Remark}

\def\A{{\bf A}}
\def\a{{\bf a}}

\def\b{{\bf b}}

\def\c{{\bf c}}
\def\e{{\bf e}}

\def\S{{\bf S}}

\def\m{{\bf m}}
\def\O{{\bf O}}
\def\o{{\bf o}}

\def\v{{\bf v}}

\def\X{{\bf X}}
\def\x{{\bf x}}

\def\W{{\bf W}}

\def\u{{\bf u}}
\def\Z{{\bf Z}}
\def\z{{\bf z}}

\def\calA{{\cal A}}

\def\calT{{\cal T}}
\def\calR{{\cal R}}

\def\mS{\mathcal S}

\def\mR{\mathbb{R}}

\def\ba{{\boldsymbol\alpha}}

\def\btheta{{\boldsymbol \theta}}
\def\bTheta{{\boldsymbol \Theta}}

\def\bb{{\boldsymbol\beta}}

\def\bpsi{\boldsymbol\psi}

\def\bg{\boldsymbol\gamma}

\def\0{{\bf 0}}
\def\trans{^{\rm T}}
\def\pr{\hbox{pr}}
\def\wh{\widehat}
\def\wt{\widetilde}
\def\wc{\widecheck}

\def\var{\hbox{var}}

\def\eff{_{\rm eff}}

\def\n{\nonumber}

\def\log{{\rm log}}

\def\squarebox#1{\hbox to #1{\hfill\vbox to #1{\vfill}}}

\def\bse{\begin{eqnarray*}}
\def\ese{\end{eqnarray*}}
\def\be{\begin{eqnarray}}
\def\ee{\end{eqnarray}}
\def\bsq{\begin{equation*}}
\def\esq{\end{equation*}}
\def\bq{\begin{equation}}
\def\eq{\end{equation}}

\def\sumi{\sum_{i=1}^n}

\def\suminone{\sum_{i=1}^{n_1}}

\def\sumIP1{\sum_{i=1, i\in P_1}^N}

\def\boxit#1{\vbox{\hrule\hbox{\vrule\kern6pt\vbox{\kern6pt#1\kern6pt}\kern6pt\vrule}\hrule}}

\def\spacingset#1{\renewcommand{\baselinestretch}%
{#1}\small\normalsize} \spacingset{1}

\begin{document}

\pagenumbering{arabic}
\thispagestyle{empty}
\begin{frontmatter}
\title{Prediction in Measurement Error Models}
\runtitle{Prediction in Measurement Error Models}
\begin{aug}
\author[A]{\fnms{Fei} \snm{Jiang}\ead[label=e1]{fei.jiang@ucsf.edu}}
and
\author[B]{\fnms{Yanyuan}
  \snm{Ma}\ead[label=e2]{yzm63@psu.edu}}
\address[A]{Department of Epidemiology and Biostatistics,
       The University of California, San Francisco,
\printead{e1}}
\address[B]{Department of Statistics, Pennsylvania State University,
\printead{e2}}
\end{aug}

\begin{abstract}
We study the well known difficult problem of prediction in measurement
error models. By targeting directly at the prediction interval instead
of the point prediction, we construct a prediction interval by
providing estimators of both
the center and the length of the 
interval which achieves a pre-determined prediction level. 
The constructing procedure requires a working model for the
distribution of the variable prone to error. If the working model is
correct, the prediction interval
estimator obtains the smallest variability in terms of assessing the
true center and length. If the working model is incorrect, the
prediction interval estimation is still consistent. We further study
how the length of the prediction interval depends on the choice of the
true prediction interval center and provide guidance on obtaining
minimal prediction interval length. Numerical experiments are
conducted to illustrate the performance and we apply our method to
predict concentration of $A\beta1-42$ in cerebrospinal
  fluid in an Alzheimer's disease data.
\end{abstract}



\end{frontmatter}
\section{Introduction}

Prediction in measurement error models is a notoriously difficult problem. 
In fact, prediction has been largely avoided in the measurement error 
literature except for a few works \citep{gaf1983, Buonaccorsi1995,
    cdh2009, ddhw2018} in special cases,
because the difficulties involved are fundamental and 
almost impossible to overcome. Instead, a different notion of
prediction is sometimes used, where although the observed 
data contain measurement errors, the ``prediction'' is conducted for
observed covariates that are error free, thus changing the true notion of
prediction \citep{zhangetal2019,zhangetal2021}. 

The difficulty involved in prediction for measurement error models is
rooted in the problem setting itself, in that the task of prediction
involves issues beyond the model estimation itself. 
Indeed, even if
the measurement error model is completely known without any unknown
parameters, prediction is still unachievable. To see this, assume we
have response variable $Y$ and covariate $X$ that is subject to
measurement error. Instead of $X$, we observe $W$, which is linked to
$X$ via $W=X+U$. A typical and simplest measurement error model specifies the
dependence of $Y$ on $X$ up to some unknown parameters, say $\bb$,  
and the relation between $W$ and $X$. Given independent and
identically distributed (iid) observations $(W_i, Y_i), i=1, \dots,
n$, various methods have been developed to estimate the unknown
parameter $\bb$. The goal of prediction  in the measurement
  error model context
is to further estimate $Y_{n+1}$ based on a
new $W_{n+1}$,  while making use of the link between $Y$ and 
  $X$ and the link between $W$ and $X$. This implies that we are
obliged to study the  
dependence of $Y$ on $W$, which is not readily available even though we
can estimate $\bb$. To see this,  $f_{Y,W}(y,w)=\int f_{Y\mid
  X}(y,x,\bb)f_{W\mid X}(w,x)f_X(x)d\mu(x)$, and without the knowledge
of the probability density function (pdf) $f_X(x)$, even after we obtain the estimation
for $\bb$, we still 
cannot obtain $f_{Y\mid W}(y,w)$. Here, we use $f_{Y\mid W}(y,w)$ to
denote the conditional pdf of $Y$ given $W$, and similarly for other
notations.  Of course,  as often raised by people who
  dismiss measurement error literatures, one can
always directly study the relation between the response $Y$ and the
errored covariates $W$.  Although it is certainly a
  natural and valid idea, this approach gives up the existing 
knowledge on the relation between the response and  the original
covariate $X$, i.e. the model $f_{Y\mid X}(y,x,\bb)$, hence giving up any inherent
relation between $Y$ and $W$ as well. Indeed, the
  information contained in a completely unspecified function $f_{Y\mid W}(y,w)$ 
and a partially specified function $\int f_{Y\mid
  X}(y,x,\bb)f_{W\mid X}(w,x)f_X(x)d\mu(x)$ is not the same, even
though on the surface both contain a single nonparametric component.
In fact, in our
  numerical experiments, we found
  that taking into account the 
  existing relation $f_{Y\mid X}(y,x,\bb)$ improves the prediction
  quite substantially in
  comparison to a direct approach to predict $Y$ based on $W$.

In this work, %borrowing the idea of conformal prediction, we propose a
%way of performing prediction in measurement error models. 
instead of
providing a 
single prediction value, we consider a prediction interval, or more
generally a prediction set, so that the probability of a future
observation belongs to this set is at least some pre-specified
value. Naturally, if we lower this probability level, the prediction interval will
shrink. Ideally, if the probability level is approaching zero, then the interval
will approach a single point which will be the classic prediction
value.

While the general idea is simple and intuitive, its treatment in
measurement error problems has its own unique features and properties,
which makes the problem interesting and far from being straightforward.
We propose a basic procedure of constructing a prediction
  interval (Section 
\ref{sec:interval}), derive the efficient (Section \ref{sec:influ}) and local
efficient (Section \ref{sec:loceff}) influence functions for estimating the
boundary values of the prediction 
interval.  We also  study
how to achieve the best prediction performance when the relationship between $Y$
and $W$ is directly taken into account (Section \ref{sec:direct}) and
when the measurement error in $W$ is ignored  (Section \ref{sec:naivepred}).
As far as we are aware, this is the first 
prediction method in the measurement error model framework, where the
characteristic of measurement errors is taken into account, and the
prediction is honestly performed under the same data feature for the
future observation as for the observed data. It is also the pioneering
  work that studies the variability of the prediction interval, which
  provides information about the reliability of a prediction.

\section{Model, goal and problem formulation}

\subsection{Model and goal}\label{sec:model}
Consider the classical measurement error model framework where
$f_{Y\mid X, \Z}(y,x,\z,\bb_1)$ is given up to the unknown parameter
$\bb_1$ and the observations are  $\O_i\equiv(W_i, \Z_i, Y_i), i=1, \dots,
n$.  Here, $\Z_i$ is
the error free covariate,  $W_i$ is an errored version of the
error prone covariate $X_i$.
We allow any relation $f_{W\mid
  X,\Z}(w,x,\z,\bb_2)$ here, while in practice,
$W$ often only depends on $X$, for example, $W_i=X_i+U_i$ for a random
noise $U_i$, 
and sometimes the dependence is completely known.
In this case,
$\bb_2$ vanishes. Let $\bb=(\bb_1\trans,\bb_2\trans)\trans\in{\calR}^p$ and assume
$\bb$ is identifiable.
Our original goal is to predict $Y_{n+1}$ given a new observation
$(W_{n+1}, \Z_{n+1})$. Practically, we aim to predict a prediction set $C(W_{n+1},
\Z_{n+1})$ so that $\pr\{Y_{n+1}\in C(W_{n+1},\Z_{n+1})\}\ge1-\alpha$
for a predetermined $\alpha$, such as $\alpha=0.05$. A familiar form
of $C(w, \z)$ is $[m(w, \z,\wh\bb)-c_\alpha
s(w, \z,\wh\bb), m(w, \z,\wh\bb)+c_\alpha
s(w, \z,\wh\bb)]$, following the practice of 
constructing confidence intervals, where $m(w,\z,\wh\bb)$ and
$s(w,\z,\wh\bb)$ are the estimated conditional mean and conditional standard
deviation of $Y$ given $w, \z$, and $c_\alpha$ is a constant that is
often the $(1-\alpha/2)$th standard normal quantile in large samples.

\begin{Rem}\label{rem:iden}
We made an assumption that $\bb$ is identifiable and all the subsequent
results will build on this assumption. Establishing the
identifiability of $\bb$ often
has to be done  case by case based on the specific model,
although many works exist in an attempt to establish
  identifiability in more general situations \citep{hs2008,hs2013,
    hsw2015,hsw2016,hs2017,hs2018,hss2022}. 

To see
why a general result is difficult to obtain, consider the simple
linear regression model with normal additive error, i.e.
$Y=\beta_{c}+\beta_{x}X+\beta_{\z}\trans\Z+\epsilon$, $W=X+U$, where
$\epsilon, U$ are both normal with zero mean  and variances $\sigma^2,
\sigma_U^2$ respectively. This corresponds to $f_{Y\mid
  X,\Z}(y,x,\z)=\phi\{(y-\beta_{c}-x\beta_{x}-\z\trans\bb_{\z})/\sigma\}/
\sigma$ and $f_{W\mid X,\Z}(w,x,\z)=\phi\{(w-x)/\sigma_U\}/\sigma_U$.
It is known that in this very simple
situation, if both $\sigma_U^2$ and $\sigma^2$ are unknown, hence
$\bb_1=(\beta_{c},\beta_{x},\bb_{\z}\trans, \sigma)\trans$,
$\beta_2=\sigma_U$, then the problem is not identifiable. However, as
soon as $\sigma$ is known, i.e.,
$\bb_1=(\beta_{c},\beta_{x},\bb_{\z}\trans)\trans$, $\beta_2=\sigma_U$, then the
problem is identifiable. Thus, in practice, it is important to first
inspect the problem to ensure identifiability before proceeding to
perform estimation and prediction. In the situation when the problem
is not identifiable, then one needs to collect additional information
such as  repeated measurements or instrumental variables to achieve identifiability.
For example, in the normal additive error case with two repeated
measurements, we can use $W_{i1}, 
W_{i2}, i=1, \dots, n$ to estimate $\sigma_U$, and then treat
$W=(W_1+W_2)/2$ with known error variance, which leads to the case of
$\bb=\bb_1$ in the discussion above. In the case of instrumental
variable, say $S$ which has the relation to $X$ captured by $f_{X\mid S,\Z}(x,s,\z,\ba)
$, then we can consider the expanded model
$f_{Y\mid X,\Z,S}(y,x,\z,s,\bb_1)
=f_{Y\mid X,\Z}(y,x,\z,\bb_1)
$ and $f_{W\mid X,\Z,S}(w,x,\z,s,\bb_2,\ba)= f_{W\mid
  X,\Z}(w,x,\z,\bb_2)f_{X\mid S,\Z}(x,s,\z,\ba)$,
where we treat $(\Z\trans,S)\trans$ as the new error free covariate
$\Z$, and treat $(\bb_2\trans,\ba\trans)\trans$ as the new $\bb_2$ in
this expanded model. 
\end{Rem}

\subsection{Estimation of $\bb$}\label{sec:estbeta}
It is natural to believe that to perform reasonable prediction, we
would first need to estimate $\bb$.
For such a simple, purely parametric measurement error model,
estimation of $\bb$ is actually not straightforward. The difficulty lies in
the presence of the conditional pdf of $X$ given $\Z$, 
denoted as $\eta_1(x,\z)$, in practically all estimation
approaches. Unfortunately, $\eta_1(x,\z)$ is  
unknown, difficult to model and even more difficult to
estimate \citep{carrollhall1988, fan1991}, hence handling
$\eta_1(x,\z)$ itself or any quantities dependent on it becomes a very
thorny issue.  Fortunately,
 \cite{tsiatisma2004} eventually bypassed this issue and
prescribed an estimator for $\bb$.  Their method requires a subjectively
posited conditional pdf $\eta_1^*(x,\z)$ which does not have to
be equal to or even approximate the true $\eta_1(x,\z)$ that governs
the data generation procedure. Specifically, the estimator $\wh\bb$
solves an estimating equation of the form
$\sumi \S\eff^*(\O_i,\bb)=\0$, where $\S\eff^*$ is the
efficient score of $\bb$ computed under the posited
conditional pdf $\eta_1^*(x,\z)$. To give a more explicit description of
$\S\eff^*(\o,\bb)$, we write out the detailed construction below.
\begin{enumerate}
\item
Posit a working conditional pdf model for $X$ given $\Z$, denote
it $\eta_1^*(x,\z)$.
\item
Compute the working score function 
\bse
\S_\bb^*(\o,\bb)=\frac{\partial\log\int f_{Y\mid
  X,\Z}(y,x,\z,\bb_1)f_{W\mid
  X,\Z}(w,x,\z,\bb_2)\eta_1^*(x,\z)d\mu(x)}{\partial\bb}.
\ese
\item
Solve the integral equation $E[E^*\{\a^*(X,\Z,\bb)\mid\O\}\mid X,\Z]
=E\{\S_\bb^*(\O,\bb)\mid X,\Z\}$, i.e.
\bse
&&\int \frac{\int \a^*(x,\z,\bb)f_{Y\mid
  X,\Z}(y,x,\z,\bb_1)f_{W\mid
  X,\Z}(w,x,\z,\bb_2)\eta_1^*(x,\z) d\mu(x)}
{\int f_{Y\mid
    X,\Z}(y,x,\z,\bb_1) f_{W\mid
    X,\Z}(w,x,\z,\bb_2)\eta_1^*(x,\z) d\mu(x)}\\
&&\times f_{Y\mid
  X,\Z}(y,x,\z,\bb_1) f_{W\mid
  X,\Z}(w,x,\z,\bb_2)d\mu(w,y)\\
&
=&\int\S_\bb^*(y,w,\z,\bb) f_{Y\mid
  X,\Z}(y,x,\z,\bb_1)f_{W\mid
  X,\Z}(w,x,\z,\bb_2)d\mu(w,y)
\ese
to obtain $\a^*(x,\z,\bb)$.
\item
Compute $\S\eff^*(\o,\bb)=\S_\bb^*(\o,\bb)-
E^*\{\a^*(X,\Z,\bb)\mid\o\}$.
\end{enumerate}

\begin{Rem}\label{rem:estbeta}
In the algorithm above, the only step that may appear nonstandard is
the integral equation solving step. This is a fredholm equation of the
first type and is well studied in numerical analysis \citep{kress1999}. In
practice, we can simply discretize the problem by writing
$a_i=\a^*(x_i,\z,\bb)$, and converting the problem into a linear
system solving problem of the form $\A\a=\b$, where $\a=(a_1, \dots,
a_m)\trans$, $\b=(b_1, \dots, b_m)\trans$
with
\bse
b_i=\int\S_\bb^*(y,w,\z,\bb) f_{Y\mid
  X,\Z}(y,x_i,\z,\bb_1)f_{W\mid
  X,\Z}(w,x_i,\z,\bb_2)d\mu(w,y),
\ese
and $\A$ is a $m\times m$ matrix with
\bse
A_{ij}&=&\int \frac{ f_{Y\mid
  X,\Z}(y,x_j,\z,\bb_1)f_{W\mid
  X,\Z}(w,x_j,\z,\bb_2)\eta_1^*(x_j,\z)}
{\sum_{j=1}^m f_{Y\mid
    X,\Z}(y,x_j,\z,\bb_1) f_{W\mid
    X,\Z}(w,x_j,\z,\bb_2)\eta_1^*(x_j,\z) }\\
&&\times f_{Y\mid
  X,\Z}(y,x_i,\z,\bb_1) f_{W\mid
  X,\Z}(w,x_i,\z,\bb_2)d\mu(w,y).
\ese
The linear system is solved at each different $\z$ value to yield
$\z$-specific $\a$. Because $m$ is often chosen to be small (8-15 in
our experience), the linear system is very easy to solve hence the
overall computation is very fast. 
\end{Rem}

\subsection{Prediction problem formulation}\label{sec:formulation}
After obtaining the estimator $\wh\bb^*$ from 
solving $\sumi\S\eff^*(\O_i,\bb)=\0$,
we proceed to consider constructing a prediction interval. 
A classical prediction interval is often of the form $C(w,\z)=[m(w,\z,\bb)-\zeta,
m(w,\z,\bb)+\zeta]$, where $m(w,\z,\bb)$ is the
center of the prediction interval and $2\zeta$ is the length. 
Here, we fix $m(w,\z,\bb)$ and consider how to obtain $\zeta$. 
In other words, $m(w,\z,\bb)$ is a pre-specified function. For example,
we can imagine 
$m(w,\z,\bb)=E(Y\mid w,\z)$ if it had been obtainable. We will
study different choices of $m(w,\z,\bb)$ in more detail in Section \ref{sec:optm}.

Let $r(\o,\bb)\equiv |y-m(w,\z,\bb)|$ be the distance of $y$ to the
center of the prediction interval. Note that
$r(\o,\bb)$ is fully specified once $m(w,\z,\bb)$ is
fixed, and it can be viewed as regression error or residual if we indeed have
$m(w,\z,\bb)=E(Y\mid w,\z)$.
 If we restrict ourselves to the prediction intervals with its center 
$m(w,\z,\bb)$ fixed, then
we can equivalently rewrite the prediction problem as
searching for $\zeta$ so that $\pr\{r(\O_{n+1},\bb)<\zeta\}
=1-\alpha$.

\section{Performing prediction}\label{sec:optm}

\subsection{Interval prediction is an estimation problem of $\zeta$}\label{sec:interval}

At the end of Section \ref{sec:formulation}, 
by fixing the functional form of the center of the prediction
interval, we have converted the problem of finding the prediction
region (interval) $C(W_{n+1}, \Z_{n+1})$ 
to the problem of finding $\zeta$ so that 
$\pr\{r(\O_{n+1},\bb)<\zeta\} 
=1-\alpha$. 
Although we motivated the $r(\o,\bb)$ function by considering it 
as a regression error or residual, the choice of $r(\o,\bb)$ can be
arbitrary. Our formulation of the prediction is also considered in the
conformal prediction literature
and $r(\o,\bb)$ is officially termed conformal score 
there \citep{leietal2018}. In fact, given an
arbitrary conformal score $r(\o,\bb)$, we can always define 
a prediction region $C(w,\z)$ by letting $C(w,\z)
\equiv\{y: r(\o,\bb)<\zeta\}$. Thus, we have
\bse
\pr\{Y_{n+1}\in C(W_{n+1},\Z_{n+1})\}
=\pr\{r(\O,\bb)<\zeta\}
= 1-\alpha.
\ese
To find the prediction region $C(w,\z)$, we only need to find
$\zeta$. Note that here $r(\o,\bb)$ can be any pre-specified conformal
score including but not limited to the form of $|y-m(w,\z,\bb)|$.

An important discovery we make here is to realize that finding  $\zeta$
such that  $\pr\{r(\O,\bb)<\zeta\} 
=1-\alpha$ is a semiparametric estimation problem, where $\bb,
\eta_1(x,\z), \eta_2(\z)$ are unknown parameters and $\zeta$ is a
functional of $\bb, \eta_1, \eta_2$ implicitly defined. 
Here, $\eta_2(\z)$ denotes the marginal pdf
of $\Z$.
We exploit this view point below to develop procedures to assess
$\zeta$ differently from any existing literature including conformal
prediction. In the following, we first derive the efficient influence
  function of $\zeta$ under the ideal setting where the joint
  distribution of  $X, \Z$ are correctly specified. This 
  leads to  the  local efficient
  influence function of $\zeta$  through replacing the true joint
  distribution of $X, \Z$ by a working model. The estimator for
  $\zeta$ is then 
  obtained by finding the root of the sum of the local efficient
  influence functions. We finally study the asymptotic properties of
  the estimator of $\zeta$.

\subsection{Efficient influence function of $\zeta$}\label{sec:influ}

In the semiparametrics literature, an important approach to
  construct efficient estimator for a parameter is through finding its
  efficient influence function. To this end,
to perform efficient estimation of $\zeta$, we note that the
likelihood of a single observation $\o$ can be written as 
$$f_\O(\o,\bb,\eta_1,\eta_2)=\eta_2(\z)\int\eta_1(x,\z)f_{Y\mid X,\Z}(y,x,\z,\bb_1)f_{W\mid
  X,\Z}(w,x,\z,\bb_2)d\mu(x), $$ and $\zeta$ depends on $\bb, \eta_1,
\eta_2$ implicitly  through 
\bse
&&\pr\{r(\O,\bb)<\zeta\} \\
&=&\int_{r(\o,\bb)< \zeta}f_\O(\o,\bb,\eta_1,\eta_2)d\mu(\o)
=1-\alpha.
\ese
We then derive the efficient influence
function for estimating $\zeta$, which we denote $\phi$. We find that
\begin{Pro}\label{pro:effinflu}
  The efficient influence function of $\zeta$ is
\bse
&&\phi(\O,\bb,\zeta)\\
&=&\c\trans\S\eff(\O,\bb)
+\frac{E\{a_1(X,\Z,\bb,\zeta)\mid\O\}+ (1-\alpha)
-E[I\{ r(\O,\bb)<\zeta\}\mid\Z]
}{E[\delta\{\zeta-r(\O,\bb)\}]},
\ese
where
\be\label{eq:c}
\c&=&
\left[E\left\{\S\eff(\O)\S\eff\trans(\O)\right\}\right]^{-1}\left(
\frac{E[\delta\{\zeta-r(\O,\bb)\}
{\partial r(\O,\bb)}/{\partial\bb}]
-E[I\{
    r(\O,\bb)<\zeta\}\S_\bb(\O,\bb)]
}{E[\delta\{\zeta-r(\O,\bb)\}]}
\right.\n\\
&&\left.-E\left[E\{a_1(X,\Z,\bb,\zeta)\mid\O\}E\{\a(X,\Z,\bb)\mid\O\}\right]\right),
\ee
with $\a(x,\z,\bb)$  the same as $\a^*(x,\z,\bb)$ defined in Section
\ref{sec:estbeta} under the true 
posited model $\eta_1(x,\z)$,
and $a_1(x,\z,\bb,\zeta)$ satisfies
\be\label{eq:a1}
E\left[E\{a_1(X,\Z,\bb,\zeta)\mid\O\}\mid X,\Z\right]
=E\left[
I\{ r(\O,\bb)<\zeta\}\mid
\Z\right]-
E\left[
I\{ r(\O,\bb)<\zeta\}
\mid
X,\Z\right].
\ee
\end{Pro}
Because the proof of Proposition \ref{pro:effinflu} is lengthy and
involved, we provide the detailed derivation and proofs in  
Appendix \ref{sec:phi}. In a nutshell, we first find the tangent space
of the model, which is formed by the tangent spaces
 associated with the parameters $\bb$, $\eta_1$ and
$\eta_2$. We then identify the space of all the influence
functions. We finally identify the intersection of the influence
function family and the tangent space to find the efficient influence
function.

Note that (\ref{eq:a1}) uniquely determines $E\{a_1(X,\Z,\bb,\zeta)\mid\O\}$
because the efficient influence function $\phi(\O,\bb,\zeta)$ is unique. 
Specifically, through solving (\ref{eq:a1}) we obtain $a_1(X,\Z,\bb,\zeta)$,
which is not necessarily unique, and then we form
$E\{a_1(X,\Z,\bb,\zeta)\mid\O\}/E[\delta\{\zeta-r(\O,\bb)\}]$, which is the
orthogonal projection of the 
efficient influence function $\phi(\O,\bb,\zeta)$ onto the nuisance tangent space
corresponding to $\eta_1$, 
i.e., the space spanned by all the score functions of 
all the parametric submodels of $\eta_1$,
hence is a unique function.

\subsection{Locally efficient influence function and the estimation of $\zeta$}\label{sec:loceff}
As expected, the efficient score involves both $\eta_1(x,\z)$ and
$\eta_2(\z)$, which are unknown. While $\eta_2(\z)$ does not involve
unobservable variable hence is standard to estimate, it is not
practical to estimate $\eta_1(x,\z)$. Fortunately, 
 we discover that we can  use two
possibly misspecified working models $\eta_1^*(x,\z), \eta_2^\star(\z)$, and the
resulting $\phi^{*\star}(\O,\bb,\zeta)$   
still has mean zero. This is
similar to the construction of the locally
efficient score $\S\eff^*$ for estimating $\bb$, which is robust to
$\eta_1^*(x,\z)$. Note that to distinguish functions and operations
affected by each working model, we used two slightly different
notations $^*$ and $^\star$ to denote the two different working models.
Specifically, let 
\be\label{eq:rel1}
&&\phi^{*\star}(\O,\bb,\zeta)\nonumber\\
&=&\c^{*\star\rm T}\S\eff^*(\O,\bb)
+\frac{E^*\{a_1^*(X,\Z,\bb,\zeta)\mid\O\}+ (1-\alpha)
-E^*[I\{ r(\O,\bb)<\zeta\}\mid\Z]
}{E^{*\star}[\delta\{\zeta-r(\O,\bb)\}]},
\ee
where $a_1^*(x,\z,\bb,\zeta)$ is a function that satisfies
\be\label{eq:a1star}
E\left[E^*\{a_1^*(X,\Z,\bb,\zeta)\mid\O\}\mid X,\Z\right]
=E^*\left[
I\{ r(\O,\bb)<\zeta\}\mid
\Z\right]-
E\left[
I\{ r(\O,\bb)<\zeta\}
\mid
X,\Z\right],
\ee
and 
\be\label{eq:cstar}
\c^{*\star}&=&
\left[E^{*\star}\left\{\S\eff^*(\O)\S\eff^{*\rm
      T}(\O)\right\}\right]^{-1}\left(
-E^{*\star}\left[E^*\{a_1(X,\Z,\bb,\zeta)\mid\O\}E^*\{\a^*(X,\Z,\bb)\mid\O\}\right]
\right.\n\\
&&+\left.
\frac{E^{*\star}[\delta\{\zeta-r(\O,\bb)\}
{\partial r(\O,\bb)}/{\partial\bb}]
-E^{*\star}[I\{
    r(\O,\bb)<\zeta\}\S_\bb(\O,\bb)]
}{E^{*\star}[\delta\{\zeta-r(\O,\bb)\}]}\right).
\ee
We can see that 
\bse
E\{\phi^{*\star} (\O,\bb,\zeta)\mid X,\Z\}=
0+
\frac{(1-\alpha)-
E\left[I\{ r(\O,\bb)<\zeta\}\mid X,\Z\right]
}{E^{*\star}[\delta\{\zeta-r(\O,\bb)\}]},
\ese
hence indeed $E\{\phi^{*\star} (\O,\bb,\zeta)\}=0$.

In fact, we can avoid positing a working model $\eta_2^\star(\z)$
by estimating all the marginal
expectations involved in $\phi^{*\star} (\O,\bb,\zeta)$ with sample average, i.e. 
we can construct
\be\label{eq:rel2}
\wh\phi^*(\O,\bb,\zeta)=\wh\c^{*\rm T}\S\eff^*(\O,\bb)
+\frac{E^*\{a_1^*(X,\Z,\bb,\zeta)\mid\O\}+ (1-\alpha)
-E^*[I\{ r(\O,\bb)<\zeta\}\mid\Z]
}{\wh E[\delta\{\zeta-r(\O,\bb)\}]},
\ee
where $a_1^*(x,\z,\bb,\zeta)$ still satisfies (\ref{eq:a1star}), 
and 
\bse
\wh\c^*&=&
\left[\wh E\left\{\S\eff(\O)\S\eff\trans(\O)\right\}\right]^{-1}\left(
\frac{\wh E[\delta\{\zeta-r(\O,\bb)\}
\frac{\partial r(\O,\bb)}{\partial\bb}]
-\wh E[I\{
    r(\O,\bb)<\zeta\}\S_\bb(\O,\bb,\eta_1,\eta_2)]
}{\wh E[\delta\{\zeta-r(\O,\bb)\}]}
\right.\n\\
&&\left.-\wh E\left[E^*\{a_1(X,\Z,\bb,\zeta)\mid\O\}E^*\{\a(X,\Z,\bb)\mid\O\}\right]\right).
\ese
We can easily verify that 
\bse
E\{\wh\phi^{*} (\O,\bb,\zeta)\mid X,\Z\}=
0+
\frac{(1-\alpha)-
E\left[I\{ r(\O,\bb)<\zeta\}\mid X,\Z\right]
}{\wh{E}[\delta\{\zeta-r(\O,\bb)\}]},
\ese
 hence $\wh\phi^*(\O,\bb,\zeta)$ still has mean
zero.

\begin{comment}
We can easily verify that $\phi^*(\O,\bb,\zeta)$ still has mean zero,
where
\bse
\phi^*(\O,\bb,\zeta)=\c^{*\rm T}\S\eff^*(\O,\bb)
+\frac{E^*\{a_1^*(X,\Z,\bb,\zeta)\mid\O\}+ (1-\alpha)
-E^*[I\{ r(\O,\bb)<\zeta\}\mid\Z]
}{E[\delta\{\zeta-r(\O,\bb)\}]},
\ese
where $a_1^*(x,\z,\bb,\zeta)$ still satisfies (\ref{eq:a1star}), 
and 
\bse
\c^*&=&
\left[E\left\{\S\eff(\O)\S\eff\trans(\O)\right\}\right]^{-1}\left(
\frac{E[\delta\{\zeta-r(\O,\bb)\}
\frac{\partial r(\O,\bb)}{\partial\bb}]
-E[I\{
    r(\O,\bb)<\zeta\}\S_\bb(\O,\bb,\eta_1,\eta_2)]
}{E[\delta\{\zeta-r(\O,\bb)\}]}
\right.\n\\
&&\left.-E\left[E^*\{a_1(X,\Z,\bb,\zeta)\mid\O\}E^*\{\a^*(X,\Z,\bb)\mid\O\}\right]\right).
\ese
\end{comment}

Finally, because the values of $c^{*\star}$ and
$E^{*\star}[\delta\{\zeta-r(\O,\bb)\}]$ does not affect the mean of
$\phi^{*\star} (\O,\bb,\zeta)$, we can actually replace them by an arbitrary vector
and constant respectively to retain its mean zero property.

We also want to point out that given we already have an estimator
$\wh\bb$ based on the procedure described in Section
\ref{sec:estbeta}, it is reasonable to insert $\wh\bb$ into
$\phi^{*\star}(\O,\bb,\zeta)$. This directly leads to the estimating equation
\bse
\sumi \phi^{*\star}(\O_i,\wh\bb,\zeta)
=\sumi \frac{E^*\{a_1^*(X_i,\Z_i,\wh\bb,\zeta)\mid\O_i\}+ (1-\alpha)
-E^*[I\{ r(\O_i,\wh\bb)<\zeta\}\mid\Z_i]
}{\wh E[\delta\{\zeta-r(\O_i,\wh\bb)\}]}=0,
\ese
which is equivalent to 
\be\label{eq:useful}
\sumi \left(E^*\{a_1^*(X_i,\Z_i,\wh\bb,\zeta)\mid\O_i\}+ (1-\alpha)
-E^*[I\{ r(\O_i,\wh\bb)<\zeta\}\mid\Z_i]\right)
=0,
\ee
and it does not involve the model $\eta_2(\z)$.
The estimating equation (\ref{eq:useful}) in combination with (\ref{eq:a1star})
suggests that for the purpose of estimating $\zeta$, we do not need to
concern ourselves with handling $\eta_2(\z)$.

\subsection{Locally efficient interval prediction and its properties}\label{sec:estzeta}

As we have discussed, 
the fact that $E\{\phi^{*\star}(\O,\bb,\zeta)\}=0$
implies that we can perform estimation of $\zeta$ under the posited
models $\eta_1^*, \eta_2^\star$, by solving
$
\sumi\phi^{*\star} (\o_i,\wt\bb,\zeta)=0,
$
where $\wt\bb$ is a consistent estimator of $\bb$. Let
the resulting estimator be $\wt\zeta$. This
is a locally efficient estimator of $\zeta$ and it
serves as an alternative procedure to the  conformal
prediction procedure in assessing $\zeta$.
Alternatively, we can use $\wh\phi^*(\o_i,\bb,\zeta)$ to construct
estimating equations  
instead of using $\phi^{*\star}(\o_i,\bb,\zeta)$. Let the resulting
estimator be $\wh\zeta$. In practice, unless we have very good
knowledge on the distribution of $\Z$, we recommend this
procedure due to its computational convenience. Further, if the
estimator $\wt\bb$ happens to be $\wh\bb$, i.e. the estimator that
satisfies $\sumi\S\eff^*(\o_i,\wh\bb)=\0$, then $\wt\zeta$ and
$\wh\zeta$ are identical and they both solve (\ref{eq:useful}).

Interestingly, to perform
conformal prediction using the locally efficient estimators $\wt\zeta$
and $\wh\zeta$, we do not
need to engage the potential new observation in estimating $\bb$,
and we do not require data splitting either. In contrast, engaging
potential new data or data splitting is a key requirement in any
 conformal prediction procedures. We next provide the
theoretical properties of $\wt\zeta$ and $\wh\zeta$, with the proofs
given in Appendix \ref{sec:proofthuseful}. We also show that the resulting
  prediction probability indeed approximates the target value by
  providing the bound on the difference in Theorem \ref{th:bound},
  with its proof given in Appendix \ref{sec:proofofthbound}.

\begin{Th}\label{th:useful}
Let $\wh\zeta$ solve (\ref{eq:useful}). Then
$\wh\zeta$ is a consistent estimator of $\zeta$ and it
satisfies 
$n^{1/2}(\wh\zeta-\zeta)\to N(0, v^*)$ in distribution
as $n\to\infty$,
where
$$v^*=[E\{{\partial\phi^*
  (\O,\bb,\zeta)}/{\partial\zeta}\}]^{-2}E\{u^* (\O)^2\}, $$
\bse
u^*(\o_i)&=&\phi^* (\o_i,\bb,\zeta)
-E\left\{
\frac{\partial\phi^*(\O,\bb,\zeta)}{\partial\bb\trans}\right\}
\left[E\left\{\frac{\partial\S\eff^*(\O,\bb)}{\partial\bb\trans}\right\}\right]^{-1}\S\eff^*(\o_i,\bb).
\ese
When $\eta_1^*(x,\z)=\eta_1(x,\z)$, we
further obtain 
$n^{1/2}(\wh\zeta-\zeta)\to N[0, E\{\phi(\O,\bb,\zeta)^2\}]$ in distribution
as $n\to\infty$, hence the estimator $\wh\zeta$ is efficient. 
\end{Th}
The consistency of $\wh\bb$ and $\wh\zeta$ directly leads
  to the consistency of the subsequent prediction probability. A direct
  prediction probability bias 
  result follows from the bias property of $\wh\bb, \wh\zeta$, as we
  stated in Corollary \ref{cor:bias}.
\begin{Cor}\label{cor:bias}
    Assume Conditions \ref{con:subGE}--\ref{con:4} and
    $\|\partial^2\pr\{r(\O,\bb)<\zeta\}/\partial \btheta\partial
    \btheta\trans\|_2 <\infty$ for any given
    $\btheta\in\bTheta$. Then we have 
\bse
|E[\pr\{r(\O,\wh\bb)<\wh\zeta\} ]- \pr\{r(\O,\bb)<\zeta\}| =
O(1/n).
\ese
\end{Cor}
We further establish the finite sample prediction error bound in
Theorem \ref{th:bound}. 
To prove Theorem \ref{th:bound}, we first make some
  definitions.
Define the norm $\|X\|_{\psi_2} \equiv \sup_{p\geq 1} p^{-1/2}
\{E(|X|^p)\}^{1/p}$, by the Definition 5.7 in \cite{vershynin2010}, we
have that $X$ is sub-Gaussian if $\|X\|_{\psi_2}
<\infty$. Furthermore, we define $\|X\|_{\psi_1} \equiv \sup_{p\geq 1} p^{-1}
\{E(|X|^p)\}^{1/p}$, by the Definition 5.13 in \cite{vershynin2010}, $X$ is sub-exponential if $\|X\|_{\psi_1} <\infty$. 
Let $\bpsi(\o_i, \bb, \zeta) = \{\S^*\eff(\o_i, {\bb})\trans,  \phi^{*\star}(\o_i, {\bb},
{\zeta})\}\trans$, and let $\btheta = (\bb\trans, \zeta)\trans \in\bTheta$. Let
$\e_j$ be a unit vector with the $j$th element 1.

We make some regularity conditions.
\begin{enumerate}[label=(C\arabic*)]
\item \label{con:subGE} Assume each $\e_j\trans\bpsi(\O_i, \bb, \zeta)$, $j = 1,
  \ldots, (p + 1)$ is a sub-Gaussian random variable with
  $\|\e_j\trans\bpsi(\O_i, \bb, \zeta)\|_{\psi_2}\leq M_1$ for a
  positive constant $M_1$.  Furthermore, assume for any 
  given $\btheta\in\bTheta$,  $\v\trans\partial \bpsi(\O_i, \bb,
  \zeta)/\partial \bb \u$ is a sub-exponential random variable for
  vectors $\v,  \u$ with $\|\v\|_2=\|\u\|_2= 1$. Thus, $\|\v\trans\partial \bpsi(\O_i, \bb,
  \zeta)/\partial \bb \u \|_{\psi_1}\leq M_2$ for a positive constant
  $M_2$.
  \item \label{con:boundp} Assume  $\|\partial
  \pr\{r(\O,\bb)<\zeta\} /\partial \btheta\|_2 \leq M_3$ for any
$\bb, \zeta$ and a
postive constant $M_3$.
\item \label{con:3} For any given $\btheta\in\bTheta$, 
$E\left\{ \partial \bpsi(\O_i, \bb,
    \zeta)/\partial \btheta\trans\right\}$
is non-singular. Let its minimum and maximum singular values be
$\lambda_{\min}, \lambda_{\max}$. Then
$0<\lambda_{\min}\le\lambda_{\max}<\infty$.
\item \label{con:4} For any given $\btheta \in \bTheta$, and
$\v$ with $\|\v \|_2= 1$, 
\bse
\|E\left\{\frac{\partial^2 \v\trans 
  \bpsi(\O_i, \bb,\zeta)}{\partial  \btheta\partial\btheta\trans}\v\right\}\|_2= O(1). 
\ese
  \end{enumerate}

\begin{Th}\label{th:bound}
  Assume Conditions \ref{con:subGE}--\ref{con:3} to hold and $n> p$. 
There are
  positive constants $c, c_1$ such that
  \bse
|\pr\{r(\O,\wh\bb)<\wh\zeta\} - (1-\alpha)|
\leq c M_1 M_3 \lambda_{\min}^{-1}  
\sqrt{\log(n)/ n}, 
\ese
with probability greater than $1 -9^{2p+2}  2 \exp\left\{-  c_1 n
  \min (\lambda_{min}^2/4, \lambda_{\min}/2)\right\} - e/n$.
Here, $M_1, M_3, \lambda_{\min}$ are defined in the
regularity conditions \ref{con:subGE}, \ref{con:boundp} and
\ref{con:3} respectively.
\end{Th}

\subsection{Conformal prediction estimator of $\zeta$}

For completeness and to facilitate comparison, we now briefly
summarize the 
estimation of $\zeta$ in the conformal prediction literature. The
conformal prediction literature includes two general approaches. One
approach involves engaging a potential new observation $W_{n+1},
\Z_{n+1}, Y_{n+1}=y$ and conducting estimation of $\bb$ under each
potential $y$ value, hence is computationally intensive and less often
used. The other approach requires data splitting and is easy to
implement, hence is more often recommended as we explain in detail
below. 

We randomly
split the data into
two parts, with sizes $n_1$ and $n_2$.  One part is used to estimate
$\bb$. Let the size of this 
part be $n_2$ and the data be $\O_{n_1+1}, \dots \O_n$.  Typically,
$n_2=o(n)$. Let the estimator be $\wt\bb$.  
The other part of size $n_1$ is used to 
estimate $\zeta$ via  solving $n_1^{-1}\suminone
I\{\zeta-r(\o_i,\wt\bb)>0\}=(1-\alpha)$. Let the resulting estimator
be $\wc\zeta$.
To obtain the asymptotic properties of the resulting estimator
$\wc\zeta$, we note that solving the equation is 
equivalent to minimizing 
$\suminone\rho_{1-\alpha}\{r(\o_i,\wt\bb)-\zeta\}$, where
$\rho_{1-\alpha}(u)\equiv u\{1-\alpha-I(u<0)\}$ is the check
function. So we directly use the standard quantile regression
asymptotic results to obtain 
$
n_1^{1/2}(\wc\zeta-\zeta)
\to N\{0, \alpha(1-\alpha)(E[\delta\{\zeta-r(\O,\bb)\}])^{-2}\}
$
in distribution, where we used $\wt\bb\to\bb$. Further note that
$n_1/n\to1$, hence
we get $
n^{1/2}(\wc\zeta-\zeta)
\to N\{0, \alpha(1-\alpha)(E[\delta\{\zeta-r(\O,\bb)\}])^{-2}\}
$. For completeness, we also provide a more detailed derivation of the
results in Appendix \ref{sec:classic}.

Compared to the locally efficient estimators 
$\wh\zeta$, we will see that $\wc\zeta$ generally leads to larger
variability. In fact, when $\eta_1^*(x,\z)=\eta_1(x,\z)$, it is
certain that $\wh\zeta$ is efficient hence has the smallest
variability. In our numerical implementations, we find that even when
$\eta_1^*(x,\z)$ is dramatically misspecified, $\wh\zeta$ still
performs similarly to the efficient estimator hence is
more efficient than $\wc\zeta$.

\section{Shortest prediction interval}\label{sec:optim}

We have fixed the center of the prediction interval $m(w,\z,\bb)$ and
focused on estimating $\zeta$ so far. As long as the functional form
of $m(w,\z,\bb)$ is fixed, the true value of $\zeta$ is well defined
hence the prediction interval length $2\zeta$ is determined. Different
estimators of $\zeta$ only leads to different estimated values of the
prediction interval length. Hence, our next question is: what choice of
$m(w,\z,\bb)$ will lead to the shortest prediction interval,
i.e. the smallest $\zeta$?

\begin{figure}[!ht]
	\centering
		\includegraphics[width=\textwidth,height=0.35\textheight]{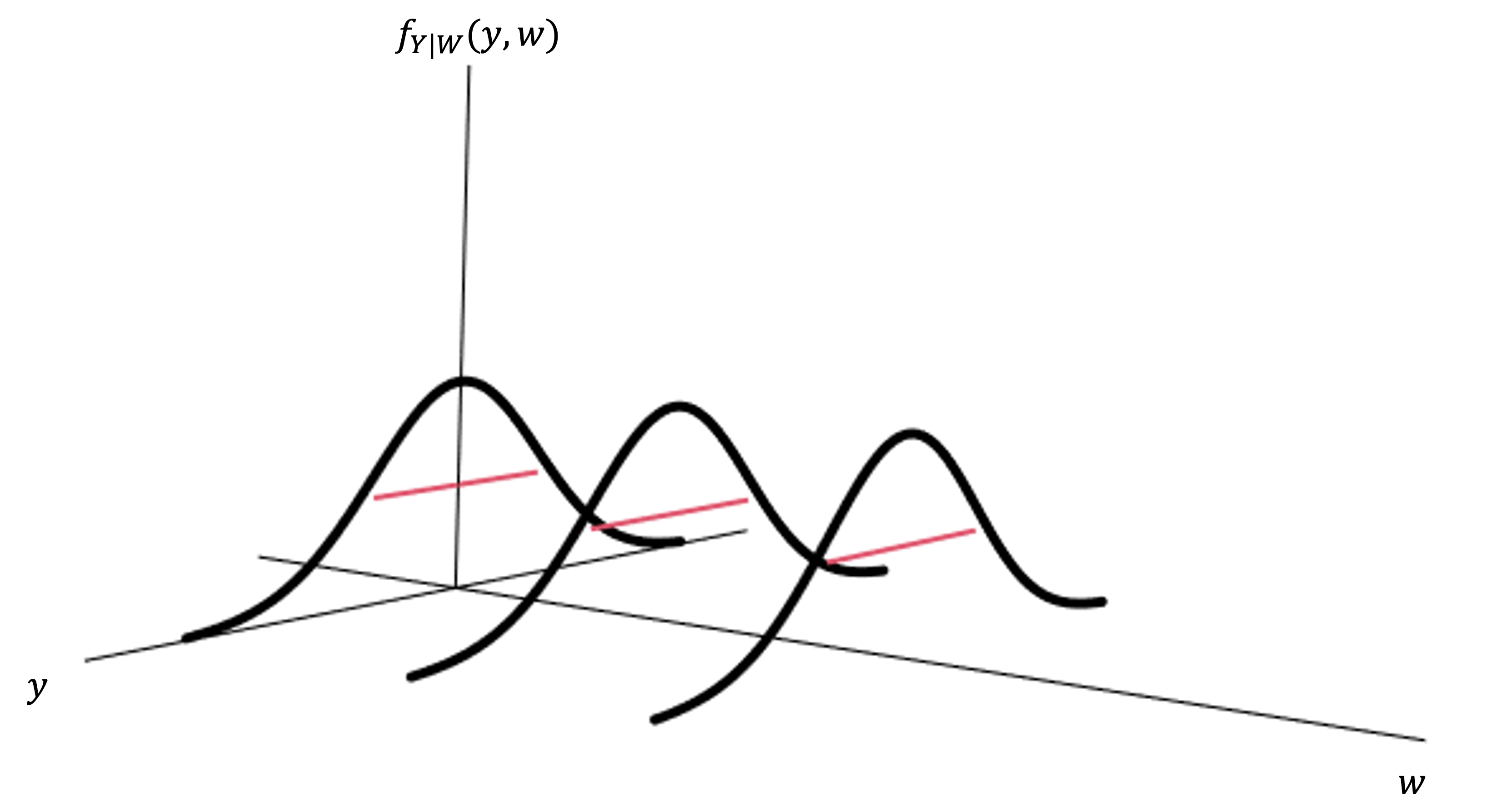}
		\caption{Illustration of the intervals with the
                  largest area under the conditional pdf curves. }
		\label{fig:interval}
              \end{figure}

To this end, note that
\bse
1-\alpha&=&\pr\{Y_{n+1}\in [m(W_{n+1},\Z_{n+1},\bb)-\zeta, m(W_{n+1},\Z_{n+1},\bb)+\zeta]\}\\
&=&E[F_{Y\mid W,\Z}\{m(W_{n+1},\Z_{n+1},\bb)+\zeta\}-
F_{Y\mid W,\Z}\{m(W_{n+1},\Z_{n+1},\bb)-\zeta\}],
\ese
hence we need to choose $m(w,\z,\bb)$ so that at any $(w,\z)$, $m(w,\z,\bb)$
is the center of the length $2\zeta$ interval so that the area under
the conditional pdf $f_{Y\mid W,\Z}(\o)$ curve restricted to this
interval is the
largest among all choices of length $2\zeta$ intervals. Figure
\ref{fig:interval}
provides an illustration of these intervals.
This interval is what is known as the highest density
  interval in the Bayesian statistics literature.
 Obviously, when $f_{Y\mid W,\Z}(\o)$ is a symmetric
unimodal function of $y$, $m(w,\z)$ is the mode/mean. 
However, interestingly, the optimal $m(w,\z)$ is not necessarily the mean
function $E(Y\mid w,\z)$ in general. In fact, 
we cannot give more descriptive statement than the above in
general.

Of course,  in practice, we are faced 
with the issue that $f_{Y\mid W,\Z}(\o)$ is unknown. We suggest to
compute $f_{Y\mid W,\Z}^*(\o,\wh\bb)$ instead, and then identify the
optimal $m(W,\Z,\bb)$ based on $f_{Y\mid W,\Z}^*(\o,\wh\bb)$  using an
iterative procedure. Specifically, we select an initial $m(w,\z,\wh\bb)$, 
for example $m(w,\z,\wh\bb)=E^*(Y\mid W,\Z,\wh\bb)$. We then estimate
$\zeta$. Based on the
estimator $\wh\zeta$ and $f_{Y\mid W,\Z}^*(\o,\wh\bb)$, we identify the
optimal $m(w,\z,\wh\bb)$ as an updated choice of $m(w,\z,\wh\bb)$. We repeat the
procedure of estimating $\zeta$ and updating $m(w,\z,\wh\bb)$, 
until $\wh\zeta$ does not decrease sufficiently large for our
application purpose.

Clearly, the whole procedure of identifying the optimal
$m(w,\z,\wh\bb)$ relies on the posited model
$\eta_1^*(x,\z,\bb,\zeta)$. If we wish, we can choose several candidate models
and retain the result associated with the smallest $\wh\zeta$. We do
not need to adjust for multiple testing here, since each
probability statement 
$1-\alpha=\pr\{Y_{n+1}\in C(W_{n+1},\Z_{n+1})\}$, regardless
the posited $\eta_1^*(x,\z)$, is valid.

\section{Alternative prediction methods}\label{sec:alt}

We now prescribe two alternative ways for prediction in the
measurement error models. The first method is a direct approach, where
the relation between $Y$ and $X$ is completely bypassed. Instead, one
directly inspects the relation between $Y$ and $W, \Z$ for
prediction. The second approach is a naive approach, where one treats
$W$ as if it is $X$. In the prediction interval context, we show that 
both methods are consistent, although they have drawbacks in
comparison to the semiparametric method. 

\subsection{Direct prediction}\label{sec:direct}
Because our goal is to predict  $Y$ based on $W, \Z$, which are
observed in our data, so a direct approach is to establish the
relation between $Y$ and $W,\Z$ and perform prediction. 
In this sense, the measurement error issue is completely dismissed.
Specifically, in the direct prediction approach, we assume 
$Y=m(w,\z)+\epsilon$ and we estimate $m(w,\z)$ nonparametrically to
form the residuals $r(\O)$. We then find the prediction interval by
estimating $\zeta$ based
on the same  relation $\pr\{r(\O)<\zeta\}=1-\alpha$.
In Appendix \ref{sec:nomodel}, we show that in this approach, the
 efficient influence function for estimating $\zeta$ is
\bse
\phi_{\rm direct}(\O,\zeta)=\frac{(1-\alpha)-I\{
  r(\O,m)<\zeta\}-\epsilon E[\delta\{\zeta-r(\O,m)\}\mid
W,\Z]
}{E[\delta\{\zeta-r(\O)\}]}.
\ese
We subsequently estimate $\zeta$ via solving 
$\sumi \phi_{\rm direct}(\O_i,\zeta)=0$.

\subsection{Naive prediction}\label{sec:naivepred}
A naive approach in measurement error models is to treat $W$ as if it
were $X$ and perform the standard analysis. While it is known that
this will lead to an inconsistent estimator for $\bb$, this is
nevertheless a valid method for prediction interval
construction. Specifically, we can estimate $\bb$ by the usual
regression methods, and then form the prediction interval by
estimating $\sumi \phi_{\rm naive}(\O_i,\wh\bb,\zeta)=0$.
In Appendix \ref{sec:naive}, we show that the 
efficient influence function for estimating $\zeta$ is
\bse
&& \phi_{\rm naive}(\O,\bb,\zeta)\\&=&
\frac{1-\alpha-
I\{
  r(\O,\bb)<\zeta\}}{E[\delta\{\zeta-r(\O,\bb)\}]}
+\epsilon \frac{E[\epsilon I\{ r(\O,\bb)<\zeta\}\mid W,\Z]
}{E[\delta\{\zeta-r(\O,\bb)\}]E(\epsilon^2\mid W,\Z) }\\
&&+\epsilon \m_\bb'(w,\z,\bb)\trans\left[E\left\{\frac{\m_\bb'(W,\Z,\bb)^{\otimes2}}{\var(\epsilon^2\mid
W,\Z)}\right\}\right]^{-1}\\
&&\times E\left(\frac{\delta\{\zeta-r(\O,\bb)\}
{\partial r(\O,\bb)}/{\partial\bb}
}{E[\delta\{\zeta-r(\O,\bb)\}]}+
\frac{\epsilon I\{ r(\O,\bb)<\zeta\}\m_\bb'(W,\Z,\bb)
}{E[\delta\{\zeta-r(\O,\bb)\}]E(\epsilon^2\mid W,\Z) }
\right).
\ese

\section{Simulation studies}\label{sec:simu}

In our first simulation, we let $\Z =
(1, Z_1, Z_2)\trans$, where $Z_1\sim U(0, 1)$, $Z_2$ is a Bernoulli random variable with
probability 0.8 to be one. We generate $X$ from a scaled and shifted
beta distribution $X\sim 2\sqrt{3} {\rm
  Beta}(2, 2) -\sqrt{3}$, and set  $W = X+ U$, where $U\sim N(0,
\sigma_U^2)$ with $\sigma_U=0.3$. We then generate $Y= m(X, \Z, \bb) + \epsilon$, where
$\epsilon \sim N(0, \sigma_{\epsilon}^2), \sigma_\epsilon=0.1$. Here
we experiment with three different mean models: 
\begin{itemize}
\item $m_1(X, \Z,  \bb)=  (X, X^2)\bb_1  +
\Z\trans\bb_2$, 
\item $m_2(X, \Z,  \bb)= \sin\{ (X, X^2)\bb_1  +
\Z\trans\bb_2\},$
\item
$m_3(X, \Z,  \bb)= \exp[-\{ (X, X^2)\bb_1 +
\Z\trans\bb_2\}^2]$,
\end{itemize}
where
$\bb = (\bb_1\trans, \bb_2\trans)\trans = (4, 1, 1, 1,
0.5)\trans$. 
We simulate the data with sample sizes
$n = 100$ and  $n=500$ respectively.

We implement  six
methods to perform prediction. 
In all these methods, 
we use $\wh{E}(Y|w, \z)$ to denote the kernel estimator of
the conditional mean of $Y$ given $w, \z$. 
In addition, in obtaining $\a(X, \Z, \bb)$ and $a_1(X,
\Z,\bb, \zeta)$, we use a K-means algorithm to group $\Z$ to two
groups, and adopt the same $\Z$ in each group.
\begin{itemize}
\item m1s: $r(\o, \bb) = |y -
    E^*(Y|w, \z, \bb)|$.  We  use semiparametric method to
    estimate both $\bb$ and $\zeta$.
\item m1c: $r(\o, \bb) = |y -
    E^*(Y|w, \z, \bb)|$. We  use semiparametric method to
    estimate $\bb$ and conformal prediction to estimate $\zeta$.
\item m2s: $r(\o, m) = |y -
    \wh{E}(Y|w, \z)|$.  We use nonparametric method to estimate $m$
    and the  direct method in Section \ref{sec:direct} to
    estimate $\zeta$.
\item  m2c: $r(\o, m) = |y -
    \wh{E}(Y|w, \z)|$. We use nonparametric method to estimate $m$
    and conformal prediction to estimate $\zeta$.
\item m3s: $r(\o,\bb) = |y -
    m(w, \z, \bb)|$. We adopt the naive model $y = m(w, \z, \bb) + \epsilon$ to
    estimate $\bb$, and use  the native method in Section
    \ref{sec:naivepred} to estimate $\zeta$.
\item m3c: $r(\o,\bb) = |y -
    m(w, \z, \bb)|$. We adopt the naive model $y = m(w, \z, \bb) + \epsilon$ to
    estimate $\bb$, and use conformal prediction estimate
    $\zeta$.
\end{itemize}
In all the implementation of the conformal prediction method, we use
the split data approach, where we use half of the data for model
estimation and the other half for prediction.
We can see that methods
m1s and m1c both take into account the model information and the
measurement issue, where
 m1s is our proposed method using semiparametrics to form prediction, 
while  m1c uses conformal prediction approach as a direct competitor.
In contrast, methods m2s and m2c completely ignores the model
information when forming the ``residual'' $r(\cdot)$. Both methods use
nonparametric approach to perform estimation of the mean function 
to form residual $r(\o,m)$, 
while m2s subsequently uses semiparametrics to estimate $\zeta$,  
 m2c uses the conformal prediction to do so. Please see Section \ref{sec:nomodel}
in the Appendix for the details on m2s.  Note that methods
  m2s and m2c correspond to the approach by those who hold
  the view point that measurement error problems do not need to be
  treated as long as one directly study the relation between the
  response and the observed variables.
Finally, methods m3s and m3c both ignore the presence of the measurement
error and treat $W$ as $X$, so are naive methods.
m3s and m3c also differ in
terms of whether prediction is carried out using semiparametric or
conformal prediction,  please see Section \ref{sec:naive}
  for the details in the semiparametric method.
In implementing all the methods that require a
working model $\eta^*$, we adopted the correct
distribution in Simulation 1. Specifically, 
we selected 30 grid points $(x_j, j = 1, \ldots, 30)$ on the support of $X$, and
let the weights be $\eta^*(x_i) = \sum_{j = 1}^{30} I(x_i =
x_j) \eta_{\x}(x_j)/ \sum_{j =
  1}^{30}\eta_{\x}(x_j)$, where $\eta_{\x}$ is the density of the
scaled and shifted
${\rm Beta}(2, 2)$ distribution.

\begin{figure}
    \centering
    \caption{Simulation 1: Boxplots of the 90\% coverage probability (CP) and
      prediction interval length (PI) of the estimated prediction
      interval in the three models using the six methods. $n=100$.}\label{fig:simu11}
    \includegraphics[scale = 0.33]{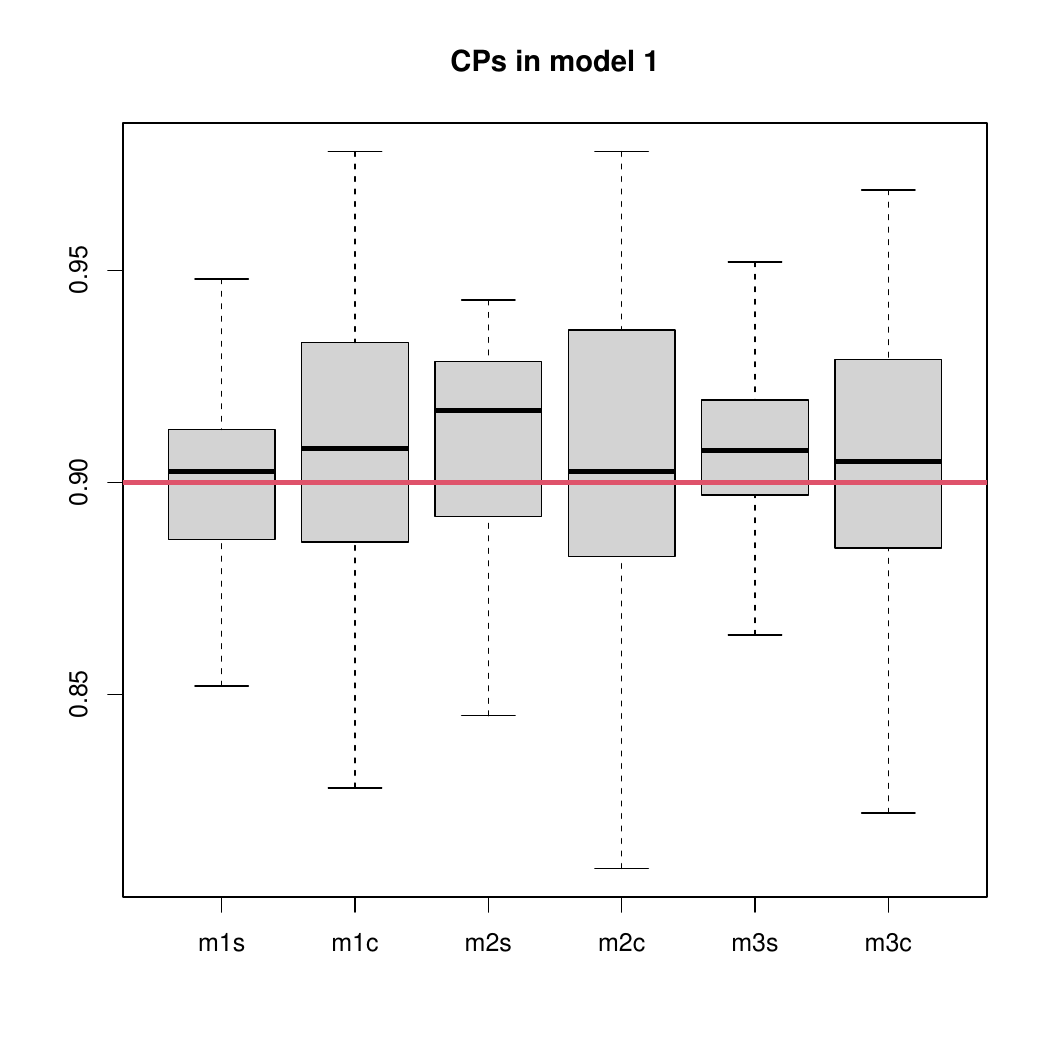}
    \includegraphics[scale = 0.33]{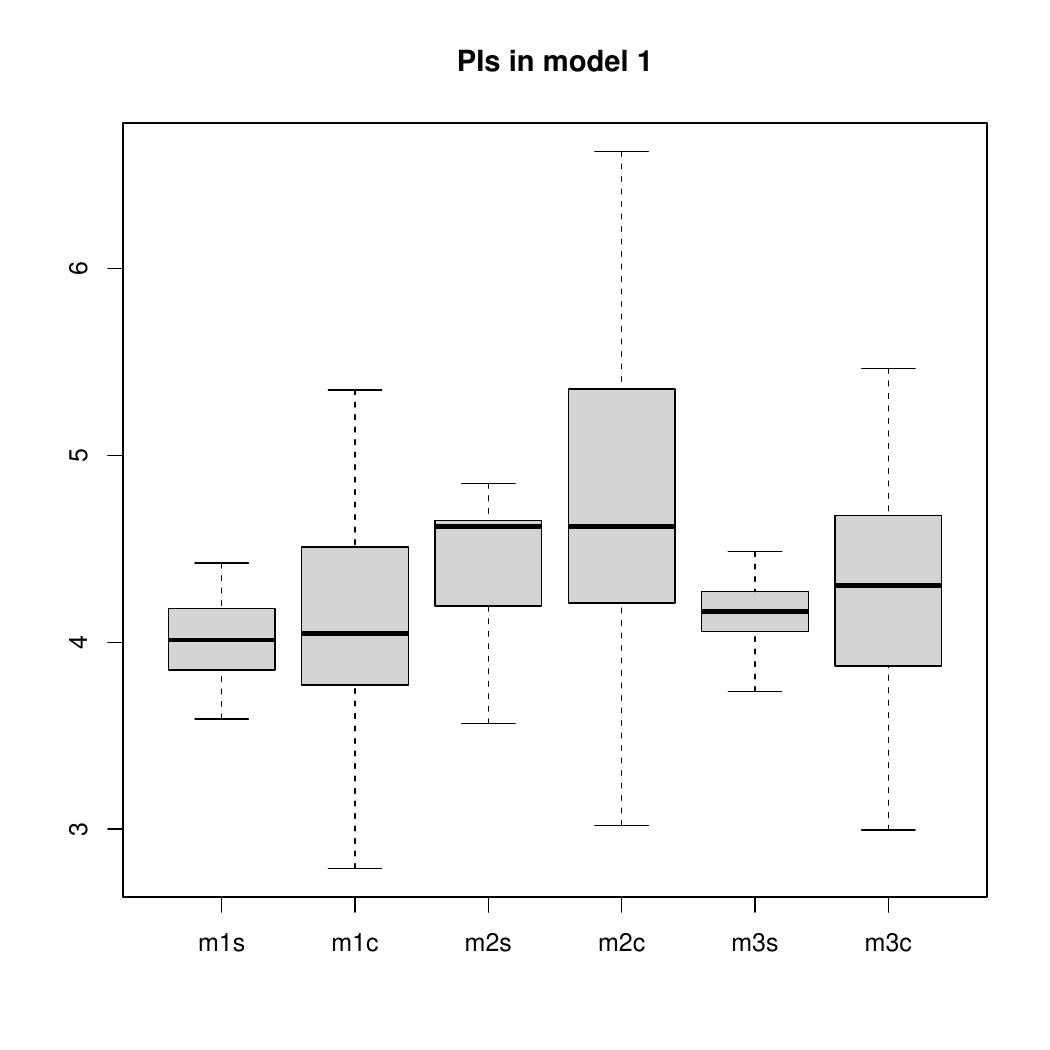}
        \includegraphics[scale = 0.33]{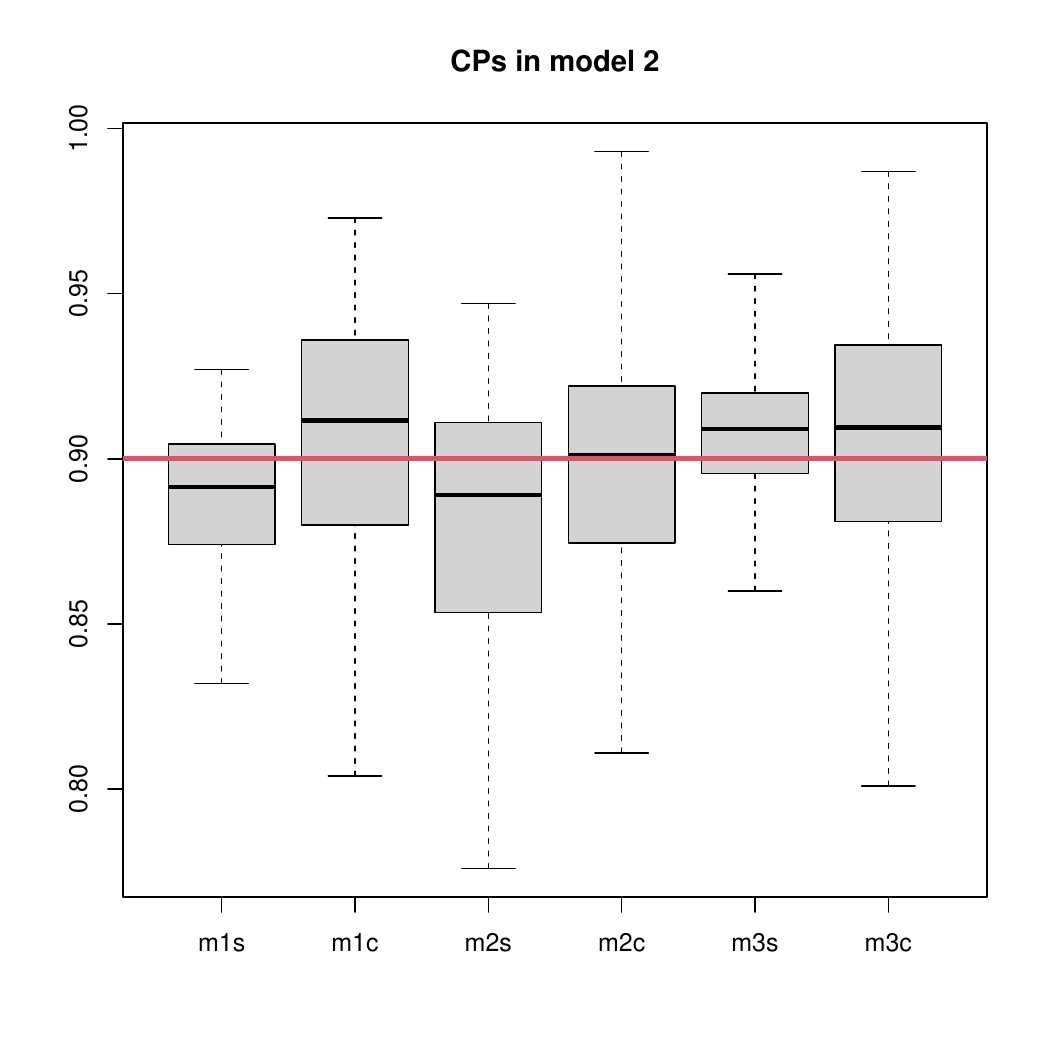}
        \includegraphics[scale = 0.33]{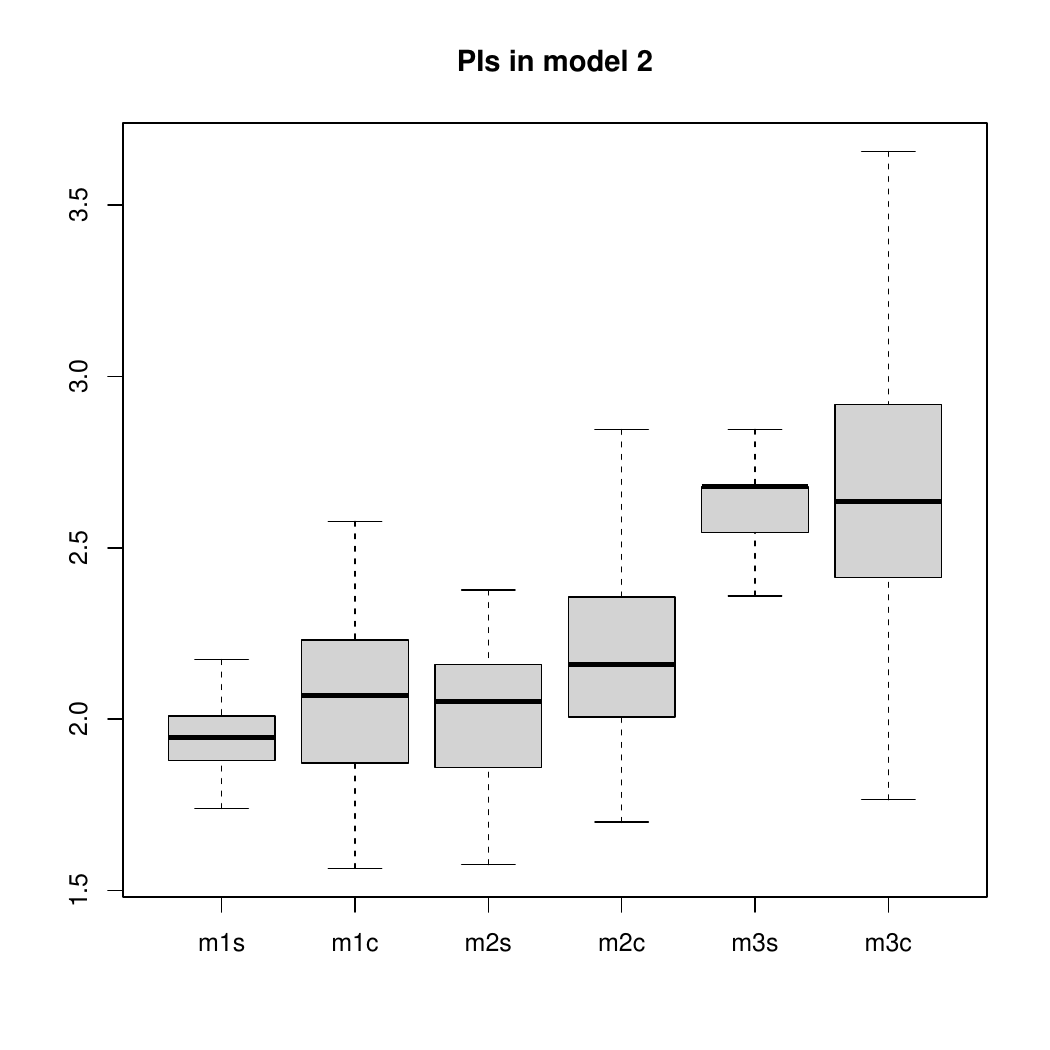}
            \includegraphics[scale = 0.33]{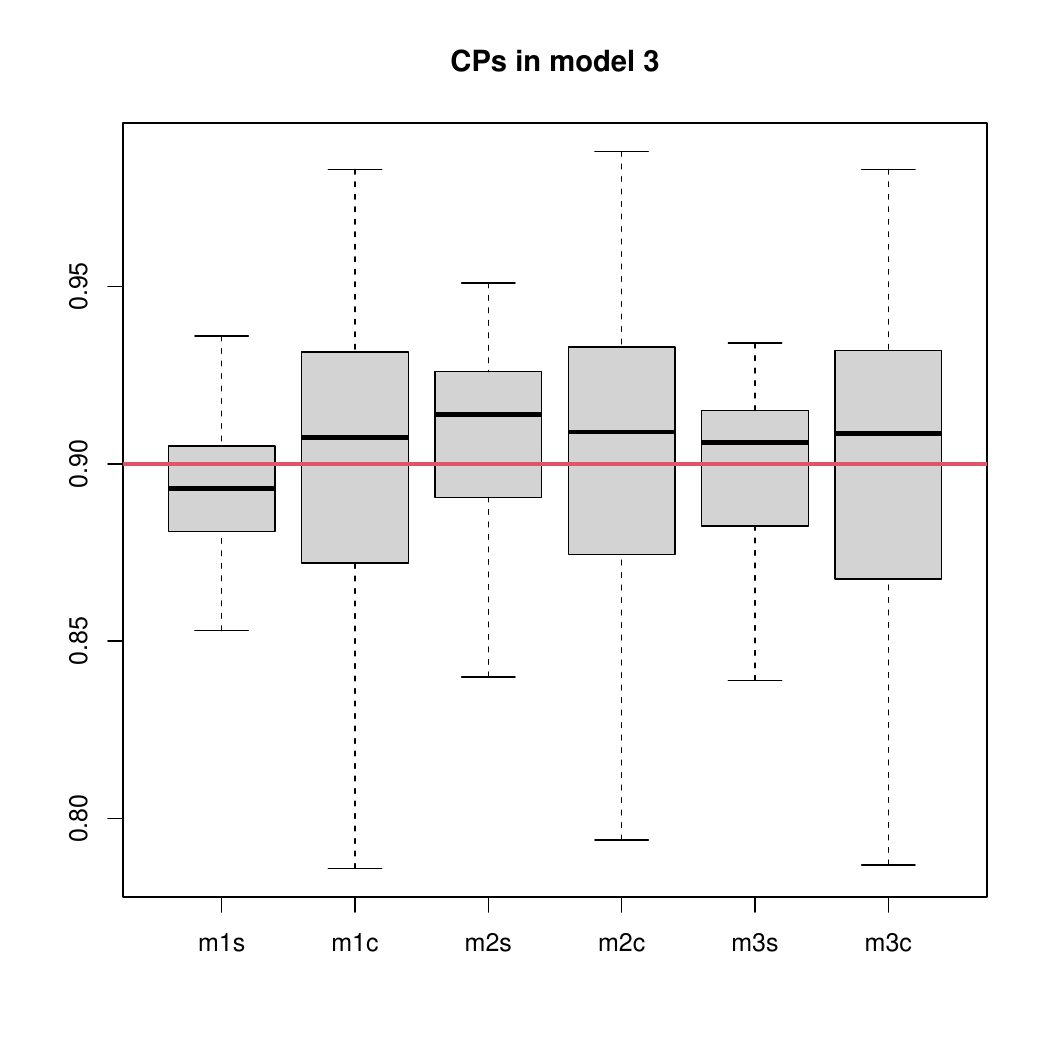}
     \includegraphics[scale = 0.33]{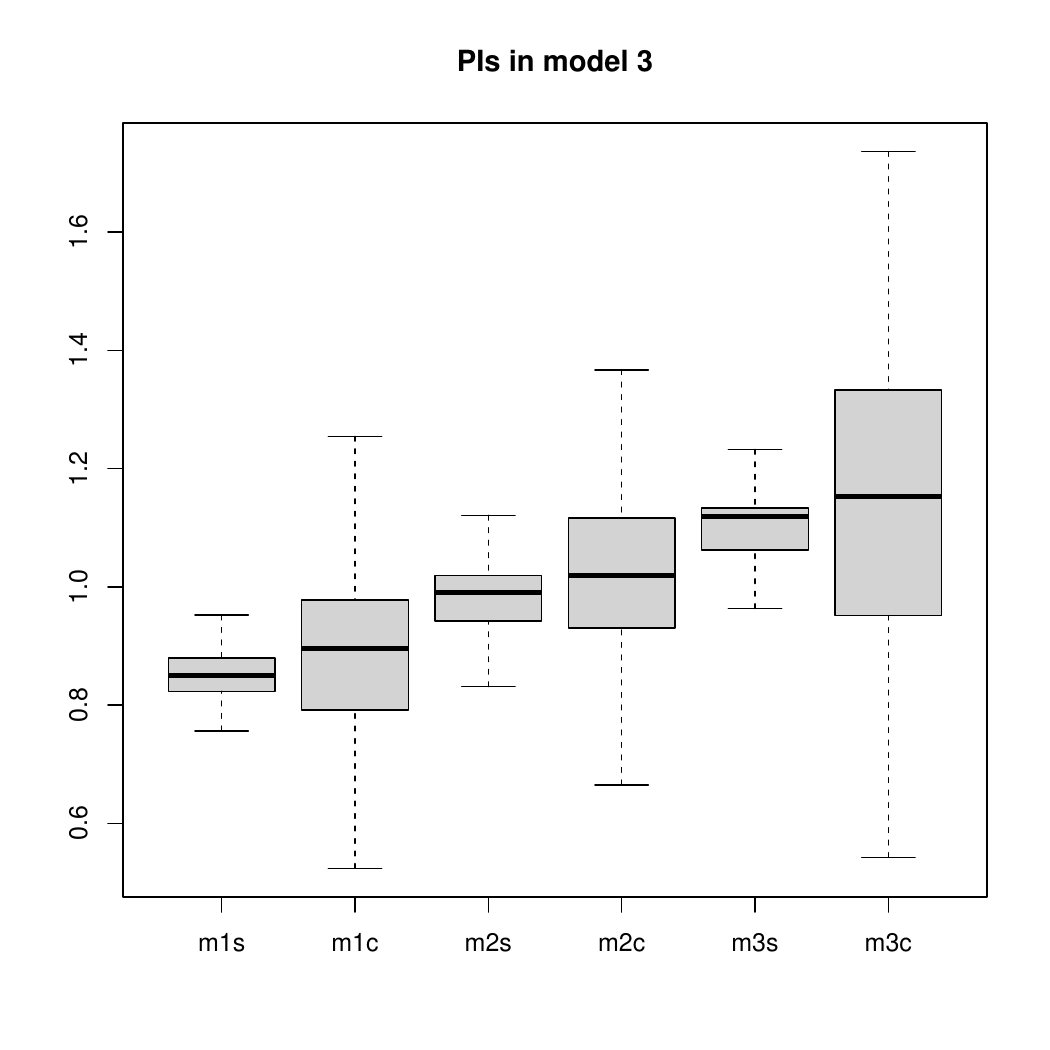}
   \end{figure}

\begin{figure}
    \centering
    \caption{Simulation 1: Boxplots of the 90\% coverage probability (CP) and
      prediction interval length (PI) of the estimated prediction
      interval in the three models using the six methods. $n=500$.}\label{fig:simu12}
    \includegraphics[scale = 0.33]{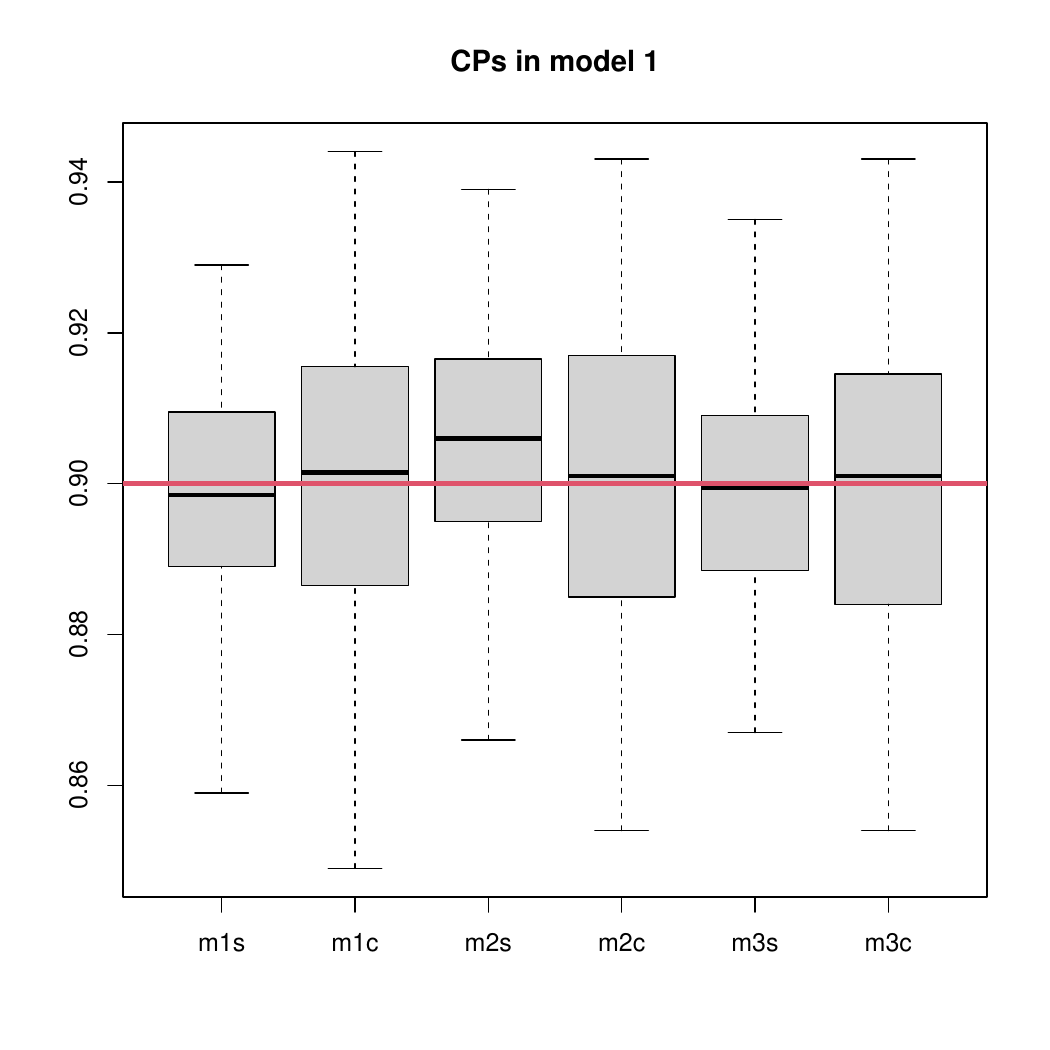}
    \includegraphics[scale = 0.33]{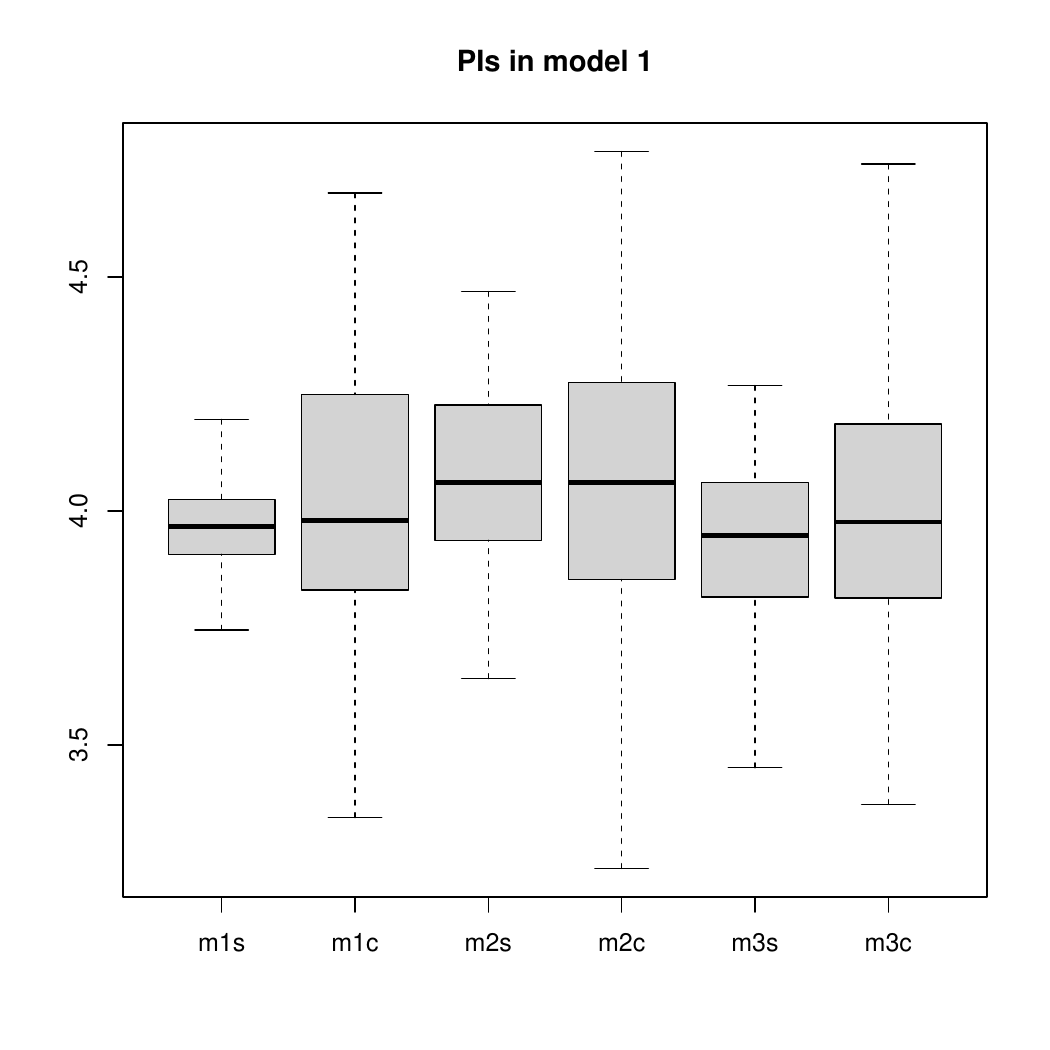}
        \includegraphics[scale = 0.33]{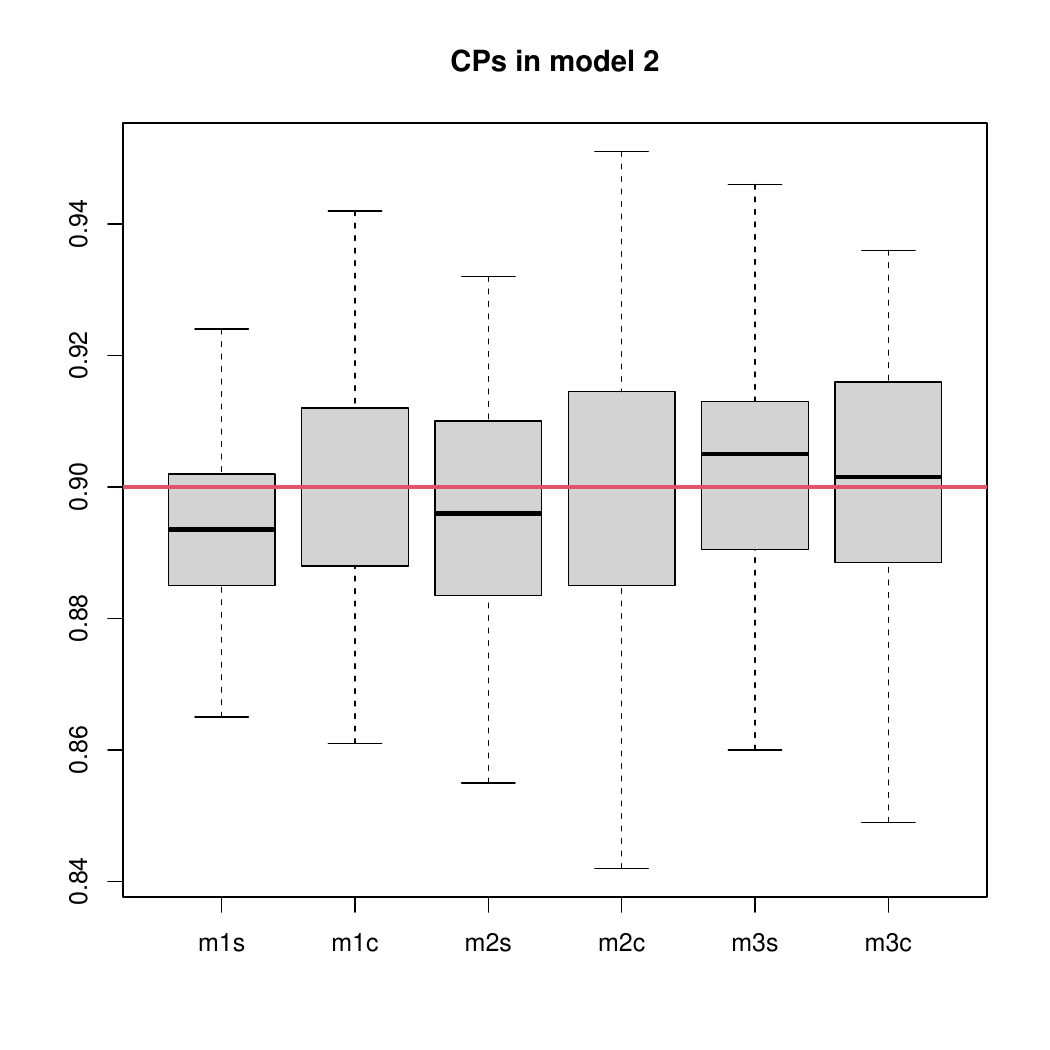}
        \includegraphics[scale = 0.33]{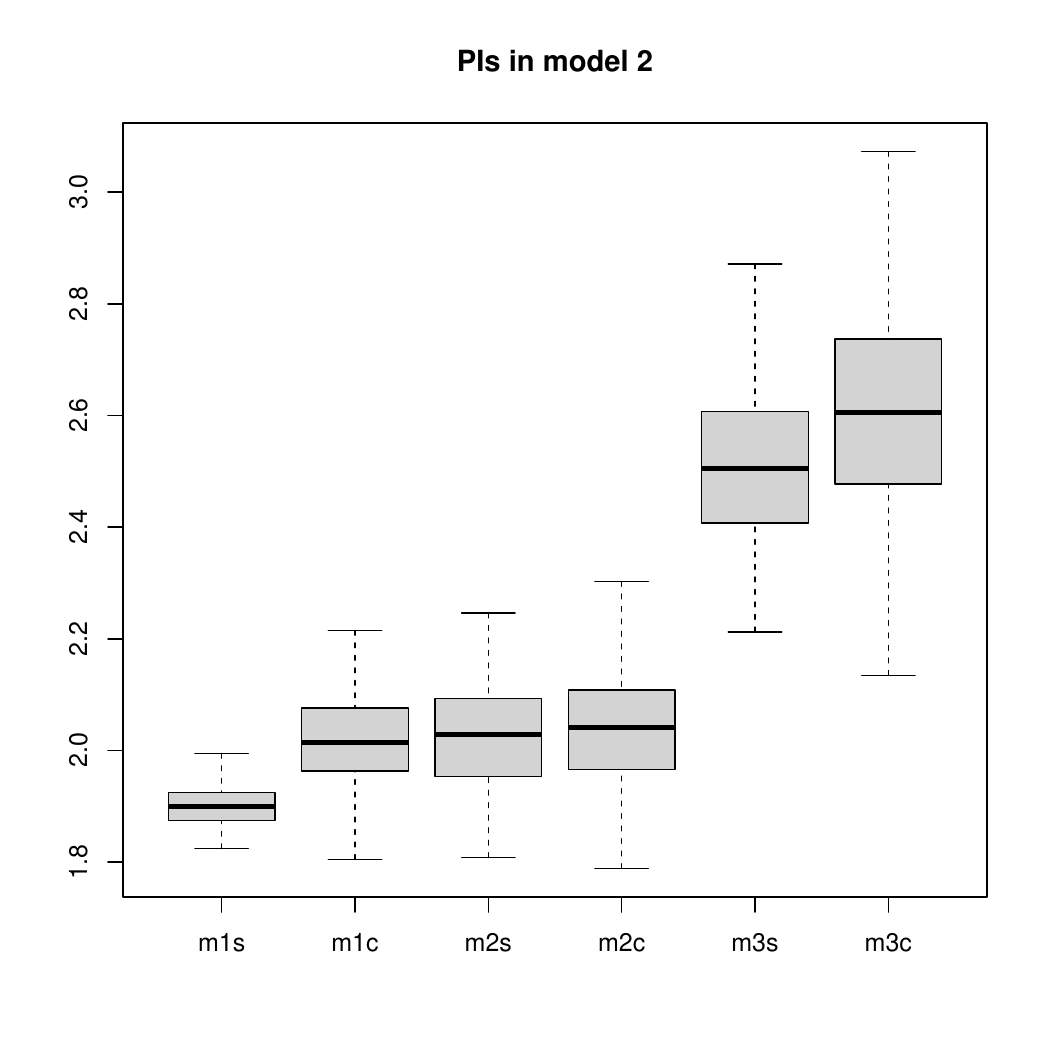}
            \includegraphics[scale = 0.33]{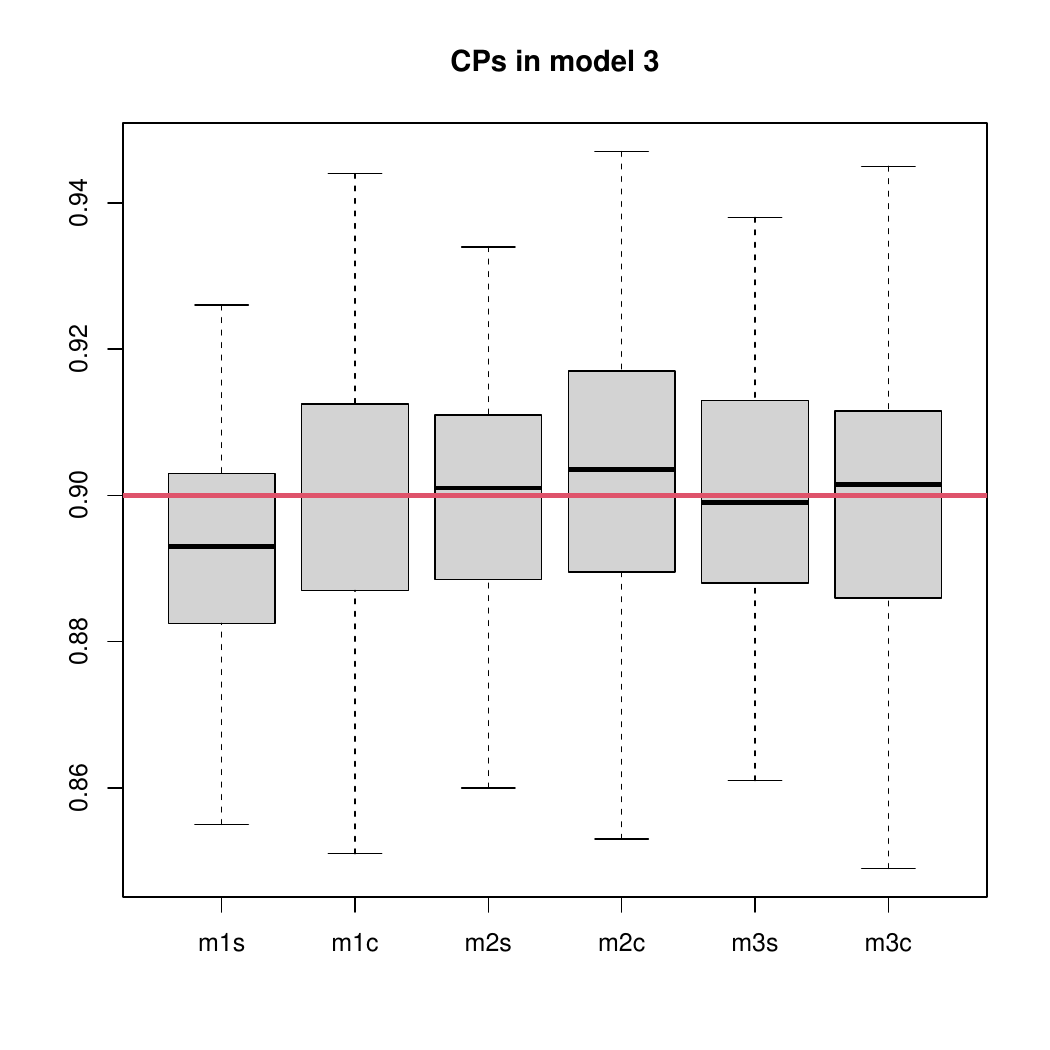}
     \includegraphics[scale = 0.33]{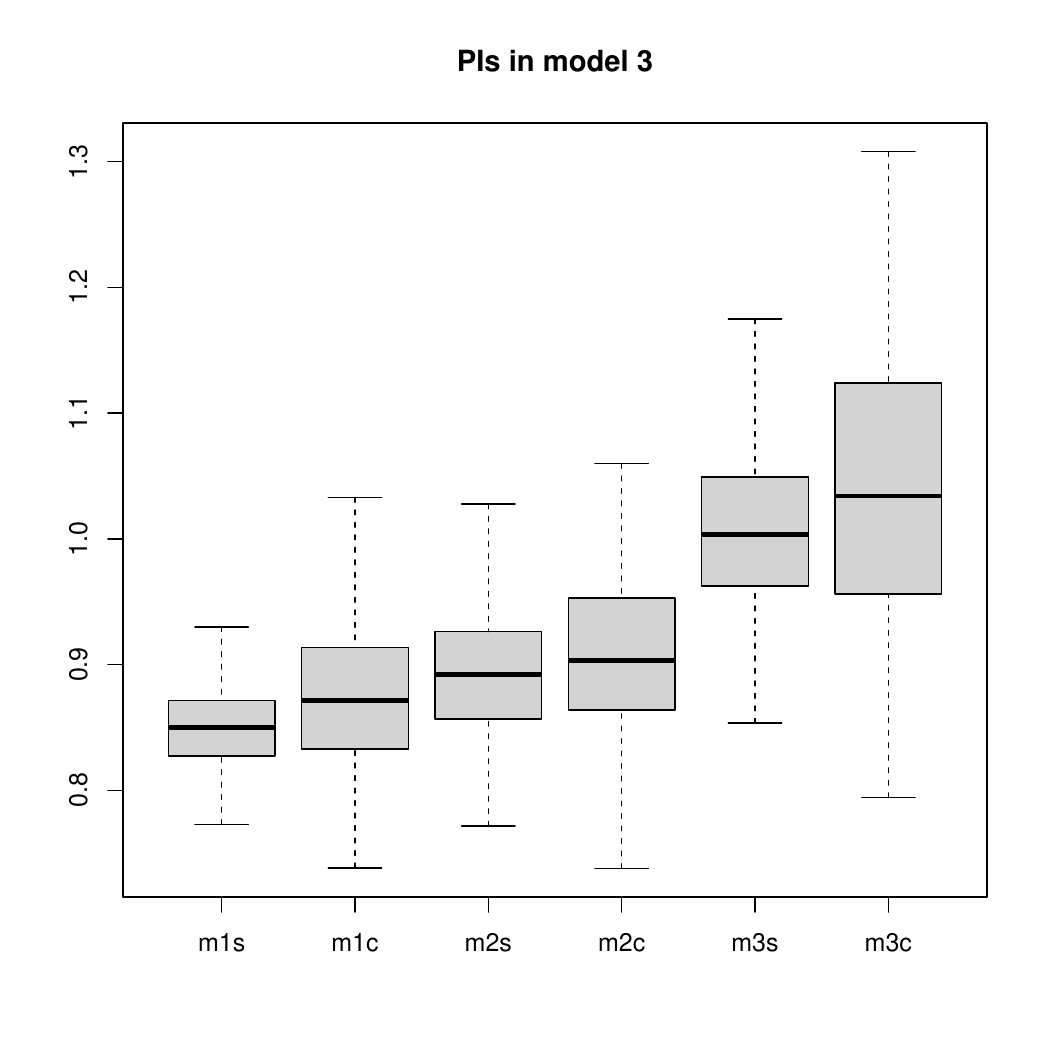}
    \end{figure}

Based on the $n$ estimated 90\% prediction intervals, 
we computed the coverage probability
(CP) and the average length (LPI), and provided the box-plots from the 100
simulations in Figures \ref{fig:simu11} 
and \ref{fig:simu12}. Further, in Table \ref{tab:sim1}, we 
present the mean and standard deviation of the 100 CPs, as well as
the mean  and standard deviation of the 100 
prediction interval lengths.
We can see that in general, all methods provide coverage close to
the nominal level 90\%, hence are all consistent.

In most cases,  the prediction intervals of all the consistent estimators
tend to be the shortest in m1s and m1c, where we used the model information.
They are the largest in m3c, where the measurement error issue is naively
ignored, and are in the middle for m2s and m2c, when the model
information is completely ignored. 
Within each of the three method classes, the
semiparametric approach leads to better performance, in that the box
is generally narrower, reflecting smaller variability. Among all six
methods, it is quite clear that m1s has the best performance, in terms
of its good coverage, short length and small variabilities.
The observed small variability agrees with our theory, because
when the working model is correct, under the same residual form, the
semiparametric method provides the most efficient estimation of the
prediction interval. Further, as we have pointed out,
among the three semiparametric methods
m1s, m2s, m3s, we can see that m1s has the best performance in
general. This indicates that it is beneficial to make use of the model
information and to take into account the measurement error issue.

We also conducted a simulation 2, where we adopted a misspecified
model for the distribution of $X$. In this simulation, we generated
$X$ via 
$X \sim N(-1, 1)$, while we choose 
$\eta^*$ to be a discrete uniform distribution function with 30
positive masses on $(\wh{\mu}_x -
3\wh{\sigma}_x, \wh{\mu}_x +
3\wh{\sigma}_x)$. Here $\wh{\mu}_x$ is the estimated mean of $X$
calculated by the sample average of $W$, and $\wh{\sigma}_x^2$ is the
estimated variance of $X$ calculated by the sample variance
of $W$ minus $\sigma_U^2$.  All other aspects of the simulation are
identical to those in Simulation 1. The results are in Table
\ref{tab:sim2}. Similar conclusions can be drawn as in Simulation 1.
Note that here because the distribution of $X$ is misspecified, there
is no theoretical guarantee that the prediction interval is optimally
estimated. However, we observe smaller variability throughout in
comparison to the corresponding conformal approach.

We also performed an additional simulation study to investigate the
performance when the model error distribution is
misspecified. The methods are very robust, while we provide the
simulation details in Appendix \ref{sec:moresimu}.

\begin{table}[!h]
  \caption{Simulation 1, true distribution for $X$. The average and
    standard deviation of the coverage
    probabilities (CP (SD)), and the average and standard deviation of 
    the  lengths  (LPI (SD)) of the estimated 90\% prediction
    intervals.}{\label{tab:sim1}}
  \tiny
  \begin{tabular}{c|c|c|c|c|c|c}
  \hline
    &m1s&m1c&m2s&m2c&m3s&m3c\\
\hline
       &\multicolumn{6}{c}{$n=100$}\\
     \hline
    &\multicolumn{6}{c}{ model 1}\\
    \hline
   CP (SD)&0.9 (0.021)&0.903 (0.042)&0.904 (0.035)&0.902 (0.046)&0.904 (0.022)&0.902 (0.041)\\
   LPI (SD)  &4.024 (0.241)&4.102 (0.596)&4.38 (0.457)&4.796 (0.945)&4.111 (0.275)&4.271 (0.607)\\
    \hline
      &\multicolumn{6}{c}{ model 2}\\
    \hline
     CP (SD)&0.889 (0.02)&0.904 (0.041)&0.88 (0.046)&0.896 (0.042)&0.906 (0.024)&0.905 (0.044)\\
     LPI (SD)&1.949 (0.095)&2.057 (0.231)&2.011 (0.202)&2.207 (0.266)&2.633 (0.197)&2.679 (0.402)\\
    \hline
        &\multicolumn{6}{c}{ model 3}\\
    \hline
     CP (SD)&0.895 (0.022)&0.897 (0.051)&0.902 (0.036)&0.9 (0.044)&0.898 (0.024)&0.899 (0.046)\\
     LPI (SD)  &0.862 (0.079)&0.881 (0.146)&0.959 (0.101)&1.035 (0.175)&1.076 (0.094)&1.133 (0.259)\\
    \hline
    &\multicolumn{6}{c}{$n=500$}\\
     \hline
    &\multicolumn{6}{c}{ model 1}\\
    \hline
   CP (SD)&0.896 (0.016)&0.9 (0.023)&0.904 (0.017)&0.9 (0.022)&0.899 (0.015)&0.9 (0.023)\\
   LPI (SD)&3.974 (0.106)&4.015 (0.29)&4.054 (0.22)&4.065 (0.309)&3.941 (0.187)&4.017 (0.301)\\
    \hline
     &\multicolumn{6}{c}{ model 2}\\
    \hline
     CP (SD)&0.894 (0.013)&0.9 (0.019)&0.895 (0.019)&0.899 (0.02)&0.903 (0.019)&0.901 (0.019)\\
     LPI (SD)  &1.901 (0.039)&2.022 (0.105)&2.026 (0.096)&2.041 (0.104)&2.511 (0.138)&2.603 (0.187)\\
    \hline
     &\multicolumn{6}{c}{model 3}\\
    \hline
     CP (SD)&0.893 (0.016)&0.899 (0.022)&0.899 (0.019)&0.902 (0.021)&0.899 (0.017)&0.899 (0.021)\\
     LPI (SD)  &0.852 (0.042)&0.873 (0.064)&0.888 (0.063)&0.906 (0.078)&1.001 (0.081)&1.044 (0.137)\\
    \hline
    \end{tabular}
  \end{table}

  \begin{table}[!h]
  \caption{ Simulation 2, misspecified distribution for $X$. 
The average and
    standard deviation of the coverage
    probabilities (CP (SD)), and the average and standard deviation of 
    the  lengths  (LPI (SD)) of the estimated 90\% prediction
    intervals.}{\label{tab:sim2}}
  \tiny
 \begin{tabular}{c|c|c|c|c|c|c}
  \hline
    &m1s& m1c &m2s&m2c&m3s&m3c\\
     \hline
      &\multicolumn{6}{c}{$n=100$}\\
     \hline
    &\multicolumn{6}{c}{ model 1}\\
    \hline
      CP (SD)&0.893 (0.017)&0.898 (0.038)&0.915 (0.026)&0.897 (0.038)&0.908 (0.019)&0.904 (0.04)\\
        LPI (SD)&2.91 (0.215)&3.087 (0.593)&3.407 (0.343)&3.673 (0.962)&2.843 (0.218)&2.952 (0.534)\\
    \hline
    &\multicolumn{6}{c}{ model 2}\\
     \hline
    CP (SD)& 0.895 (0.019)&0.91 (0.036)&0.906 (0.032)&0.907 (0.041)&0.913 (0.02)&0.909 (0.034)\\
        LPI (SD) &  1.544 (0.086)&1.66 (0.215)&1.818 (0.192)&1.975 (0.372)&1.98 (0.201)&2.052 (0.343)\\
   \hline
    &\multicolumn{6}{c}{model 3}\\
      \hline
    CP (SD)& 0.907 (0.021)&0.906 (0.038)&0.905 (0.041)&0.904 (0.039)&0.904 (0.029)&0.902 (0.039)\\
 LPI (SD) &   0.907 (0.056)&0.92 (0.125)&1.012 (0.12)&1.091 (0.164)&1.141 (0.085)&1.185 (0.241)\\
   \hline
       &\multicolumn{6}{c}{$n=500$}\\
     \hline
    &\multicolumn{6}{c}{ model 1}\\
    \hline
    CP (SD)&0.893 (0.013)&0.898 (0.021)&0.894 (0.02)&0.898 (0.021)&0.897 (0.014)&0.896 (0.022)\\
        LPI (SD)&2.79 (0.11)&2.907 (0.293)&2.68 (0.208)&2.802 (0.28)&2.618 (0.16)&2.636 (0.241)\\
   \hline
 &\multicolumn{6}{c}{ model 2}\\
     \hline
    CP (SD)&0.901 (0.013)&0.898 (0.023)&0.891 (0.022)&0.897 (0.023)&0.902 (0.016)&0.898 (0.022)\\
         LPI (SD)  &   1.613 (0.042)&1.607 (0.103)&1.555 (0.129)&1.632 (0.125)&1.771 (0.145)&1.893 (0.253)\\
      
   \hline
   &\multicolumn{6}{c}{ model 3}\\
 \hline
    CP (SD)&0.9 (0.012)&0.897 (0.024)&0.896 (0.02)&0.898 (0.024)&0.897 (0.022)&0.897 (0.021)\\
         LPI (SD) & 0.884 (0.027)&0.887 (0.067)&0.883 (0.059)&0.916 (0.067)&1.022 (0.086)&1.054 (0.131)\\
   \hline

    \end{tabular}
  \end{table}

   \begin{figure}
    \centering
    \caption{Simulation 2: Boxplots of the 90\% coverage probability (CP) and
      prediction interval length (PI) of the estimated prediction
      interval in the three models using the six methods. $n=100$.}\label{fig:simu21}
    \includegraphics[scale = 0.33]{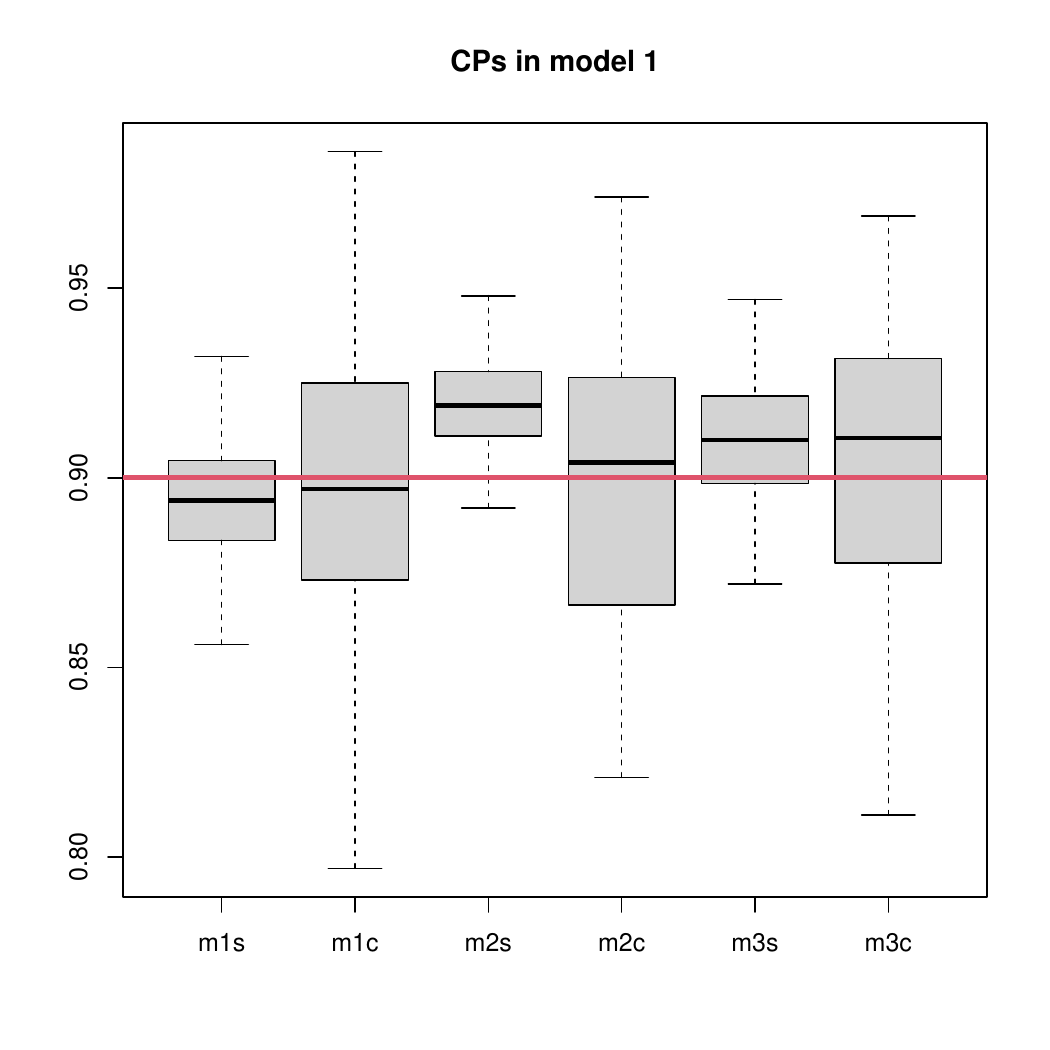}
    \includegraphics[scale = 0.33]{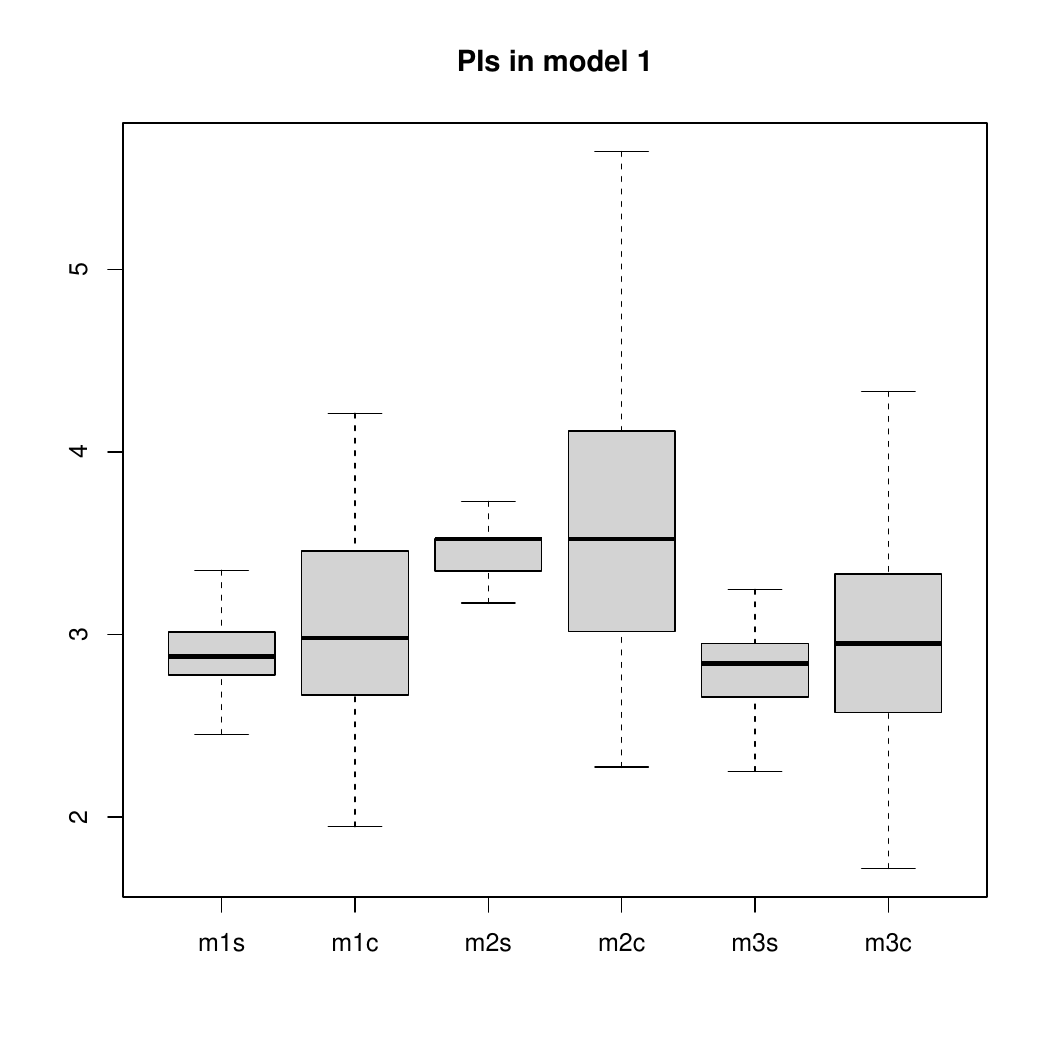}
        \includegraphics[scale = 0.33]{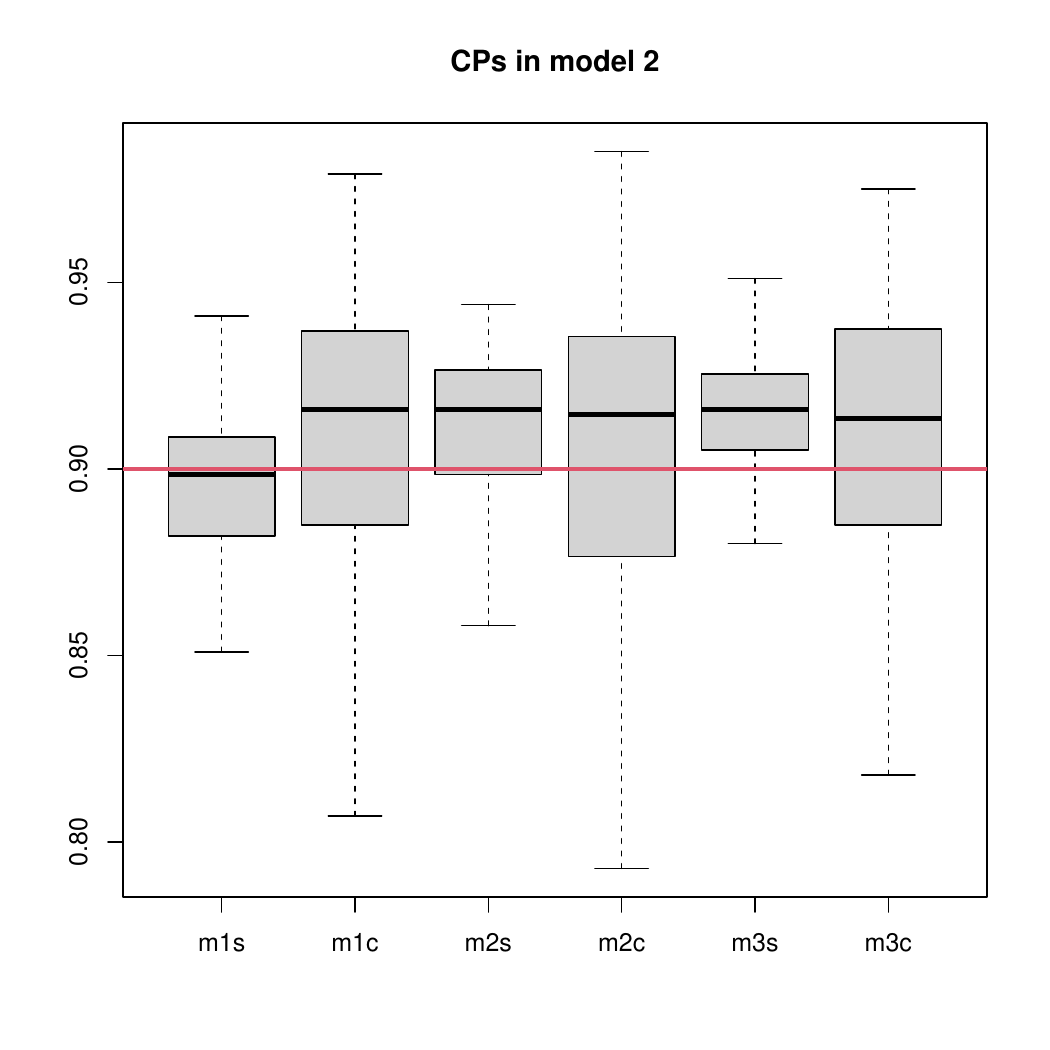}
        \includegraphics[scale = 0.33]{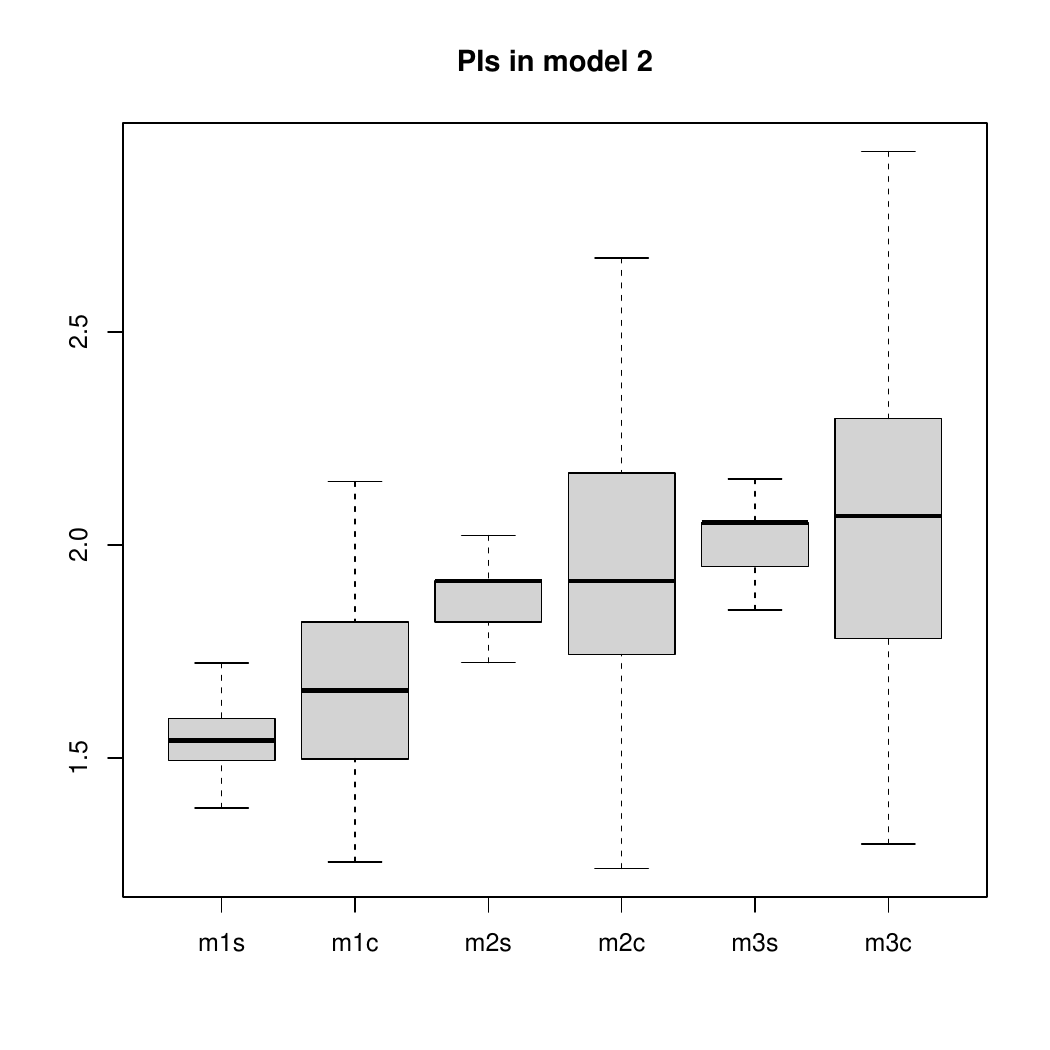}
            \includegraphics[scale = 0.33]{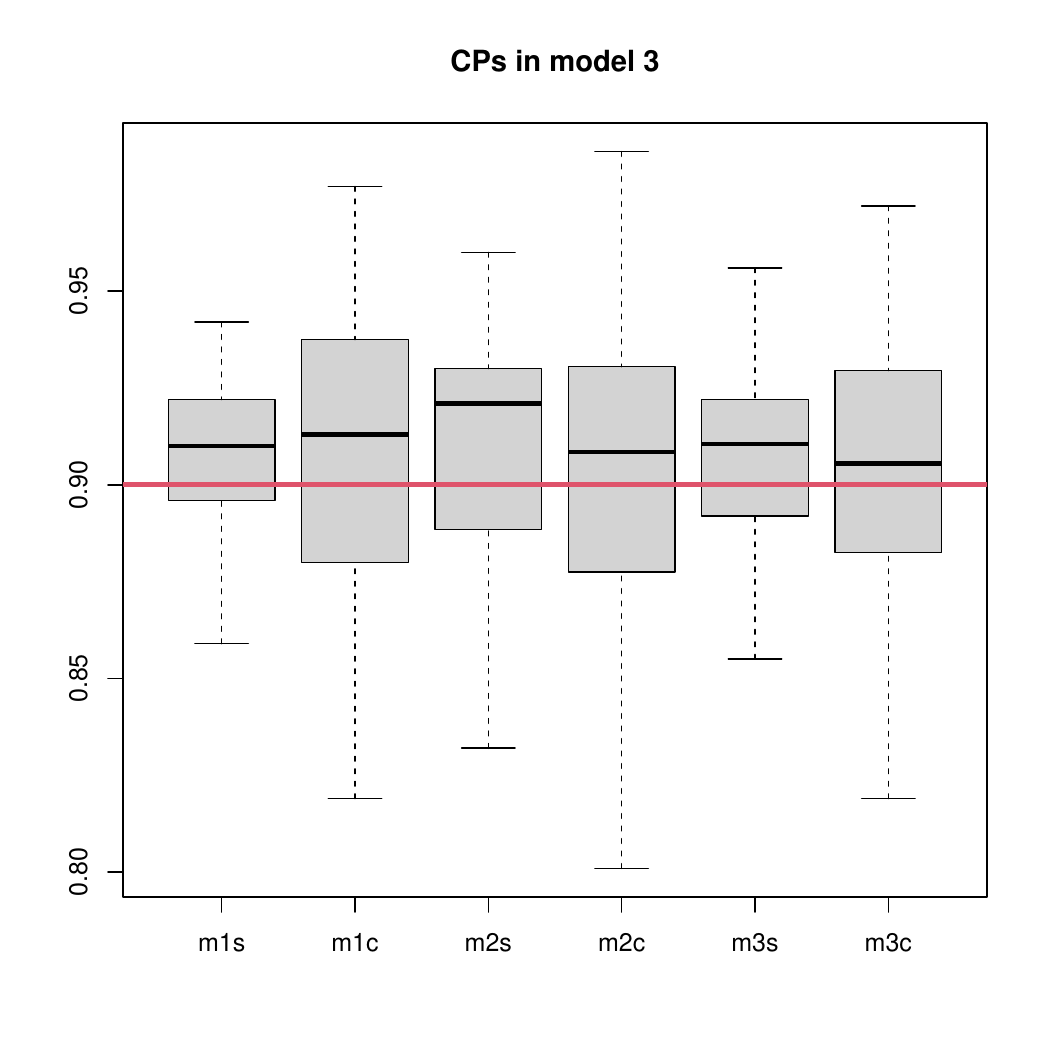}
     \includegraphics[scale = 0.33]{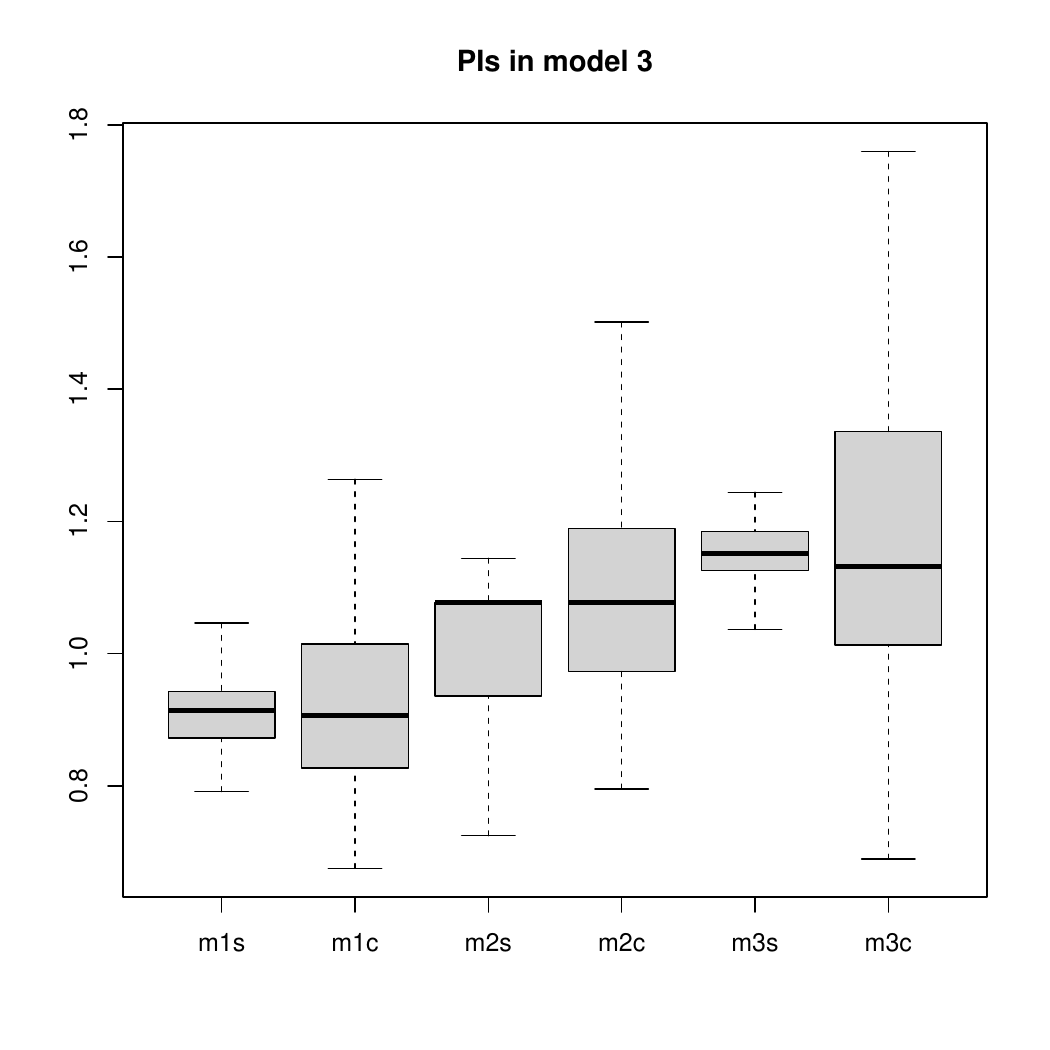}
   \end{figure}

     \begin{figure}
    \centering
    \caption{Simulation 2: Boxplots of the 90\% coverage probability (CP) and
      prediction interval length (PI) of the estimated prediction
      interval in the three models using the six methods. $n=500$.}\label{fig:simu22}
    \includegraphics[scale = 0.33]{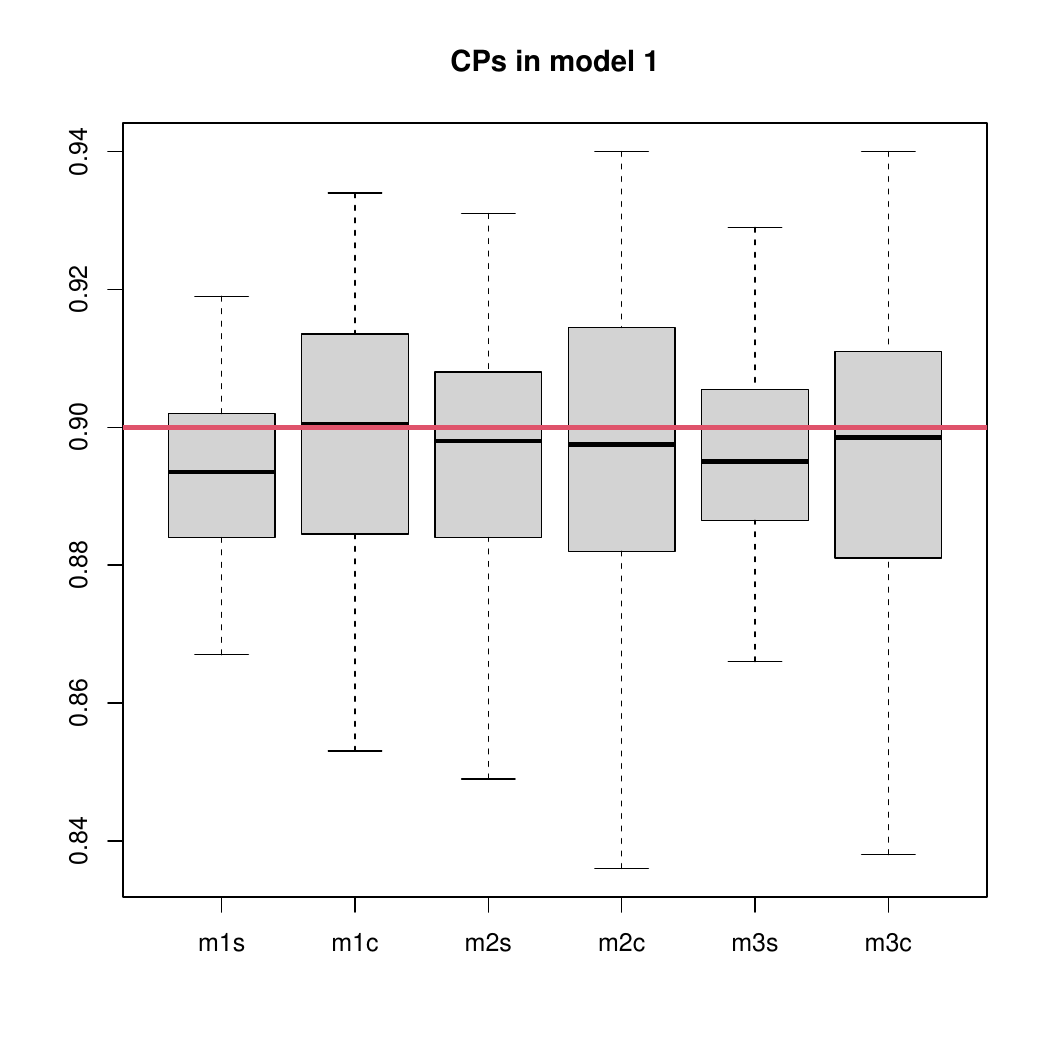}
    \includegraphics[scale = 0.33]{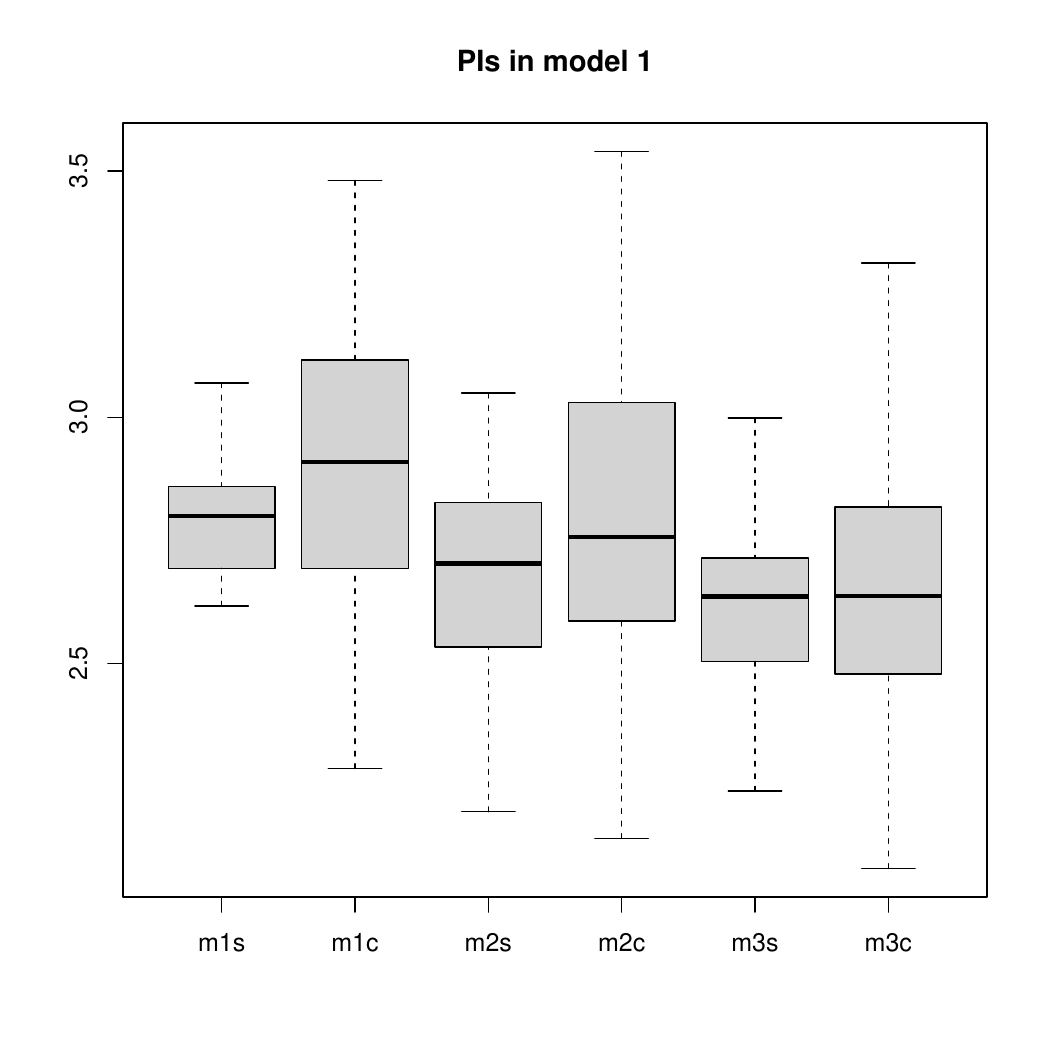}
        \includegraphics[scale = 0.33]{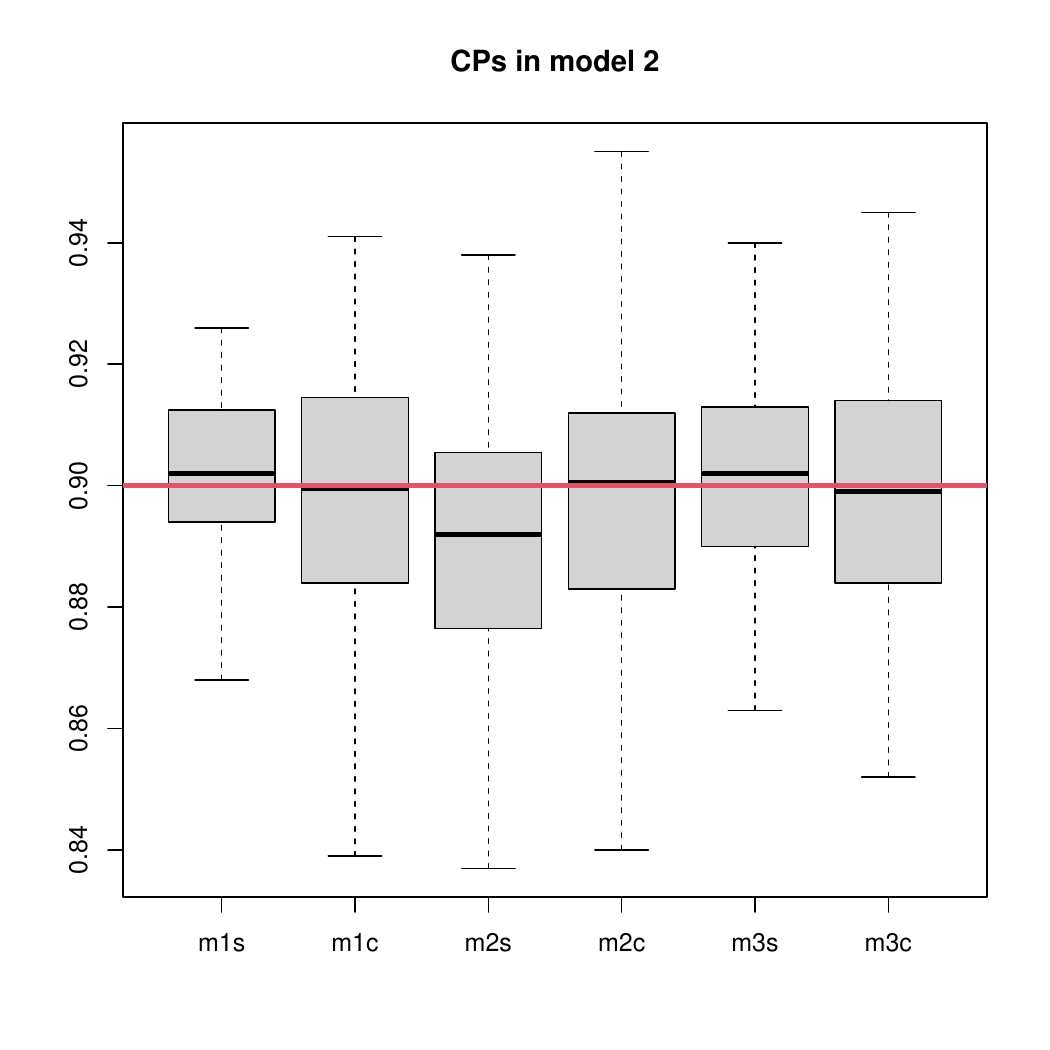}
        \includegraphics[scale = 0.33]{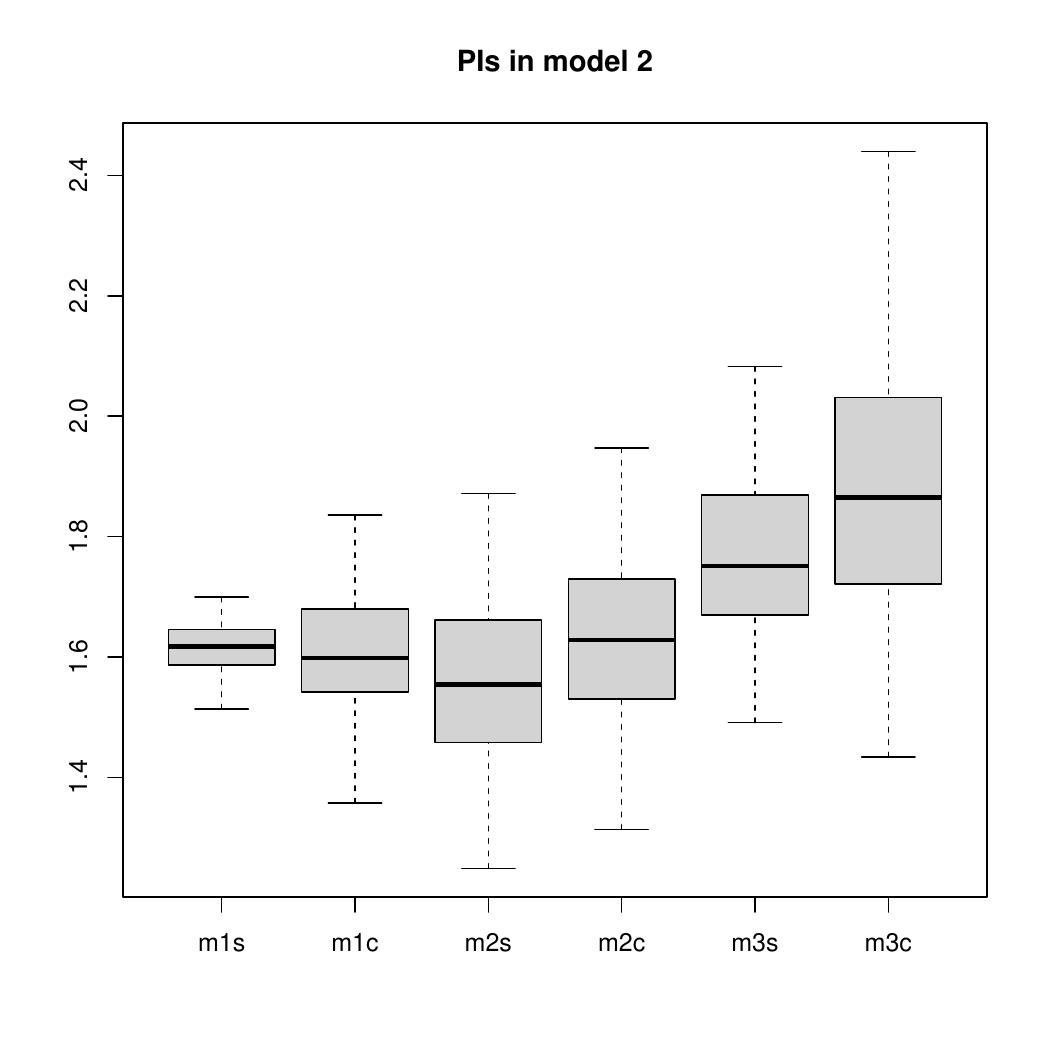}
            \includegraphics[scale = 0.33]{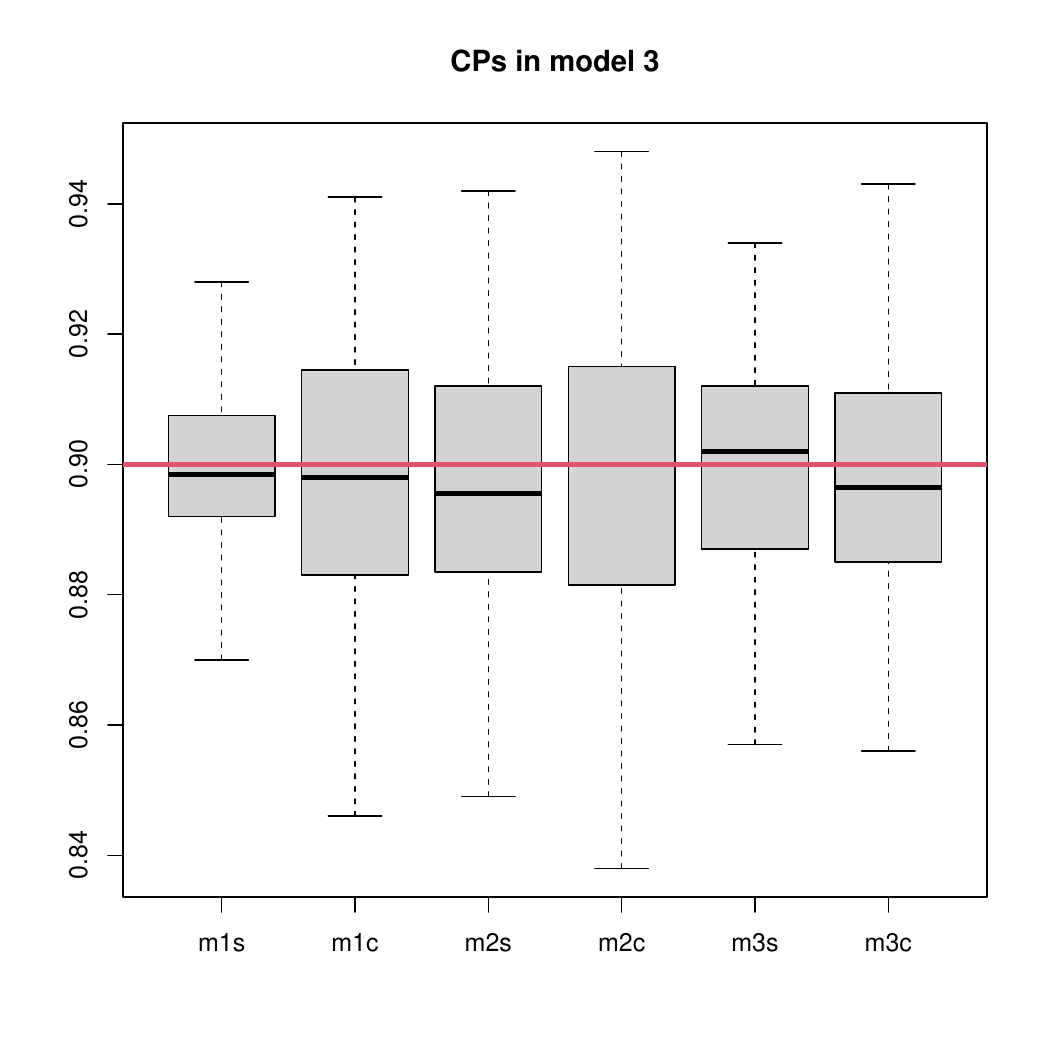}
     \includegraphics[scale = 0.33]{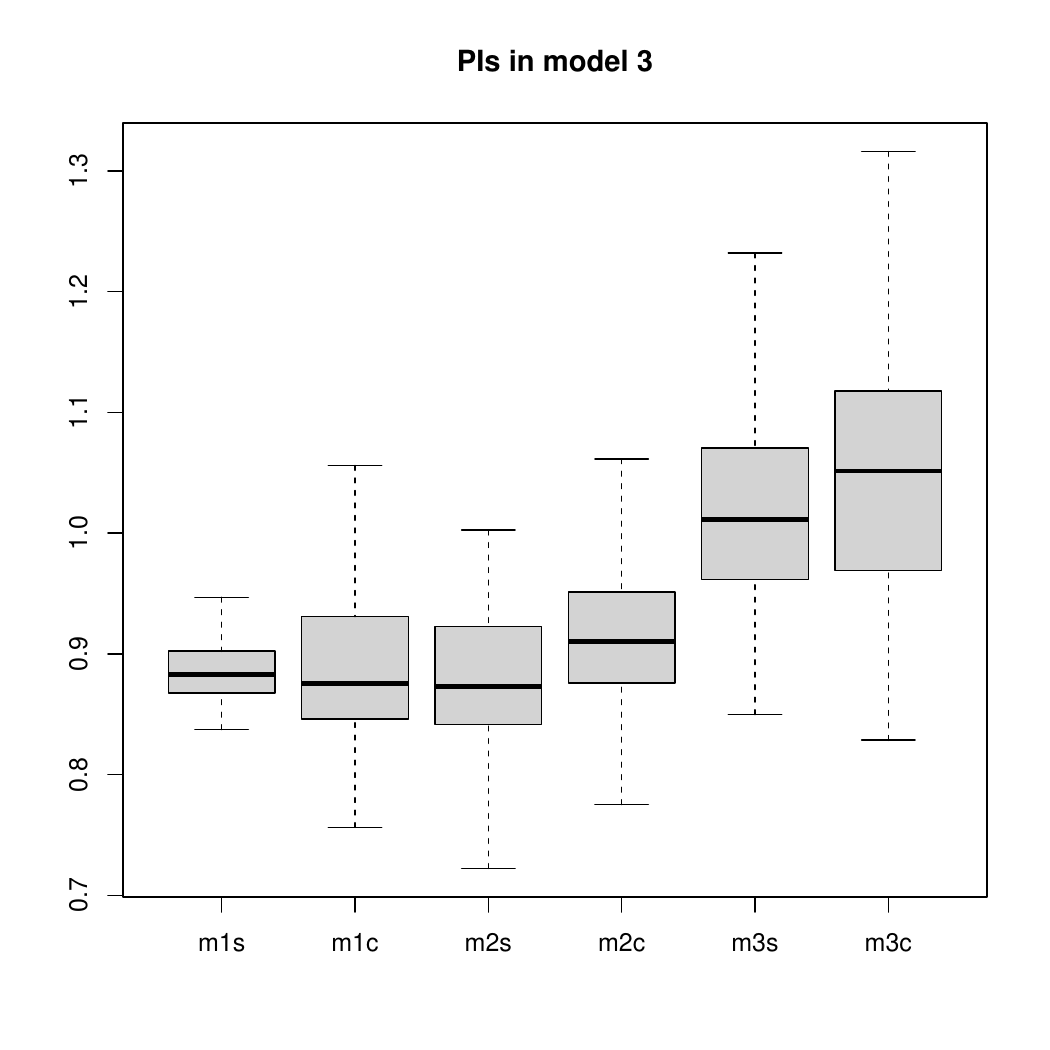}
   \end{figure}

   \section{Real data analysis}
   Alzheimer's disease (AD) is the most common cause of dementia and
   is characterized by accumulation of amyloid-$\beta$ ($A\beta$)
   plaques in the earliest phase of the disease
   \citep{masters2015blennow, scheltens2016alzheimer}. Two established
   methods for detecting the presence of $A\beta$  pathology are
   reduced concentrations of $A\beta 1-42$ ($A\beta42$) in
cerebrospinal fluid (CSF)  and increased retention of $A\beta$
positron emission tomography (PET)   tracers
\citep{mattsson2017clinical}.  It is often
assumed that CSF $A\beta42$ and $A\beta$  PET can be used
interchangeably, because there are mounting evidences showing that 
PET and CSF biomarkers are strongly associated
\citep{schipke2017correlation, leuzy2016pittsburgh,
  palmqvist2015detailed}. 
It is natural to ask how accurate it
would be if we use one marker, for example $A\beta$ PET  to predict
the other. Ideally, if one perfectly predicts the other, patients will no
longer need to take  multiple examinations, which
reduces the chance of side
effect and lowers patients' psychological stress. However, prediction based on 
neuroimage data is challenging because neuroimage data
are often subject to measurement errors resulting from  the data acquisition and
processing steps.

\begin{figure}
  \includegraphics[scale = 0.33]{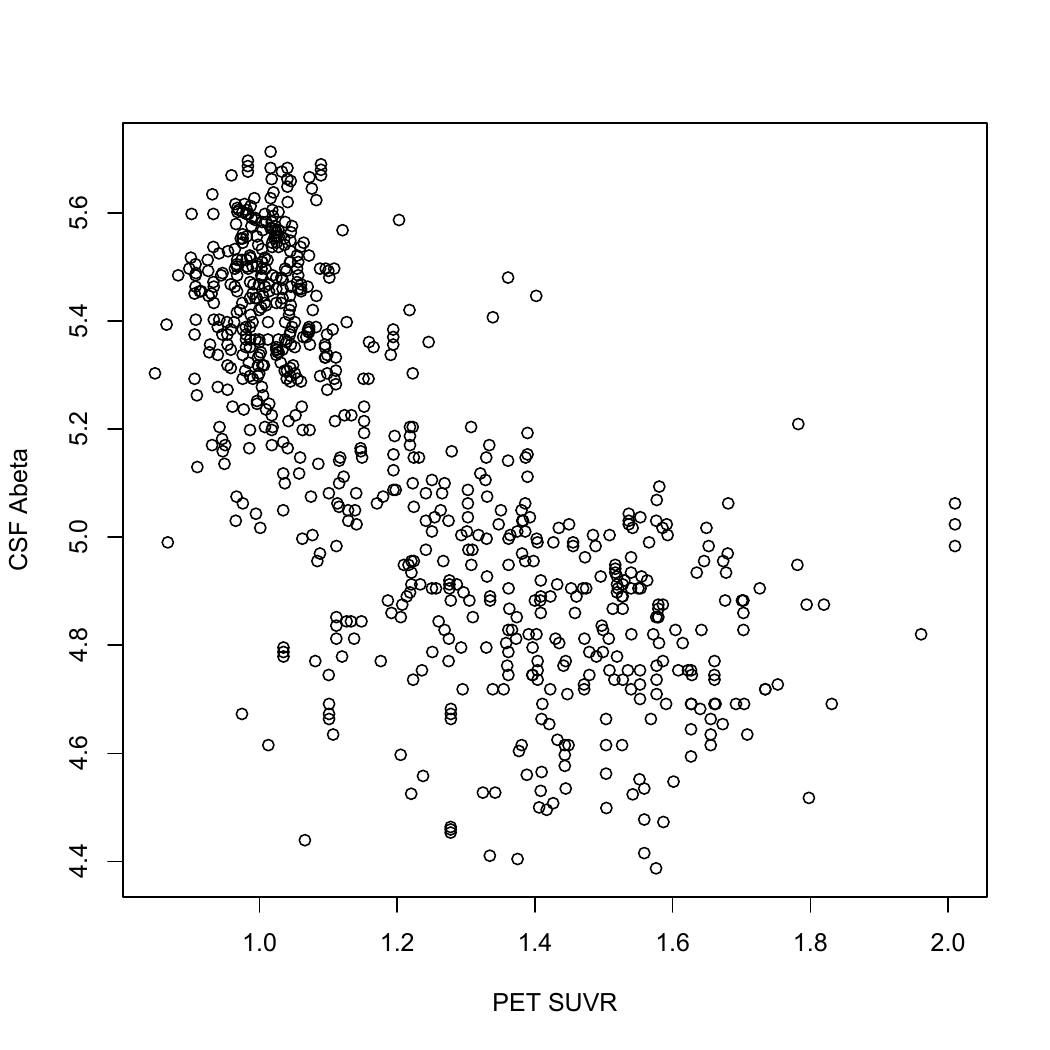}
  \includegraphics[scale = 0.33]{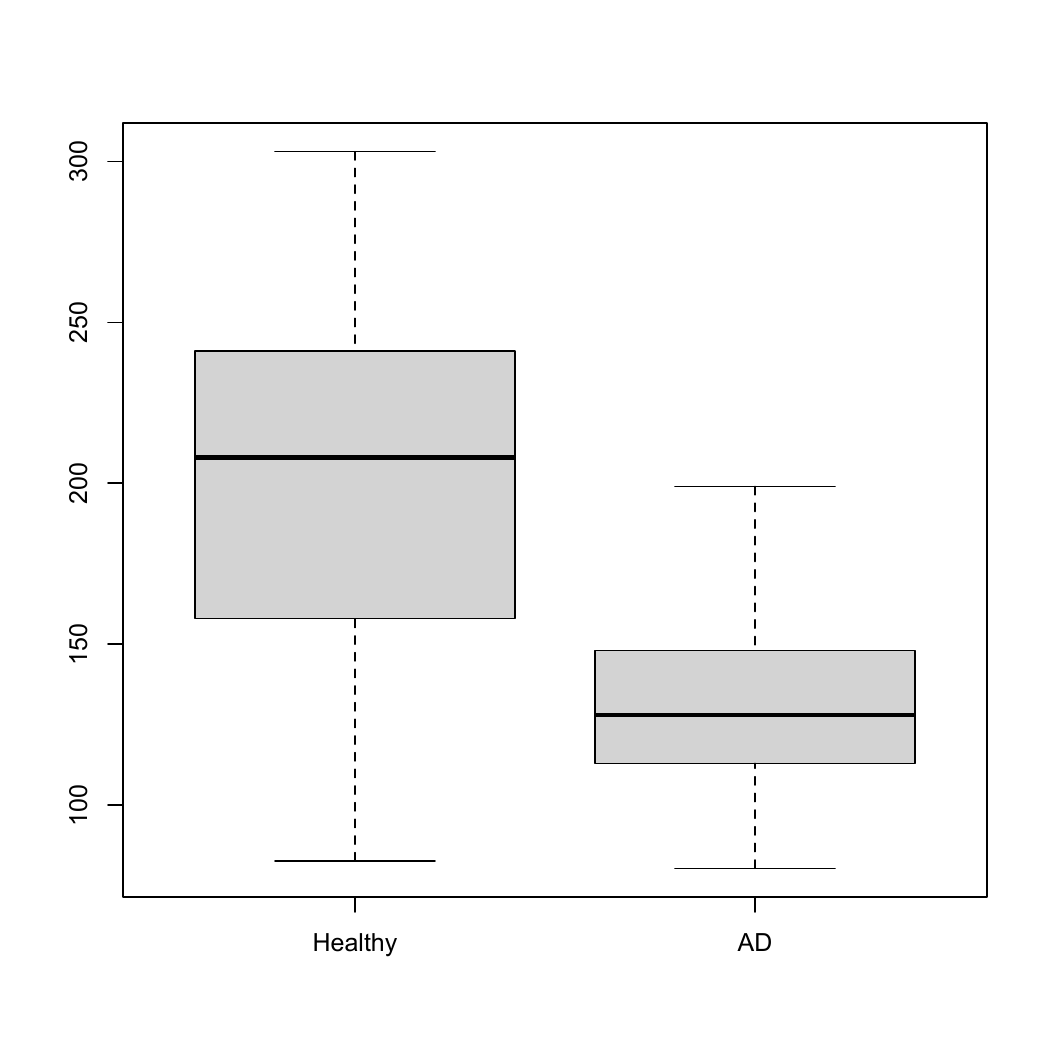}
  \centering
  \caption{Left: The scatterplots of CSF $A\beta$ and SUVR from
    amyloid PET. Right: The distribution of CSF values in healthy and
    AD subjects. }\label{fig:dataplots}
\end{figure}

We utilize the prediction interval methods m1s--m3c discussed in
Section \ref{sec:simu} to study the performance of using the standardized
uptake value ratio  (SUVR) of florbetapir (radiotracer) from the  $A\beta$ PET
to predict the 
$A\beta42$ from CSF. We download the preprocessed CSF 
\citep{shaw2016overview} and florbetapir  PET
\citep{landau2021} data  from the Alzheimer's Disease Neuroimaging
Initiative (ADNI) phase 2/Go study, from which we obtain a
subsample with $A\beta42$  values from the CSF tests that were taken
within 30 days of the PET
scans. Furthermore, we only retain the healthy subjects and the subjects with
Alzheimer's disease who had their diagnoses
within 30 days of the CSF tests. After removing missing data, we
obtain  669 
subjects (48.5\% female) with an average age of 74.5 years,  where
388 are classified as CSF negative subjects at
 the risk of cognitive impairment, and among which 
220 are subjects with AD. Let $Y$ be 
the logarithm of  $A\beta42$ from CSF, $W$ be the SUVR from PET, $\Z = \{1,
{\rm gender}, \log({\rm age})\}\trans$.  The scatter plot
  in the left panel of Figure \ref{fig:dataplots}
indicates an approximate
quadratic relationship between  $Y$ and $W$, therefore we
assume $m(X, \Z, \bb)= (X, X^2)\bb_1 + \Z\trans\bb_2$,
where $X$ is the underlying error free covariate. 
The box plots in the right panel of
Figure
\ref{fig:dataplots} indicate that there is a significant
difference in CSF values 
between healthy (199.748 $\pm$ 51.30)  and AD groups (135.465 $\pm$
37.57), with the
p-value from a student t-test to be less than 0.0001.

\begin{figure}
  \centering
  \includegraphics[scale = 0.33]{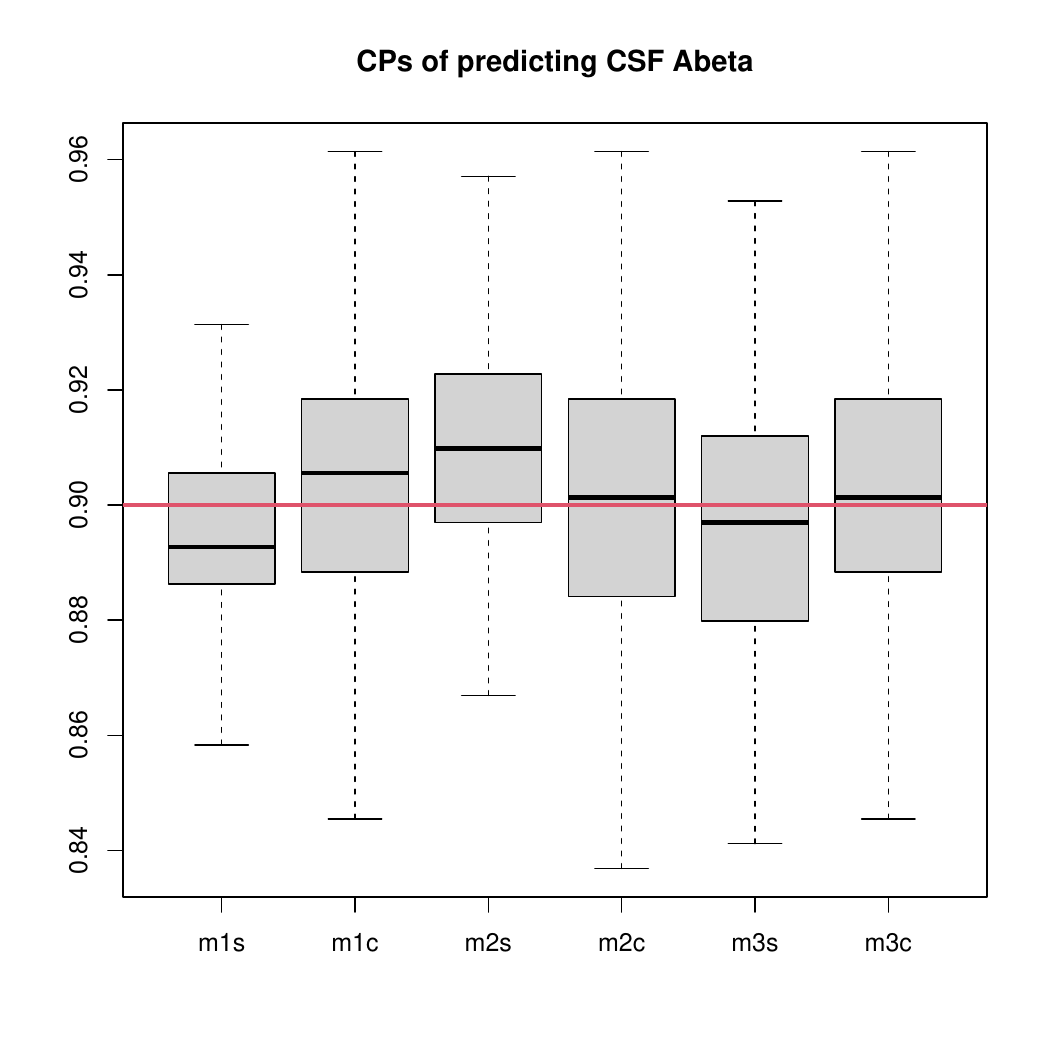}
  \includegraphics[scale = 0.33]{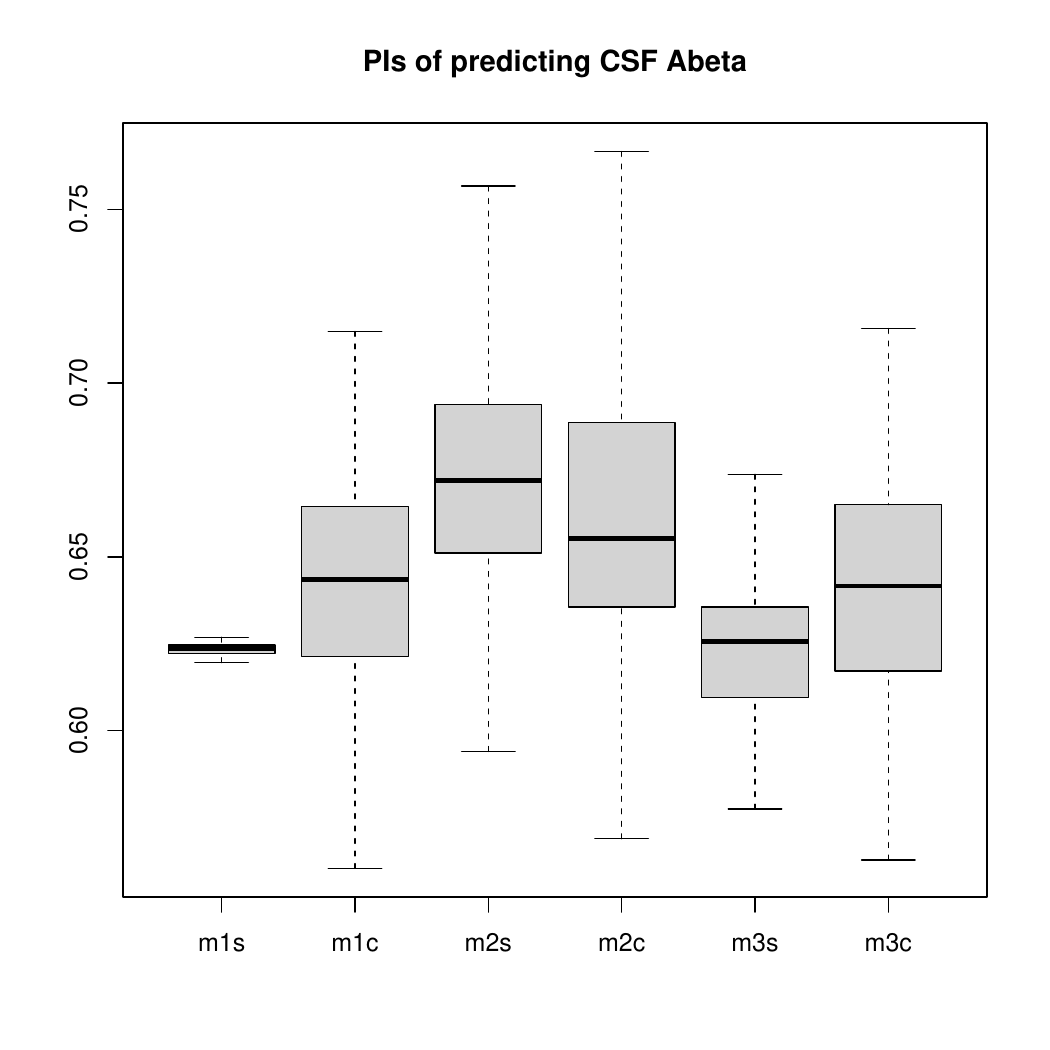}
  \caption{Boxplots of the 90\% coverage probability (CP) and
      prediction interval length (PI) of the estimated prediction
      interval using the six methods m1s--m3c over 100 cross validations.  }\label{fig:realCPPI}
  \end{figure}

We randomly sample two thirds of the subjects to construct
  the prediction interval  and 
evaluate the coverage probability on the remaining one third. We
repeat
the above cross validation process 100 times to
compare the performance of m1s--m3c.   In m1s and
m1c,  we choose 
$\eta^*$ to be a normal probability density function and assign 30
positive masses evenly spaced on $(\wh{\mu}_x -
r_1\wh{\sigma}_x, \wh{\mu}_x +
r_2\wh{\sigma}_x)$. Here $\wh{\mu}_x$ is the estimated mean of $X$
calculated by the sample average of $W$, and $\wh{\sigma}_x^2$ is the
estimated variance of $X$ calculated by the sample variance
of $W$ minus $\sigma_U^2$. We set $\sigma_U$ to be 10\% of the
standard deviation of $W$ based on empirical knowledge and experience.
 Furthermore, following \cite{romano2019conformalized} we calibrate
 the tuning parameters 
$\sigma_{\epsilon}, r_1, r_2$ (in m1s, m1c) and the kernel bandwidths (in
m2s, m2c) to allow the coverage probabilities in the test samples to
achieve the nominal 90\%  level on average in the cross
validation. 
  Figure \ref{fig:realCPPI} indicates that all
six methods are consistent, which yield the coverage probabilities
close to
90\%, while the semiparametric methods generally have much smaller
variations in CP and PI. 


To illustrate how to use the prediction interval in practice, we propose two strategies to
predict whether a subject has AD or is healthy  based on his/her SUVR value.
In strategy 1 (interval strategy), we categorize a subject to 
be an AD patient if the lower bound of his/her 90\% PI is less than the
estimated 
mean of  $\log(A\beta 42)$ in the AD group.
In strategy  2 (mean strategy), we categorize a subject to
AD if the  estimated
conditional mean of $Y$ given $W$ is less than the 90\% upper
confidence bound of mean $\log(A\beta 42)$ in the AD group. 
Here,
the specific form of the conditional mean for m1s, m1c 
is $E^*(Y|w,
  \z, \wh\bb)$, for m2s, m2c is $\wh{E}(Y|w, \z)$,
and for m3s, m3c is
 $m(w, \z, \wh\bb)$ as described in  the definitions of these methods in Section \ref{sec:simu}.

\begin{figure}
  \centering
  \includegraphics[scale = 0.8]{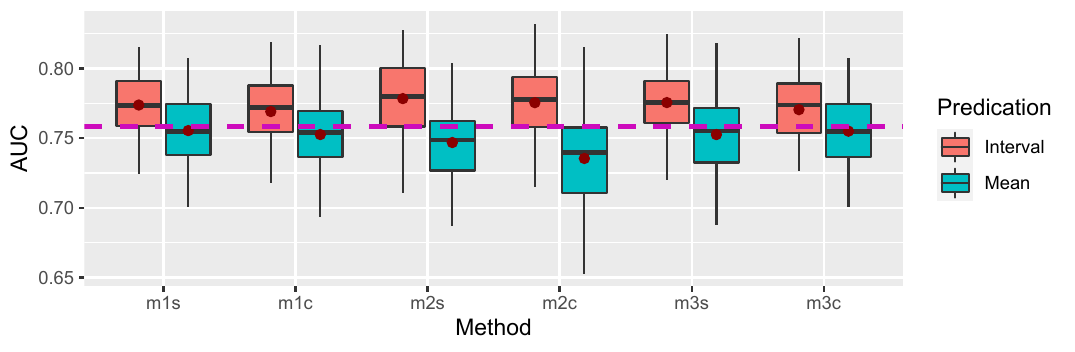}
  \caption{Boxplots over 100 cross validation of AUC from interval and mean strategies of
    predicting CSF $A\beta$ postive and negative samples. The
    dash-line represents the AUC of using CSF positive/negative to
  classify AD and healthy subject (AUC$=0.7583$). }\label{fig:auc}
\end{figure}

We show the area under the receiver operating
characteristic  (AUC)  curve of the two prediction strategies
 in Figure \ref{fig:auc}.  As shown in
Figure \ref{fig:auc}, the interval strategy outperforms the mean
strategy  in all methods. 
  Furthermore, in the interval
strategy, the AUCs from the  semiparametric methods (m1s,
m2s and m3s) have smaller variations than those from their conformal 
counterparts (m1c, m2c and m3c) as shown in the upper part of 
Table \ref{tab:auc}, 
which likely attributes
to the fact that the semiparametric methods have smaller variations in 
PI. Moreover, by considering the measurement
errors, the combination of the
interval strategy and m1s method achieves the smallest
variability in AUC. It is worth mentioning that the interval strategy also
outperforms the standard clinical practice of using CSF
positive/negative to  diagnose  AD as shown in Figure \ref{fig:auc}. 

\begin{table}[ht]
  \centering
  \caption{The mean and standard deviation over 100 cross validation of AUC from interval and mean strategies of
    predicting CSF $A\beta$ postive and negative samples.}\label{tab:auc}
\begin{tabular}{c|rrrrrr}
  \hline
 & m1s & m1c & m2s & m2c & m3s & m3c \\ 
  \hline
  &\multicolumn{6}{c}{Interval strategy}\\
  \hline
Mean & 0.7738 & 0.7691 & 0.7784 & 0.7755 & 0.7755 & 0.7704 \\ 
  \makecell{Standard deviation} &  0.0204 & 0.0238 & 0.0247 & 0.0252 & 0.0224 & 0.0236 \\ 
  \hline
  &\multicolumn{6}{c}{Mean strategy}\\
  \hline
Mean & 0.7554 & 0.7526 & 0.7468 & 0.7354 & 0.7525 & 0.7550 \\ 
\makecell{Standard deviation} &0.0259 & 0.0291 & 0.0279 & 0.0321 & 0.0283 & 0.0261 \\ 
   \hline
\end{tabular}
\end{table}

\section{Discussion}\label{sec:diss}
Many interesting problems are worth further research in the prediction
issue for measurement error models. We highlight some here.

We have used the method in \cite{tsiatisma2004} as the
  estimator due to its generality and root-$n$ rate. Other estimation
  methods can certainly be used. When the original
  model is nonparametric and semiparametric, various other methods
  exist based on deconvolution or Sieve \citep{chl2008,chl2009,cch2010},
  and some even handle high dimensional covariates
\citep{jm2022, jzlm2023,jmc2024}. See 
\cite{lm2024} for a recent review. These methods can
also be incorporated with the prediction procedures proposed here. When different
estimators are combined with the prediction interval construction,
their subsequent analysis will be different and some could be
challenging.

As we have mentioned, more general choice of the conformal score
$r(\o,\bb)$ is possible 
beyond the form $|y-m(w,\z,\bb)|$. Under the general conformal score
$r(\o,\bb)$, one can also ask the question: what choice of the
conformal score will lead to the smallest prediction set 
$C(W_{n+1},\Z_{n+1})$, where it satisfies $\pr\{Y_{n+1}\in C(W_{n+1},\Z_{n+1})\}
=\pr\{r(\o,\bb)<\zeta\}=1-\alpha$.
Note that  as soon as the functional form of
$r(\o,\bb)$ is chosen, 
different estimators of $\bb$ and $\zeta$
only affect the estimated version of $C(W_{n+1},\Z_{n+1})$. So
identifying the smallest conformal set $C(W_{n+1},\Z_{n+1})$ is not a
statistical problem, but a mathematical problem, which involves
inverting 
$r(\O_{n+1},\bb)<\zeta$ to obtain $C(W_{n+1},\Z_{n+1})$ and finding
the optimal $r$ which yields a smallest $C(W_{n+1},\Z_{n+1})$ under a
predefined measure.
Unfortunately, this is a very difficult question to answer in
general. To see the potential difficulties, as an example, if we have
set $r(\o,\bb)=-\log f_{Y\mid W,\Z}^*(\o,\bb)$, then we get 
\bse
1-\alpha&=&\pr\{-\log f_{Y\mid W,\Z}^*(\O_{n+1},\bb)<\zeta\}\\
&=&E[\pr\{-\log f_{Y\mid W,\Z}^*(\O_{n+1},\bb)<\zeta\mid W_{n+1},\Z_{n+1}\}].
\ese
From here, we cannot further conclude that $\zeta$ is
directly linked to the size of $C(W_{n+1},\Z_{n+1})$, unless we assume
special features of $f_{Y\mid W, \Z}^*(\o,\bb)$ as a function of
$y$. Thus we suspect this problem has to be studied case-by-case in
specific concrete models.

Lastly, we comment on the original issue of predicting a single response
$Y_{n+1}$ given $W_{n+1}, \Z_{n+1}$. 
Indeed, at any $m(w,\z,\bb)$, we can estimate the prediction
interval associated with any $\eta_1^*(x,\z)$. Thus, we can use the
smallest non-empty
intersection of any of these intervals as a most aggressive prediction
interval. Now if we consider an increasing series of $\alpha$ values
that approach 1, we can obtain a series of
intervals, which approximates the single prediction value. This of
course does not necessarily lead to a single value. In fact, we cannot guarantee
that the interval length will approach zero.

\bibliographystyle{agsm}
\bibliography{errpred}

\clearpage
\pagenumbering{arabic}
{\centering
\section*{Appendix}}
\setcounter{equation}{0}\renewcommand{\theequation}{A.\arabic{equation}}
\setcounter{subsection}{0}\renewcommand{\thesubsection}{A.\arabic{subsection}}
\setcounter{table}{0}\renewcommand{\thetable}{A.\arabic{table}}
\setcounter{figure}{0}\renewcommand{\thefigure}{A.\arabic{figure}}
\subsection{Proof of Proposition \ref{pro:effinflu}: the efficient influence function for $\zeta$}\label{sec:phi}

Let the dimension of $\bb$ be $d_\bb$.
Let $f_{X\mid\Z}(x,\z,\bg_1), f_\Z(\z,\bg_2)$ be parametric sub-models
of $\eta_1(x,\z), \eta_2(\z)$. We can easily see that the score
functions are
\bse
\S_{\bb}(\o,\bb)&\equiv&
\frac{\partial\log\int \eta_1(x,\z) \eta_2(\z) f_{W\mid
  X,\Z}(w,x,\z)f_{Y\mid X,\Z}(y,x,\z,\bb)dx}{\partial\bb}\\
&=&\frac{\int {\partial \log f_{Y\mid X,\Z}(y,x,\z,\bb)}/{\partial\bb}\eta_1(\x,\z)\eta_2(\z)f_{W\mid
  X,\Z}(w,x,\z)f_{Y\mid X,\Z}(y,x,\z,\bb)dx}{
\int \eta_2(\z)\eta_1(x,\z)f_{W\mid
  X,\Z}(w,x,\z)f_{Y\mid X,\Z}(y,x,\z,\bb)dx}\\
&=&E\left\{\frac{\partial \log f_{Y\mid X,\Z}(y,X,\z,\bb)}{\partial\bb}\mid\O=\o\right\},\\
\S_{\bg_1}(\o,\bb)&\equiv&
\frac{\partial\log\int \eta_2(\z)f_{X\mid\Z}(x,\z,\bg_1)f_{W\mid
  X,\Z}(w,x,\z)f_{Y\mid X,\Z}(y,x,\z,\bb)dx}{\partial\bg_1}\\
&=&\frac{\int {\partial \log f_{X\mid\Z}(x,\z,\bg_1)}/{\partial\bg_1}\eta_2(\z)f_{X\mid\Z}(x,\z,\bg_1)f_{W\mid
  X,\Z}(w,x,\z)f_{Y\mid X,\Z}(y,x,\z,\bb)dx}{
\int \eta_2(\z)\eta_1(x,\z)f_{W\mid
  X,\Z}(w,x,\z)f_{Y\mid X,\Z}(y,x,\z,\bb)dx}\\
&=&E\left\{\frac{\partial \log f_{X\mid\Z}(X,\z,\bg_1)}{\partial\bg_1}\mid
Y=y,W= w,\Z=\z\right\},\\
\S_{\bg_2}(\z,\bg_2)&\equiv&
\frac{\partial\log \int \eta_1(\x,\z) f_\Z(\z,\bg_2) f_{W\mid
  X,\Z}(w,x,\z)f_{Y\mid X,\Z}(y,x,\z,\bb)dx}{\partial\bg_2}\\
&=&\frac{\partial\log f_\Z(\z,\bg_2)}{\partial\bg_2}.
\ese
Thus,  the tangent space $\calT$, which is the span of all
  the score functions 
derived above associated with all the possible parametric submodels,  is
$\calT=\Lambda_\bb+(\Lambda_1\oplus\Lambda_2)
=\Lambda\eff\oplus\Lambda_1\oplus\Lambda_2$,
where
\bse
\Lambda_\bb&=&\left\{\c\trans \S_\bb(\o,\bb): \c\in\mR^{d_\bb}\right\}\\
\Lambda_1&=&\left[E\{a(X,\z)\mid\O=\o\}: E\{a(X,\z)\mid\z\}=0\right]\\
\Lambda_2&=&[a(\z): E\{a(\Z)\}=0],\\
\Lambda\eff&=&\left\{\c\trans\S\eff(\o,\bb): \c\in\mR^{d_\bb}\right\}.
\ese
Here, we rewrite $\Lambda_\bb+(\Lambda_1\oplus\Lambda_2)$ into
$\Lambda\eff\oplus\Lambda_1\oplus\Lambda_2$ to ease the utility of
$\calT$ later when we compute the efficient influence function.

Next,  we compute the derivatives of $\zeta$ with respect to
  the parameters $\bb, \bg_1, \bg_2$ associated with an arbitrary
  parametric submodel.
Consider 
\bse
1-\alpha&=&\pr\{r(\O,\bb)<\zeta(\bb,\eta_1,\eta_2)\}\\
&=&\int_{\o:r(\o,\bb)<\zeta(\bb,\eta_1,\eta_2)}f_\O(\o,\bb,\eta_1,\eta_2)d\o\\
&=&
\int_{\o:r(\o,\bb)<\zeta(\bb,\eta_1,\eta_2)}\eta_1(x,\z) \eta_2(\z) f_{W\mid
  X,\Z}(w,x,\z)f_{Y\mid X,\Z}(y,x,\z,\bb)d\mu(x,\o),
\ese
where we write $\zeta$ as $\zeta(\bb,\eta_1,\eta_2)$ to emphasize that
$\zeta$ is a functional of $\bb, \eta_1,\eta_2$.
Taking derivative with respect to $\bb$ leads to
\bse
\0&=&\int I\{r(\o,\bb)<\zeta(\bb,\eta_1,\eta_2)\}
\S_\bb(y,w,\z,\bb,\eta_1,\eta_2)f_\O(\o,\bb,\eta_1,\eta_2)d\o\\
&&+\int\delta\{\zeta(\bb,\eta_1,\eta_2)-r(\o,\bb)\}
\left\{\frac{\partial\zeta(\bb,\eta_1,\eta_2)}{\partial\bb}
-\frac{\partial r(\o,\bb)}{\partial\bb}\right\}
f_\O(\o,\bb,\eta_1,\eta_2)d\o\\
&=&E\left[I\{ r(\O,\bb)<\zeta(\bb,\eta_1,\eta_2)\}\S_\bb(\O,\bb,\eta_1,\eta_2)\right]\\
&&+\frac{\partial\zeta(\bb,\eta_1,\eta_2)}{\partial\bb}
\int\delta\{\zeta(\bb,\eta_1,\eta_2)-r(\o,\bb)\}
f_\O(\o,\bb,\eta_1,\eta_2)d\o\\
&&-\int\delta\{\zeta(\bb,\eta_1,\eta_2)-r(\o,\bb)\}
\frac{\partial r(\o,\bb)}{\partial\bb}
f_\O(\o,\bb,\eta_1,\eta_2)d\o,
\ese
hence
\bse
\frac{\partial\zeta(\bb,\eta_1,\eta_2)}{\partial\bb}
&=&\frac{\int\delta\{\zeta-r(\o,\bb)\}
\frac{\partial r(\o,\bb)}{\partial\bb}
f_\O(\o,\bb,\eta_1,\eta_2)d\o
}{\int\delta\{\zeta(\bb,\eta_1,\eta_2)-r(\o,\bb)\}
f_\O(\o,\bb,\eta_1,\eta_2)d\o}\\
&&-\frac{ E\left[I\{
    r(\O,\bb)<\zeta\}\S_\bb(\O,\bb,\eta_1,\eta_2)\right] 
}{\int\delta\{\zeta(\bb,\eta_1,\eta_2)-r(\o,\bb)\}
f_\O(\o,\bb,\eta_1,\eta_2)d\o}\\
&=&\frac{E[\delta\{\zeta-r(\O,\bb)\}
\frac{\partial r(\O,\bb)}{\partial\bb}]
-E[I\{
    r(\O,\bb)<\zeta\}\S_\bb(\O,\bb,\eta_1,\eta_2)]
}{E[\delta\{\zeta-r(\O,\bb)\}]}.
\ese
Similarly, replacing $\eta_1$ with its parametric submodel,
and taking derivative with respect to $\bg_1$, we have
\bse
\0&=&\int I\{r(\o,\bb)<\zeta(\bb,\bg_1,\eta_2)\}
\S_{\bg_1}(\o,\bb,\bg_1,\eta_2)f_\O(\o,\bb,\bg_1,\eta_2)d\o\\
&&+\int\delta\{\zeta(\bb,\bg_1,\eta_2)-r(\o,\bb)\}
\frac{\partial\zeta(\bb,\bg_1,\eta_2)}{\partial\bg_1}
f_\O(\o,\bb,\bg_1,\eta_2)d\o\\
&=&E\left[I\{ r(\O,\bb)<\zeta\}\S_{\bg_1}(\O,\bb,\bg_1,\eta_2)\right]
+\frac{\partial\zeta(\bb,\bg_1,\eta_2)}{\partial\bg_1}
\int\delta\{\zeta-r(\o,\bb)\}
f_\O(\o,\bb,\bg_1,\eta_2)d\o,
\ese
hence
\bse
\frac{\partial\zeta(\bb,\bg_1,\eta_2)}{\partial\bg_1}
=-
\frac{
E[I\{ r(\O,\bb)<\zeta\}\S_{\bg_1}(\O, \bb,\bg_1,\eta_2)]
}{E[\delta\{\zeta-r(\O,\bb)\}]}.
\ese
Finally, replacing $\eta_2$ with its parametric submodel, and 
taking derivative with respect to $\bg_2$, we get
\bse
\0&=&\int I\{r(\o,\bb)<\zeta(\bb,\eta_1,\bg_2)\}
\S_{\bg_2}(\z,\bg_2)f_\O(\o,\bb,\eta_1,\bg_2)d\o\\
&&+\int\delta\{\zeta(\bb,\eta_1,\bg_2)-r(\o,\bb)\}
\left\{\frac{\partial\zeta(\bb,\eta_1,\bg_2)}{\partial\bg_2}
-\frac{\partial r(\o,\bb)}{\partial\bg_2}\right\}
f_\O(\o,\bb,\eta_1,\bg_2)d\o\\
&=&E\left[I\{ r(\O,\bb)<\zeta\}\S_{\bg_2}(\Z,\bg_2)\right]
+\frac{\partial\zeta(\bb,\eta_1,\bg_2)}{\partial\bg_2}
\int\delta\{\zeta-r(\o,\bb)\}
f_\O(\o,\bb,\eta_1,\bg_2)d\o,
\ese
hence
\bse
\frac{\partial\zeta(\bb,\eta_1,\bg_2)}{\partial\bg_2}
=\frac{-E[I\{ r(\O,\bb)<\zeta\}\S_{\bg_2}(\Z,\bg_2)]
}{E[\delta\{\zeta-r(\O,\bb)\}]}.
\ese

We are now ready to characterize te efficient influence
  function.
Let $\phi(\o,\bb,\zeta)$ be the efficient influence function for estimating
$\zeta$. An efficient influence function much belong to
  $\calT$, hence it has the form
$\phi(\o,\bb,\zeta)=\c\trans\S\eff(\o,\bb)+E\{\wt a_1(X,\z,\bb,\zeta)\mid\O=\o\}+a_2(\z)$, where 
$E\{\wt a_1(X,\z,\bb,\zeta)\mid\z)=0$ and $E\{a_2(\Z)\}=0$.
In addition, at the true
parameter values, any influence function, including the
  efficient influence function 
$\phi(\o,\bb,\zeta)$, must satisfy 
\bse
\frac{\partial\zeta(\bb,\eta_1,\eta_2)}{\partial\bb}&=&E\{\phi(\O,\bb,\zeta)\S_\bb(\O)\},\\
\frac{\partial\zeta(\bb,\bg_1,\eta_2)}{\partial\bg_1}&=&E\{\phi(\O,\bb,\zeta)\S_{\bg_1}(\O)\},\\
\frac{\partial\zeta(\bb,\eta_1,\bg_2)}{\partial\bg_2}&=&E\{\phi(\O,\bb,\zeta)\S_{\bg_2}(\Z)\}.
\ese

This implies that $\c, \wt a_1(x,\z,\bb,\zeta),a_2(\z)$ must satisfy 
\bse
&&\frac{E[\delta\{\zeta-r(\O,\bb)\}
{\partial r(\O,\bb)}/{\partial\bb}]
-E[I\{
    r(\O,\bb)<\zeta\}\S_\bb(\O,\bb,\eta_1,\eta_2)]
}{E[\delta\{\zeta-r(\O,\bb)\}]}\\
&=&E\left\{[\c\trans\S\eff(\O)+E\{\wt a_1(X,\Z,\bb,\zeta)\mid\O\}+a_2(\Z)]
  [\S\eff(\O)+E\{\a(X,\Z,\bb)\mid\O\}]\right\}\\
&=&E\left\{[\c\trans\S\eff(\O)+E\{\wt a_1(X,\Z,\bb,\zeta)\mid\O\}]
  [\S\eff(\O)+E\{\a(X,\Z,\bb)\mid\O\}]\right\}\\
&=&E\left\{\S\eff(\O)\S\eff\trans(\O)\right\}\c+
E\left[E\{\wt a_1(X,\Z,\bb,\zeta)\mid\O\}E\{\a(X,\Z,\bb)\mid\O\}\right],\\
&&
\frac{
-E[I\{ r(\O,\bb)<\zeta\}\S_{\bg_1}(\O, \bb,\bg_1,\eta_2)]
}{E[\delta\{\zeta-r(\O,\bb)\}]}\\
&=&E\left([\c\trans\S\eff(\O)+E\{\wt a_1(X,\Z,\bb,\zeta)\mid\O\}+a_2(\Z)]
\S_{\bg_1}(\O)\right)\\
&=&E\left[E\{\wt a_1(X,\Z,\bb,\zeta)\mid\O\}\S_{\bg_1}(\O)\right],\\
&&\frac{
-E[I\{ r(\O,\bb)<\zeta\}\S_{\bg_2}(\Z,\bg_2)]
}{E[\delta\{\zeta-r(\O,\bb)\}]}\\
&=&E\left([\c\trans\S\eff(\O)+E\{\wt a_1(X,\Z,\bb,\zeta)\mid\O\}+a_2(\Z)]
\S_{\bg_2}(\Z)\right)\\
&=&E\left\{a_2(\Z)
\S_{\bg_2}(\Z)\right\}.
\ese
These requirements directly lead to
\bse
a_2(\Z)&=&\frac{(1-\alpha)
-E[I\{ r(\O,\bb)<\zeta\}\mid\Z]
}{E[\delta\{\zeta-r(\O,\bb)\}]},
\ese
$\wt a_1(X,\Z,\bb,\zeta)$ satisfies
\bse
&&E\left[E\{\wt a_1(X,\Z,\bb,\zeta)\mid\O\}+
\frac{I\{ r(\O,\bb)<\zeta\}}{E[\delta\{\zeta-r(\O,\bb)\}]}
\mid
X,\Z\right]\\
&=&
E\left[E\{\wt a_1(X,\Z,\bb,\zeta)\mid\O\}+
\frac{I\{ r(\O,\bb)<\zeta\}}{E[\delta\{\zeta-r(\O,\bb)\}]}\mid
\Z\right],
\ese
and $E\{\wt a_1(X,\Z,\bb,\zeta)\mid\Z\}=0$. For notational convenience later
on, we let  $a_1(X,\z,\bb,\zeta)=\wt a_1(X,\Z,\bb,\zeta)
E[\delta\{\zeta-r(\O,\bb)\}]$. Then the above requirements on
$\wt a_1(X,\Z,\bb,\zeta)$ are equivalent to
\bse
E\left[E\{a_1(X,\Z,\bb,\zeta)\mid\O\}\mid X,\Z\right]
=E\left[
I\{ r(\O,\bb)<\zeta\}\mid
\Z\right]-
E\left[
I\{ r(\O,\bb)<\zeta\}
\mid
X,\Z\right],
\ese
which is the same as in (\ref{eq:a1}), 
and 
\bse
\c&=&
\left[E\left\{\S\eff(\O)\S\eff\trans(\O)\right\}\right]^{-1}\left(
\frac{E[\delta\{\zeta-r(\O,\bb)\}
{\partial r(\O,\bb)}/{\partial\bb}]
-E[I\{
    r(\O,\bb)<\zeta\}\S_\bb(\O,\bb)]
}{E[\delta\{\zeta-r(\O,\bb)\}]}
\right.\n\\
&&\left.-E\left[E\{a_1(X,\Z,\bb,\zeta)\mid\O\}E\{\a(X,\Z,\bb)\mid\O\}\right]\right),
\ese
which is identical to (\ref{eq:c}). Here $\a(x,\z,\bb)$ is
$\a^*(x,\z,\bb)$ defined in
Section \ref{sec:estbeta} under the situation $\eta_1^*(x,\z)=\eta_1(x,\z)$.

In summary, our efficient influence function for estimating $\zeta$ is
\bse
\phi(\O,\bb,\zeta)=\c\trans\S\eff(\O,\bb)
+\frac{E\{a_1(X,\Z,\bb,\zeta)\mid\O\}+ (1-\alpha)
-E[I\{ r(\O,\bb)<\zeta\}\mid\Z]
}{E[\delta\{\zeta-r(\O,\bb)\}]},
\ese
where $\c$ is given in (\ref{eq:c}) and $a_1(X,\Z,\bb,\zeta)$ satisfies
(\ref{eq:a1}).

\subsection{Proof of Theorem \ref{th:useful}}\label{sec:proofthuseful}

The definition of $\wh\zeta$ leads to
\be\label{eq:mid}
0&=&n^{-1/2}\sumi \phi^* (\o_i,\wh\bb,\wh\zeta)\n\\
&=&n^{-1/2}\sumi \phi^* (\o_i,\wh\bb,\zeta)
+n^{-1}\sumi \frac{\partial\phi^* (\o_i,\wh\bb,\zeta)}{\partial\zeta}n^{1/2}(\wh\zeta-\zeta)
+O_p\{n^{1/2}(\wh\zeta-\zeta)^2\}\n\\
&=&n^{-1/2}\sumi \phi^* (\o_i,\bb,\zeta)
+n^{-1}\sumi
\frac{\partial\phi^* (\o_i,\bb,\zeta)}{\partial\bb\trans}n^{1/2}(\wh\bb-\bb) +O_p(n^{-1/2})\n\\
&&+\left[E\left\{\frac{\partial\phi^* (\O,\bb,\zeta)}{\partial\zeta}\right\}+o_p(1)\right]n^{1/2}(\wh\zeta-\zeta)
+O_p\{n^{1/2}(\wh\zeta-\zeta)^2\}\n\\
&=&n^{-1/2}\sumi \phi^* (\o_i,\bb,\zeta)
+E\left\{
\frac{\partial\phi^* (\O,\bb,\zeta)}{\partial\bb\trans}\right\}n^{1/2}(\wh\bb-\bb) +o_p(1)\n\\
&&+\left[E\left\{\frac{\partial\phi^* (\O,\bb,\zeta)}{\partial\zeta}\right\}+o_p(1)\right]n^{1/2}(\wh\zeta-\zeta)
+O_p\{n^{1/2}(\wh\zeta-\zeta)^2\}\n\\
&=&n^{-1/2}\sumi
\phi^*(\o_i,\bb,\zeta)+n^{-1/2}E\left\{
\frac{\partial\phi^* (\O,\bb,\zeta)}{\partial\bb\trans}\right\}
\sumi \left[-E\left\{\frac{\partial\S\eff^*(\O,\bb)}{\partial\bb\trans}\right\}\right]^{-1}\S\eff^*(\o_i,\bb)\n\\
&&+\left[E\left\{\frac{\partial\phi^* (\O,\bb,\zeta)}{\partial\zeta}\right\}+o_p(1)\right]
n^{1/2}(\wh\zeta-\zeta)
+O_p\{n^{1/2}(\wh\zeta-\zeta)^2\} +o_p(1)\n\\
&=&n^{-1/2}\sumi\left(
\phi^* (\o_i,\bb,\zeta)-E\left\{
\frac{\partial\phi^* (\O,\bb,\zeta)}{\partial\bb\trans}\right\}
\left[E\left\{\frac{\partial\S\eff^*(\O,\bb)}{\partial\bb\trans}\right\}\right]^{-1}\S\eff^*(\o_i,\bb)\right)\n\\
&&+E\left\{\frac{\partial\phi^* (\O,\bb,\zeta)}{\partial\zeta}\right\}
n^{1/2}(\wh\zeta-\zeta)
+o_p(1)\n\\
&=&n^{-1/2}\sumi u(\o_i) +E\left\{\frac{\partial\phi^* (\O,\bb,\zeta)}{\partial\zeta}\right\}
n^{1/2}(\wh\zeta-\zeta)
+o_p(1).
\ee
Note that $E\{\phi^* (\O_i,\bb,\zeta)\}=0$ and
$E\{\S\eff^*(\O_i,\bb)\}=\0$.
This leads to $n^{1/2}(\wh\zeta-\zeta)\to N(0, v^*)$ in distribution
as $n\to\infty$.

When $\eta_1^*=f_{X\mid\Z}(x,\z), \eta_2^*=f_\Z(\z)$, 
\bse
&&E\left\{\frac{\partial\phi^* (\O,\bb,\zeta)}{\partial\zeta}\right\}\\
&=&E\left\{\frac{\partial}{\partial\zeta}\left(\c\trans\S\eff(\O,\bb)
+E\{a_1(X,\Z,\bb,\zeta)\mid\O\}+\frac{(1-\alpha)
-E[I\{ r(\O,\bb)<\zeta\}\mid\Z]
}{E[\delta\{\zeta-r(\O,\bb)\}]}\right)\right\}\\
&=&-E\left(\frac{
E[{\partial} I\{ r(\O,\bb)<\zeta\}/{\partial\zeta}\mid\Z]
}{E[\delta\{\zeta-r(\O,\bb)\}]}\right)\\
&=&-1, 
\ese
and
\bse
u^* (\o_i)
&=&\left(\phi(\o_i,\bb,\zeta)-E\left\{
\frac{\partial\phi(\O,\bb,\zeta)}{\partial\bb\trans}\right\}
\left[E\left\{\frac{\partial\S\eff(\O,\bb)}{\partial\bb\trans}\right\}\right]^{-1}\S\eff(\o_i,\bb)\right)\\
&=&\left\{\phi(\o_i,\bb,\zeta)-
\left(\c\trans\left[E\left\{\frac{\partial\S\eff(\O,\bb)}{\partial\bb\trans}\right\}\right]
-E[E\{a_1(X,\Z,\bb,\zeta)\mid\O\}E\{\a(X,\Z,\bb)\mid\O\}]\trans\right.\right.\\
&&\left.\left.-\frac{E[I\{ r(\O,\bb)<\zeta\}\S_\bb(\O)]
-E\left[\delta\{\zeta-r(\O,\bb)\}{\partial r(\O,\bb)}/{\partial\bb\trans}\right]
}{E[\delta\{\zeta-r(\O,\bb)\}]}
\right)\right.
\\
&&\left.\times\left[E\left\{\frac{\partial\S\eff(\O,\bb)}{\partial\bb\trans}\right\}\right]^{-1}
\S\eff(\o_i,\bb)\right\}\\
&=&\phi(\o_i,\bb,\zeta),
\ese
where we used the definition of $\c$ in (\ref{eq:c}).
\qed

\subsection{Proof of Corollary \ref{cor:bias}}
A taylor expansion leads to
\be\label{eq:convE0}
  &&E[\pr\{r(\O,\wh\bb)<\wh\zeta\} ]- \pr\{r(\O,\bb)<\zeta\}\nonumber \\
  &=&  
\frac{\partial \pr\{r(\O,\bb)<\zeta\}}{\partial \btheta\trans}
 E(\wh{\btheta} - \btheta)+
E\left[(\wh{\btheta} - \btheta)\trans\frac{\partial^2\pr\{r(\O,\bb^*)<\zeta^*\}}{\partial \btheta \btheta\trans}
  (\wh{\btheta} - \btheta)\right],
\ee
where $\btheta^*\equiv(\bb^{*\rm T},\zeta^*)\trans$ is on
  the line connecting $\btheta$ and $\wh\btheta$.
Now  
\be\label{eq:convE1}
&& |E\left[(\wh{\btheta} - \btheta)\trans\frac{\partial^2\pr\{r(\O,\bb^*)<\zeta^*\}}{\partial \btheta \btheta\trans}
  (\wh{\btheta} - \btheta)\right]|\nonumber\\
&\leq&  E\left[\|\wh{\btheta} - \btheta)\|_2^2
  \|\frac{\partial^2\pr\{r(\O,\bb^*)<\zeta^*\}}{\partial \btheta
  \partial  \partial \btheta\trans}\|_{2}\right]\nonumber\\
&=&O(1/n)
\ee
by the assumption that $|\partial^2 \pr\{r(\O,\bb)<\zeta\}/\partial \btheta
    \partial \btheta\trans\|_{2} <\infty$ for any $\bb, \zeta$ and the fact
  that $\|\wh{\btheta} - \btheta\|_2 = O_p(1/\sqrt{n})$.

  Furthermore, 
\bse
&& - n^{-1}\sumi
\bpsi(\O_i, \bb, \zeta) \\
&=& n^{-1}\sumi \frac{\partial
  \bpsi(\O_i, \bb,\zeta)}{\partial  \btheta\trans} (\wh{\btheta} -
\btheta) +  n^{-1}\sumi \frac{\partial^2 (\wh{\btheta} -
\btheta) \trans
  \bpsi(\O_i, \bb^*,\zeta^*)}{2\partial  \btheta\partial\btheta\trans} (\wh{\btheta} -
\btheta)  \\
&=& \left[ n^{-1}\sumi \frac{\partial
  \bpsi(\O_i,\bb,\zeta)}{\partial  \btheta\trans} - E \left\{\frac{\partial
  \bpsi(\O_i, \bb,\zeta)}{\partial  \btheta\trans} \right\} \right]  (\wh{\btheta} -
\btheta)\\
&&  + E \left\{\frac{\partial
  \bpsi(\O_i, \bb,\zeta)}{\partial  \btheta\trans} \right\} (\wh{\btheta} -
\btheta) + O_p(1/n), 
\ese
where $(\bb^*, \zeta^*)$ is a  point between $(\wh{\bb}, \wh\zeta)$
and $({\bb}, \zeta)$. The last equality holds by the law of large
numbers, Condition \ref{con:4} and the fact that $\|\wh{\btheta} - \btheta\|_2^2 = O_p(1/n)$.
This  implies
\bse
&&(\wh{\btheta} -
\btheta) \\
&=& - \left[E\left\{\frac{\partial
  \bpsi(\O_i, \bb,\zeta)}{\partial  \btheta\trans} \right\}
\right]^{-1} n^{-1}\sumi
\bpsi(\O_i, \bb, \zeta) \\
&& {-} \left[E\left\{\frac{\partial
  \bpsi(\O_i, \bb,\zeta)}{\partial  \btheta\trans} \right\}\right]
^{-1}  \\
&& \times \left[ n^{-1}\sumi \frac{\partial
  \bpsi(\O_i,\bb,\zeta)}{\partial  \btheta\trans} - E \left\{\frac{\partial
  \bpsi(\O_i, \bb,\zeta)}{\partial  \btheta\trans} \right\}\right]  (\wh{\btheta} -
\btheta)+ O_p(1/n)\\
&=&- \left[E\left\{\frac{\partial
  \bpsi(\O_i, \bb,\zeta)}{\partial  \btheta\trans} \right\}
\right]^{-1} n^{-1}\sumi
\bpsi(\O_i, \bb, \zeta) +O_p(1/n),
\ese
where the last step used
  $\|\wh\btheta-\btheta\|=O_p(n^{-1/2})$.
Taking expectation on both sides, we have
\bse
E (\wh{\btheta} -
\btheta) =0 + O(1/n).
\ese

Combining with (\ref{eq:convE0}) and
(\ref{eq:convE1}), we have
\bse
|E[\pr\{r(\O,\wh\bb)<\wh\zeta\} ]- \pr\{r(\O,\bb)<\zeta\}| =O(1/n). 
\ese
This proves the result. 
\qed

\subsection{Proof of Theorem \ref{th:bound}}\label{sec:proofofthbound}

We first establish several lemmas.

\begin{Lem}(Hoeffding-type inequality)\label{lem:Hoeff}. Let $X_1, \ldots, X_n$ be
  independent centered sub-gaussian random variables, and let $K =
  \max_i\|X_i\|_{\psi_2}$. Then for every $\a = (a_1, \ldots, a_n)\trans \in
  \mathbb{R}^n$ and every $t\geq 0$, we have
  \bse
  \pr\left(|\sumi a_i X_i| \geq t\right) \leq e \exp\left(-\frac{ct^2}{K^2
      \|\a\|_2^2}\right), 
  \ese
  where $c>0$ is an absolute constant, and
$e\approx 2.71$ is the Euler's number.
\end{Lem}
\noindent Proof: This lemma follows Proposition 5.10 in
\cite{vershynin2010}.

\begin{Lem}(Bernstein-type inequality)\label{lem:Bern}. Let $X_1,
  \ldots, X_n$ be independent centered sub-exponential random
  variables, and $K = \max_i\|\X_i\|_{\psi_1}$. Then for every $\a = (a_1, \ldots, a_n)\trans \in
  \mathbb{R}^n$  and every $t\geq 0$, we have
 \bse
  \pr\left(|\sumi a_i X_i| \geq t\right) \leq 2\exp\left[-c \min\left(\frac{t^2}{K^2
      \|\a\|_2^2}, \frac{t}{K\|\a\|_{\infty}}\right)\right], 
  \ese
   where $c>0$ is an absolute constant. 
 \end{Lem}
 \noindent Proof: This lemma follows Proposition 5.16 in
\cite{vershynin2010}.

\begin{Lem}\label{lem:first}
 Assume Condition \ref{con:subGE}. There is a constant
 $c>0$ such that 
\bse
\pr\left\{\|n^{-1}\sumi \bpsi(\O_i, \bb, \zeta)\|_{\infty}\geq
  \sqrt{2M_1^2\log\{\max(n, p +1)\}/(nc)}\right\} \leq e \exp[-\log\{\max(n, p+1)\}]. 
\ese
\end{Lem}
\noindent Proof:
Because $\e_j\trans\bpsi(\O_i, \bb, \zeta)$, $i = 1,
\ldots, d$ is a sub-Gaussian random variable, by Lemma \ref{lem:Hoeff}
\bse
\pr\left\{|n^{-1}\sumi \e_j\trans\bpsi(\O_i, \bb, \zeta)|\geq
  t\right\} \leq e \exp(-\frac{cnt^2}{M_1^2}), 
\ese
and
\bse
\pr\left\{\|n^{-1}\sumi \bpsi(\O_i, \bb, \zeta)\|_{\infty}\geq
  t\right\} \leq e (p + 1) \exp(-\frac{cnt^2}{M_1^2}). 
\ese
Let $t = \sqrt{2M_1^2\log\{\max(n, p +1)\}/(nc)}$, we have
\bse
\pr\left\{\|n^{-1}\sumi \bpsi(\O_i, \bb, \zeta)\|_{\infty}\geq
  \sqrt{2M_1^2\log\{\max(n, p +1)\}/(nc)}\right\} \leq e \exp[-\log\{\max(n, p+1)\}]. 
\ese
This proves the result.\qed

\begin{Lem}\label{lem:second}
  Assume Conditions \ref{con:subGE}--\ref{con:3}.  Define
  $\mS = \{\v \in {\mathbb R}^{p+1}, \|\v\|_2= 1\}$. There is
  a constant $c_1>0$ such that for any $\v_1, \v_2 \in \mS$ and given $\btheta\in\bTheta$,
  \bse
n^{-1}|\v_1\trans \sumi \frac{\partial \bpsi(\O_i, \bb,
  \zeta)}{\partial \btheta\trans} \v_2| \geq  \lambda_{\min}/2
\ese
with probability greater than $1 - 9^{2p+2}   2 \exp\left\{-  c_1 n
  \min (\lambda_{min}^2/4, \lambda_{\min}/2)\right\}$. 
  \end{Lem}
{\noindent Proof:} Define
$\calA= \{\u_1, \ldots, \u_m\}\subset \mS $   to be  a $\delta$-cover of
$\mS$, if for every $\v \in \mS$, there is some $\u_i \in
\calA$ such that $\|\v - \u_i\|_2 \leq \delta$. Define
$\Delta\v=\v-\u_j$ where $\u_j = \arg\min_{\u_i\in \calA}\|\v -\u_i\|_2$. Now
because the $\delta$-covering number of $\mS$ is
$\le(3/\delta)^{p+1}$ for
$\delta<1$ (Lemma 5.2 of \cite{vershynin2010}). We select 
$\delta= 1/3$, and then 
the $1/3$-covering number of $\mS$ is no greater
than 
$9^{p+1}$. That is $|\mathcal{A}| \leq 9^{p+1}$.  For
 any  given $\btheta\in\bTheta$,  let
\bse
\Phi (\v_1, \v_2) \equiv \v_1 \trans\left[n^{-1}\sumi \frac{\partial \bpsi(\O_i, \bb,
  \zeta)}{\partial \btheta\trans}  - E\left\{ \frac{\partial \bpsi(\O_i, \bb,
  \zeta)}{\partial \btheta\trans}\right\}\right]\v_2. 
\ese 
We have 
\bse
&& |\Phi (\v_1, \v_2)| \\
&= & |\Phi (\Delta \v_1 + \u_i, \Delta \v_2
+ \u_j )|  \\
&\le&  \max_{i, j}  |\Phi (\u_i,
\u_j)| + \max_j  |\Phi (\Delta
\v_1,\u_j)  | + \max_i | \Phi (\u_i,\Delta
\v_2)|  +   | \Phi (\Delta
\v_1,\Delta
\v_2)  |.
\ese 
Hence, 
\bse
 && \sup_{\v_1, \v_2 \in \mS} |\Phi (\v_1, \v_2)| \\
&\leq& \max_{i, j} |\Phi (\u_i,
\u_j)| +  \sup_{\v_1 \in \mS} \max_j |\Phi (\Delta
\v_1,\u_j)  | +\sup_{\v_2 \in \mS} \max_i |\Phi (\u_i, \Delta
\v_2) |+   \sup_{\v_1, \v_2 \in \mS} | \Phi (\Delta
\v_1,\Delta
\v_2)  |. 
\ese
Since $\|3\Delta\v_1\|_2 \leq 1$ and $\|3\Delta\v_2\|_2 \leq 1$, 
$3\Delta\v_1  \in \mS$ and $3\Delta\v_2  \in \mS$. 
It follows that 
\bse
&&\sup_{\v_1, \v_2 \in \mS}|\Phi (\v_1, \v_2)| \\
&\leq&
\max_{i, j} |\Phi (\u_i,
\u_j)| + 1/3 \sup_{\v_1 \in  \mS} \max_j |\Phi (3\Delta
\v_1,\u_j)  | + 1/3 \sup_{\v_2 \in  \mS} \max_i |\Phi (\u_i,  3\Delta
\v_2)  |  \\
&&+ 1/9\sup_{\v_1, \v_2 \in \mS} | \Phi (3\Delta
\v_1,3\Delta
\v_2)  |\\
&\le&
\max_{i, j} |\Phi (\u_i,
\u_j)| +2/3 \sup_{\v_1, \v_2 \in  \mS} |\Phi (\v_1, \v_2)  | 
+ 1/9\sup_{\v_1, \v_2 \in \mS} | \Phi (\v_1,
\v_2)  |\\
&\le&
 \max_{i, j} |\Phi (\u_i,
\u_j)| + 7/9\sup_{\v_1, \v_2 \in\mS } |\Phi (\v_1, \v_2)|. 
\ese
Hence,  
$\sup_{\v_1, \v_2 \in \mS}|\Phi (\v_1, \v_2)|\leq 9/2
\max_{i, j} |\Phi (\u_{i},
\u_j)|$. By Lemma \ref{lem:Bern}, Condition \ref{con:subGE} and a
union bound, we have that there is a constant $c_0>0$ such that  
\bse
&& \pr\left(\sup_{\v_1, \v_2 \in \mS} |\v_1\trans\left[\sumi \frac{\partial \bpsi(\O_i, \bb,
  \zeta)}{\partial \btheta\trans}  - n E\left\{ \frac{\partial \bpsi(\O_i, \bb,
    \zeta)}{\partial \btheta\trans}\right\}\right] \v_2| > 9/2 nt \right)\\
&\leq& \pr\left(\max_{i, j} |\u_i\trans\left[\sumi \frac{\partial \bpsi(\O_i, \bb,
  \zeta)}{\partial \btheta\trans}  - n E\left\{ \frac{\partial \bpsi(\O_i, \bb,
  \zeta)}{\partial \btheta\trans}\right\}\right] \u_j| > nt \right)\\
&\leq&  9^{2p+2}   2 \exp\left\{-  c_0 \min \left(\frac{ n  t^2}{M_2^2 },\frac{n
      t}{M_2 } \right) \right\}. 
\ese
Replacing $t$ with $2/9 t$, we get that there is a constant $c_1>0$ such that  
\bse
&& \pr\left(\sup_{\v_1, \v_2 \in \mS} | n^{-1}\v_1\trans\left[\sumi \frac{\partial \bpsi(\O_i, \bb,
  \zeta)}{\partial \btheta\trans}  - E\left\{ \frac{\partial \bpsi(\O_i, \bb,
    \zeta)}{\partial \btheta\trans}\right\}\right] \v_2| > t \right)\\
&\leq&  9^{2p+2}   2 \exp\left\{-  c_1 n \min (t^2, t)\right\}.
\ese
Now, let $t = \lambda_{\min}/2$. Then for any $\v_1,
  \v_2\in \mS$, 
\bse
&&| n^{-1}\v_1\trans\left[\sumi \frac{\partial \bpsi(\O_i, \bb,
  \zeta)}{\partial \btheta\trans}  - E\left\{ \frac{\partial \bpsi(\O_i, \bb,
    \zeta)}{\partial \btheta\trans}\right\}\right] \v_2| {\le}\lambda_{\min}/2\\
\ese
with probability greater than $1 - 9^{2p+2}   2 \exp\left\{-  c_1 n
  \min (\lambda_{min}^2/4, \lambda_{\min}/2)\right\}$. And hence by
Condition \ref{con:3}, we have 
\bse
n^{-1}|\v_1\trans \sumi \frac{\partial \bpsi(\O_i, \bb,
  \zeta)}{\partial \btheta\trans} \v_2 |\geq  \lambda_{\min}/2
\ese
with probability greater than $1 - 9^{2p+2}   2 \exp\left\{-  c_1 n
  \min (\lambda_{min}^2/4, \lambda_{\min}/2)\right\}$. 
\qed

We now prove Theorem \ref{th:bound}.

{\noindent Proof:}
First note that
\bse
n^{-1}\sumi \bpsi(\O_i, \wh\bb, \wh\zeta)-  n^{-1}\sumi \bpsi(\O_i,
\bb, \zeta)= n^{-1}\sumi \frac{\partial \bpsi(\O_i, \bb^*,
  \zeta^*)}{\partial \btheta\trans} (\wh{\btheta} - \btheta), 
\ese
where $\bb^*$ is a point between $\wh{\bb}$ and $\bb$ and $\zeta^*$ is
a point between $\wh{\zeta}$ and $\zeta$. Multiplying both sides by $(\wh{\btheta} - \btheta)\trans$, we have 
\bse
- (\wh{\btheta} - \btheta)\trans  n^{-1}\sumi \bpsi(\O_i,
\bb, \zeta)= (\wh{\btheta} - \btheta)\trans n^{-1}\sumi \frac{\partial \bpsi(\O_i, \bb^*,
  \zeta^*)}{\partial \btheta\trans} (\wh{\btheta} - \btheta),
\ese
so
\bse
 \|\wh{\btheta} - \btheta\|_{1}  \|n^{-1}\sumi \bpsi(\O_i,
\bb, \zeta)\|_\infty \geq |(\wh{\btheta} - \btheta)\trans n^{-1}\sumi \frac{\partial \bpsi(\O_i, \bb^*,
  \zeta^*)}{\partial \btheta\trans} (\wh{\btheta} - \btheta)|. 
\ese

%For convenience, we assume $(\wh{\btheta}- \btheta)\trans \sumi \partial \bpsi(\O_i, \bb,
%  \zeta)/\partial \btheta\trans (\wh{\btheta}- \btheta) >0$, if  it is
%  negative,  the
%  following derivations follow by replacing $\bpsi$ by $-\bpsi$.  
Now by Lemma \ref{lem:second} and 
  and the fact that  $\|\wh{\btheta} -
\btheta\|_{1}\leq \sqrt{p+1} \|\wh{\btheta} -
\btheta\|_2$, we have
\bse
 2 \lambda_{\min}^{-1} \sqrt{p+1}\|n^{-1}\sumi \bpsi(\O_i,
\bb, \zeta)\|_\infty \geq \|\wh{\btheta} - \btheta\|_2
\ese
with probability greater than $1 - 9^{2p+2}   2 \exp\left\{-  c_1 n
  \min (\lambda_{min}^2/4, \lambda_{\min}/2)\right\}$. 
Furthermore, by Lemma \ref{lem:first}, we have 
\bse
\|\wh{\btheta} - \btheta\|_2 \leq  2 \lambda_{\min}^{-1}  \sqrt{p+1}
\sqrt{2M_1^2\log\{\max(n, p +1)\}/(nc)}, 
\ese
with probability great than $1 - 9^{2p+2}   2 \exp\left\{-  c_1 n
  \min (\lambda_{min}^2/4, \lambda_{\min}/2)\right\} - e
\exp[-\log\{\max(n, p+1)\}]$. 
Combine with the fact that 
\bse
\pr\{r(\O,\wh\bb)<\wh\zeta\} &=&  \pr\{r(\O,\bb)<\zeta\}  +
\frac{\partial \pr\{r(\O,\bb^{**})<\zeta^{**}\}}{\partial \btheta\trans}
(\wh{\btheta} - \btheta), 
\ese
where $\bb^{**}$ and $\zeta^{**}$ is a point between $\wh{\bb}$ and
$\bb$, we have by Condition \ref{con:boundp}
\bse
|\pr\{r(\O,\wh\bb)<\wh\zeta\} - (1-\alpha)|&\leq& M_3 \|\wh{\btheta} -
\btheta\|_2 \\
&\leq& 2 M_3  \lambda_{\min}^{-1}  \sqrt{p+1}
\sqrt{2M_1^2\log\{\max(n, p +1)\}/(nc)}, 
\ese
with probability greater than $1 - 9^{2p+2}   2 \exp\left\{-  c_1 n
  \min (\lambda_{min}^2/4, \lambda_{\min}/2)\right\} - e
\exp[-\log\{\max(n, p+1)\}]$. Now because $p<n$,
  combining all the constants, we obtain that there exists constants,
  which we still name  $c, c_1$,
  so that
\bse
|\pr\{r(\O,\wh\bb)<\wh\zeta\} - (1-\alpha)|
\leq c M_1M_3 \lambda_{\min}^{-1}  
\sqrt{\log(n)/(n)}, 
\ese
with probability greater than $1 - 9^{2p+2}   2 \exp\left\{-  c_1 n
  \min (\lambda_{min}^2/4, \lambda_{\min}/2)\right\} - e/n$. 
This proves the result. 
\qed

\subsection{Derivation of the properties of $\wc\zeta$}\label{sec:classic}

Note that because of the data splitting, we can write 
\bse 
0&=&n_1^{-1/2}\suminone 
[I\{\wc\zeta-r(\o_i,\wt\bb)\ge0\}-(1-\alpha)]\\
&=&n_1^{-1/2}\suminone 
[I\{\zeta-r(\o_i,\wt\bb)\ge0\}-(1-\alpha)]+E [\delta\{\zeta-r(\O,\wt\bb)\}]n^{1/2}(\wc\zeta-\zeta)+o_p(1)\\
&=&n_1^{-1/2}\suminone 
[I\{\zeta-r(\o_i,\wt\bb)\ge0\}-(1-\alpha)]+E [\delta\{\zeta-r(\O,\bb)\}]n^{1/2}(\wc\zeta-\zeta)+o_p(1), 
\ese 
hence 
$E[\delta\{\zeta-r(\O,\bb)\}] n_1^{1/2}(\wc\zeta-\zeta) 
\to N\{E[I\{\zeta-r(\O,\wt\bb)\ge0\}-(1-\alpha), 
\var[I\{\zeta-r(\O,\wt\bb)\ge0\}]\}$
in distribution,  where $\wt\bb$ is viewed at nonrandom 
  because it does not involve the $n_1$ observations. Now 
$E[I\{\zeta-r(\O,\wt\bb)\ge0\}-(1-\alpha)]\to0$ and 
$\var[I\{\zeta-r(\O,\wt\bb)\ge0\}]\to 
\var[I\{\zeta-r(\O,\bb)\ge0\}]=\alpha(1-\alpha)$, hence 
$E[\delta\{\zeta-r(\O,\bb)\}] n_1^{1/2}(\wc\zeta-\zeta) 
\to N\{0, \alpha(1-\alpha)\}$ in distribution. This leads to the 
result.

\subsection{Derivation of the efficient influence function for $\zeta$
in no model}\label{sec:nomodel}

In this case, the model is $f_{W,\Z}(w,\z)f_{\epsilon\mid
  W,\Z}\{y-m(w,\z),w,\z\}$, where $\int \epsilon f_{\epsilon\mid
  W,\Z}(\epsilon,w,\z)d\epsilon=0$.
Let $f_{\epsilon\mid W,\Z}(\epsilon,w,\z,\bg_1)$,
$f_{W,\Z}(w,\z,\bg_2)$ and $m(w,\z,\bg_3)$  be sub-models of 
$\eta_1(\epsilon,w,\z)\equiv f_{\epsilon\mid
  W,\Z}(\epsilon,w,\z)$, $\eta_2(w,\z)\equiv f_{W,\Z}(w,\z)$ and
$m(w,\z)$. 
Now the scores are
\bse
\S_{\bg_3}(\o,\bg_3,\eta_1,\eta_2)&=&-d\log f_{\epsilon\mid
  W,\Z}(\epsilon,w,\z)/d\epsilon \frac{\partial
m(w,\z,\bg_3)}{\partial\bg_3},\\
\S_{\bg_1}(\o,m,\bg_1,\eta_2)&=&\frac{\partial f_{\epsilon\mid
  W,\Z}\{y-m(w,\z),w,\z,\bg_1\}/\partial\bg_1 }{f_{\epsilon\mid
  W,\Z}\{y-m(w,\z),w,\z,\bg_1\}},\\
\S_{\bg_2}(w,\z,\bg_2)&=& \frac{\partial f_{W,\Z}(w,\z,\bg_2)/\partial\bg_2}{f_{W,\Z}(w,\z)}.
\ese
We can easily see that
the tangent space is
$\calT=\Lambda_m+(\Lambda_1\oplus\Lambda_2)
=\Lambda_1\oplus\Lambda_2\oplus\Lambda_3$,
where
\bse
\Lambda_m&=&\{a(w,\z) d\log f_{\epsilon\mid
  W,\Z}(\epsilon,w,\z)/d\epsilon \},\\
\Lambda_1&=&\left[a(\epsilon,w,\z): E\{a(\epsilon,w,\z)\mid w,\z\}=0, 
E\{\epsilon a(\epsilon,w,\z)\mid w,\z\}=0
\right]\\
\Lambda_2&=&[a(w,\z): E\{a(W,\Z)\}=0],\\
\Lambda_3&=&\{a(w,\z)\epsilon\}.
\ese

Next, consider 
\bse
1-\alpha&=&\pr\{r(\O,m)<\zeta(m,\eta_1,\eta_2)\}\\
&=&
\int_{\o:r(\o,m)<\zeta(m,\eta_1,\eta_2)}\eta_2(w,\z) \eta_1\{y-m(w,\z),w,\z\}d\mu(\o),
\ese
where we write $\zeta$ as $\zeta(m,\eta_1,\eta_2)$ to emphasize that
$\zeta$ is a functional of $m, \eta_1,\eta_2$.
Taking derivative with respect to $\bg_3$ leads to
\bse
\0&=&\int I\{r(\o,\bg_3)<\zeta(\bg_3,\eta_1,\eta_2)\}
\S_{\bg_3}(y,w,\z,\bg_3,\eta_1,\eta_2)f_\O(\o,\bg_3,\eta_1,\eta_2)d\o\\
&&+\int\delta\{\zeta(\bg_3,\eta_1,\eta_2)-r(\o,\bg_3)\}
\left\{\frac{\partial\zeta(\bg_3,\eta_1,\eta_2)}{\partial\bg_3}
-\frac{\partial r(\o,\bg_3)}{\partial\bg_3}\right\}
f_\O(\o,\bg_3,\eta_1,\eta_2)d\o\\
&=&E\left[I\{ r(\O,\bg_3)<\zeta(\bg_3,\eta_1,\eta_2)\}\S_{\bg_3}(\O,\bg_3,\eta_1,\eta_2)\right]\\
&&+\frac{\partial\zeta(\bg_3,\eta_1,\eta_2)}{\partial\bg_3}
\int\delta\{\zeta(\bg_3,\eta_1,\eta_2)-r(\o,\bg_3)\}
f_\O(\o,\bg_3,\eta_1,\eta_2)d\o\\
&&-\int\delta\{\zeta(\bg_3,\eta_1,\eta_2)-r(\o,\bg_3)\}
\frac{\partial r(\o,\bg_3)}{\partial\bg_3}
f_\O(\o,\bg_3,\eta_1,\eta_2)d\o,
\ese
hence
\bse
\frac{\partial\zeta(\bg_3,\eta_1,\eta_2)}{\partial\bg_3}
&=&\frac{\int\delta\{\zeta-r(\o,\bg_3)\}
\frac{\partial r(\o,\bg_3)}{\partial\bg_3}
f_\O(\o,\bg_3,\eta_1,\eta_2)d\o
}{\int\delta\{\zeta(\bg_3,\eta_1,\eta_2)-r(\o,\bg_3)\}
f_\O(\o,\bg_3,\eta_1,\eta_2)d\o}\\
&&-\frac{ E\left[I\{
    r(\O,\bg_3)<\zeta\}\S_{\bg_3}(\O,\bg_3,\eta_1,\eta_2)\right] 
}{\int\delta\{\zeta(\bg_3,\eta_1,\eta_2)-r(\o,\bg_3)\}
f_\O(\o,\bg_3,\eta_1,\eta_2)d\o}\\
&=&\frac{E[\delta\{\zeta-r(\O,\bg_3)\}
\frac{\partial r(\O,\bg_3)}{\partial\bg_3}]
-E[I\{
    r(\O,\bg_3)<\zeta\}\S_{\bg_3}(\O,\bg_3,\eta_1,\eta_2)]
}{E[\delta\{\zeta-r(\O,\bg_3)\}]}.
\ese
Similarly, replacing $\eta_1$ with its parametric submodel,
and taking derivative with respect to $\bg_1$, we have
\bse
\0&=&\int I\{r(\o,m)<\zeta(m,\bg_1,\eta_2)\}
\S_{\bg_1}(\o,m,\bg_1,\eta_2)f_\O(\o,m,\bg_1,\eta_2)d\o\\
&&+\int\delta\{\zeta(m,\bg_1,\eta_2)-r(\o,m)\}
\frac{\partial\zeta(m,\bg_1,\eta_2)}{\partial\bg_1}
f_\O(\o,m,\bg_1,\eta_2)d\o\\
&=&E\left[I\{ r(\O,m)<\zeta\}\S_{\bg_1}(\O,m,\bg_1,\eta_2)\right]
+\frac{\partial\zeta(m,\bg_1,\eta_2)}{\partial\bg_1}
\int\delta\{\zeta-r(\o,m)\}
f_\O(\o,m,\bg_1,\eta_2)d\o,
\ese
hence
\bse
\frac{\partial\zeta(m,\bg_1,\eta_2)}{\partial\bg_1}
=-
\frac{
E[I\{ r(\O,m)<\zeta\}\S_{\bg_1}(\O, m,\bg_1,\eta_2)]
}{E[\delta\{\zeta-r(\O,m)\}]}.
\ese
Finally, replacing $\eta_2$ with its parametric submodel, and 
taking derivative with respect to $\bg_2$, we get
\bse
\0&=&\int I\{r(\o,m)<\zeta(m,\eta_1,\bg_2)\}
\S_{\bg_2}(w,\z,\bg_2)f_\O(\o,m,\eta_1,\bg_2)d\o\\
&&+\int\delta\{\zeta(m,\eta_1,\bg_2)-r(\o,m)\}
\frac{\partial\zeta(m,\eta_1,\bg_2)}{\partial\bg_2}
f_\O(\o,m,\eta_1,\bg_2)d\o\\
&=&E\left[I\{ r(\O,m)<\zeta\}\S_{\bg_2}(W,\Z,\bg_2)\right]
+\frac{\partial\zeta(m,\eta_1,\bg_2)}{\partial\bg_2}
\int\delta\{\zeta-r(\o,m)\}
f_\O(\o,m,\eta_1,\bg_2)d\o,
\ese
hence
\bse
\frac{\partial\zeta(m,\eta_1,\bg_2)}{\partial\bg_2}
=\frac{-E[I\{ r(\O,m)<\zeta\}\S_{\bg_2}(W,\Z,\bg_2)]
}{E[\delta\{\zeta-r(\O,m)\}]}.
\ese

Let $\phi(\o,m,\zeta)$ be the efficient influence function for estimating
$\zeta$. Then it has the form
$\phi(\o,m,\zeta)=a_1(\o)+a_2(w,\z)+\epsilon a_3(w,\z)$, where 
$E\{\epsilon a_1(\O)\mid w,\z\}=E\{a_1(\O)\mid w,\z\}=0$ and $E\{a_2(W,\Z)\}=0$.
In addition, at the true
parameter values, 
$\phi(\o,m,\zeta)$ satisfies
\bse
\frac{\partial\zeta(\bg_3,\eta_1,\eta_2)}{\partial\bg_3}&=&E\{\phi(\O,m,\zeta)\S_{\bg_3}(\O)\},\\
\frac{\partial\zeta(m,\bg_1,\eta_2)}{\partial\bg_1}&=&E\{\phi(\O,m,\zeta)\S_{\bg_1}(\O)\},\\
\frac{\partial\zeta(m,\eta_1,\bg_2)}{\partial\bg_2}&=&E\{\phi(\O,m,\zeta)\S_{\bg_2}(\W,\Z)\}.
\ese
This implies that $a_1(\o,\bg_3,\zeta),a_2(w,\z), a_3(w,\z)$ must satisfy 
\bse
&&\frac{E[\delta\{\zeta-r(\O,\bg_3)\}
{\partial r(\O,\bg_3)}/{\partial\bg_3}]
-E[I\{
    r(\O,\bg_3)<\zeta\}\S_{\bg_3}(\O,\bg_3,\eta_1,\eta_2)]
}{E[\delta\{\zeta-r(\O,\bg_3)\}]}\\
&=&E\left\{\{a_1(\O,\bg_3,\zeta)+a_2(W,\Z)+\epsilon a_3(W,\Z)\}
  [
-d\log f_{\epsilon\mid
  W,\Z}(\epsilon,w,\z)/d\epsilon \frac{\partial
m(w,\z,\bg_3)}{\partial\bg_3}]\right\}\\
&=&E\left\{\{a_1(\O,\bg_3,\zeta)+\epsilon a_3(W,\Z)\}
  [
-d\log f_{\epsilon\mid
  W,\Z}(\epsilon,w,\z)/d\epsilon \frac{\partial
m(w,\z,\bg_3)}{\partial\bg_3}]\right\}\\
&=&-E\left\{a_1(\O,\bg_3,\zeta)
  d\log f_{\epsilon\mid
  W,\Z}(\epsilon,w,\z)/d\epsilon \frac{\partial
m(w,\z,\bg_3)}{\partial\bg_3}\right\}+E\left\{ a_3(W,\Z) \frac{\partial
m(w,\z,\bg_3)}{\partial\bg_3}
\right\},\\
&&
\frac{
-E[I\{ r(\O,m)<\zeta\}\S_{\bg_1}(\O, m,\bg_1,\eta_2)]
}{E[\delta\{\zeta-r(\O,m)\}]}\\
&=&E\left[\{a_1(\O,m,\zeta)+a_2(W,\Z)+\epsilon a_3(W,\Z)\}
\S_{\bg_1}(\O)\right]\\
&=&E\{a_1(\O,m,\zeta)\S_{\bg_1}(\O)\},\\
&&\frac{
-E[I\{ r(\O,m)<\zeta\}\S_{\bg_2}(W,\Z,\bg_2)]
}{E[\delta\{\zeta-r(\O,m)\}]}\\
&=&E\left[\{a_1(\O,m,\zeta)+a_2(W,\Z)+\epsilon a_3(W,\Z)\}
\S_{\bg_2}(W,\Z)\right]\\
&=&E\left\{a_2(W,\Z)
\S_{\bg_2}(W,\Z)\right\}.
\ese
Because $\S_{\bg_2}(W,\Z)$ can  be an arbitrary mean zero
  function of $W,\Z$, $\S_{\bg_1}(\O)$ can be any function that
  satisfies $E\{\S_{\bg_1}(\O)\mid W,\Z\}=
  E\{\epsilon\S_{\bg_1}(\O)\mid W,\Z\}=\0$ and $\partial m(W,\Z,\bg_3)/\partial\bg_3$ can
  be any function of $W,\Z$, 
these requirements directly lead to
\bse
a_2(W,\Z)&=&\frac{(1-\alpha)
-E[I\{ r(\O,m)<\zeta\}\mid W,\Z]
}{E[\delta\{\zeta-r(\O)\}]},\n\\
a_1(\O)&=&\frac{
E[I\{
  r(\O,m)<\zeta\}\mid W,\Z]-
I\{
  r(\O,m)<\zeta\}}{E[\delta\{\zeta-r(\O,m)\}]}
+\frac{\epsilon E[\epsilon I\{
  r(\O,m)<\zeta\}\mid W,\Z ]}{ E[\delta\{\zeta-r(\O,m)\}]
  E(\epsilon^2\mid W,\Z)},\n\\
a_3(W,\Z)
&=&-\frac{E[\epsilon I\{
  r(\O,m)<\zeta\}\mid W,\Z ]}{ E[\delta\{\zeta-r(\O,m)\}]
  E(\epsilon^2\mid W,\Z)}-\frac{E[\delta\{\zeta-r(\O,m)\}\mid
W,\Z]}{E[\delta\{\zeta-r(\O,m)\}]}.
\ese
In summary, our efficient influence function for estimating $\zeta$ is
\bse
\phi_{\rm direct}(\O,\zeta)=\frac{(1-\alpha)-I\{
  r(\O,m)<\zeta\}-\epsilon E[\delta\{\zeta-r(\O,m)\}\mid
W,\Z]
}{E[\delta\{\zeta-r(\O)\}]}.
\ese

\subsection{Derivation of the efficient influence function for $\zeta$
in naive model}\label{sec:naive}

In this case, the model is $f_{W,\Z}(w,\z)f_{\epsilon\mid
  W,\Z}\{y-m(w,\z,\bb),w,\z\}$, where $\int \epsilon f_{\epsilon\mid
  W,\Z}(\epsilon,w,\z)d\epsilon=0$.
Let $f_{\epsilon\mid W,\Z}(\epsilon,w,\z,\bg_1)$ and
$f_{W,\Z}(w,\z,\bg_2)$  be sub-modelssub-models of 
$\eta_1(\epsilon,w,\z)\equiv f_{\epsilon\mid
  W,\Z}(\epsilon,w,\z)$ and $\eta_2(w,\z)\equiv f_{W,\Z}(w,\z)$.
Now the scores are
\bse
\S_{\bb}(\o,\bb,\eta_1,\eta_2)&=&-d\log f_{\epsilon\mid
  W,\Z}(\epsilon,w,\z)/d\epsilon 
\m'_\bb(w,\z,\bb),\\
\S_{\bg_1}(\o,\bb,\bg_1,\eta_2)&=&\frac{\partial f_{\epsilon\mid
  W,\Z}\{y-m(w,\z,\bb),w,\z,\bg_1\}/\partial\bg_1 }{f_{\epsilon\mid
  W,\Z}\{y-m(w,\z,\bb),w,\z,\bg_1\}},\\
\S_{\bg_2}(w,\z,\bg_2)&=& \frac{\partial f_{W,\Z}(w,\z,\bg_2)/\partial\bg_2}{f_{W,\Z}(w,\z)}.
\ese
We can easily see that
the tangent space is
$\calT=\Lambda_\bb+(\Lambda_1\oplus\Lambda_2)
=\Lambda_1\oplus\Lambda_2\oplus\Lambda_3$,
where
\bse
\Lambda_\bb&=&\{\a\trans\m'_\bb(w,\z,\bb) d\log f_{\epsilon\mid
  W,\Z}(\epsilon,w,\z)/d\epsilon :\a\in\calR^{d_\bb}\},\\
\Lambda_1&=&\left[a(\epsilon,w,\z): E\{a(\epsilon,w,\z)\mid w,\z\}=0, 
E\{\epsilon a(\epsilon,w,\z)\mid w,\z\}=0
\right]\\
\Lambda_2&=&[a(w,\z): E\{a(W,\Z)\}=0],\\
\Lambda_3&=&\{\a\trans\m'_\bb(w,\z,\bb) \epsilon/\var(\epsilon^2\mid w,\z):\a\in\calR^{d_\bb}\}.
\ese
Next, consider 
\bse
1-\alpha&=&\pr\{r(\O,\bb)<\zeta(\bb,\eta_1,\eta_2)\}\\
&=&
\int_{\o:r(\o,\bb)<\zeta(\bb,\eta_1,\eta_2)}\eta_2(w,\z) \eta_1\{y-m(w,\z,\bb),w,\z\}d\mu(\o),
\ese
where we write $\zeta$ as $\zeta(\bb,\eta_1,\eta_2)$ to emphasize that
$\zeta$ is a functional of $\bb, \eta_1,\eta_2$.
Taking derivative with respect to $\bb$ leads to
\bse
\0&=&\int I\{r(\o,\bb)<\zeta(\bb,\eta_1,\eta_2)\}
\S_{\bb}(y,w,\z,\bb,\eta_1,\eta_2)f_\O(\o,\bb,\eta_1,\eta_2)d\o\\
&&+\int\delta\{\zeta(\bb,\eta_1,\eta_2)-r(\o,\bb)\}
\left\{\frac{\partial\zeta(\bb,\eta_1,\eta_2)}{\partial\bb}
-\frac{\partial r(\o,\bb)}{\partial\bb}\right\}
f_\O(\o,\bb,\eta_1,\eta_2)d\o\\
&=&E\left[I\{ r(\O,\bb)<\zeta(\bb,\eta_1,\eta_2)\}\S_{\bb}(\O,\bb,\eta_1,\eta_2)\right]\\
&&+\frac{\partial\zeta(\bb,\eta_1,\eta_2)}{\partial\bb}
\int\delta\{\zeta(\bb,\eta_1,\eta_2)-r(\o,\bb)\}
f_\O(\o,\bb,\eta_1,\eta_2)d\o\\
&&-\int\delta\{\zeta(\bb,\eta_1,\eta_2)-r(\o,\bb)\}
\frac{\partial r(\o,\bb)}{\partial\bb}
f_\O(\o,\bb,\eta_1,\eta_2)d\o,
\ese
hence
\bse
\frac{\partial\zeta(\bb,\eta_1,\eta_2)}{\partial\bb}
&=&\frac{\int\delta\{\zeta-r(\o,\bb)\}
\frac{\partial r(\o,\bb)}{\partial\bb}
f_\O(\o,\bb,\eta_1,\eta_2)d\o
}{\int\delta\{\zeta(\bb,\eta_1,\eta_2)-r(\o,\bb)\}
f_\O(\o,\bb,\eta_1,\eta_2)d\o}\\
&&-\frac{ E\left[I\{
    r(\O,\bb)<\zeta\}\S_{\bb}(\O,\bb,\eta_1,\eta_2)\right] 
}{\int\delta\{\zeta(\bb,\eta_1,\eta_2)-r(\o,\bb)\}
f_\O(\o,\bb,\eta_1,\eta_2)d\o}\\
&=&\frac{E[\delta\{\zeta-r(\O,\bb)\}
\frac{\partial r(\O,\bb)}{\partial\bb}]
-E[I\{
    r(\O,\bb)<\zeta\}\S_{\bb}(\O,\bb,\eta_1,\eta_2)]
}{E[\delta\{\zeta-r(\O,\bb)\}]}.
\ese
Similarly, replacing $\eta_1$ with its parametric submodel,
and taking derivative with respect to $\bg_1$, we have
\bse
\0&=&\int I\{r(\o,\bb)<\zeta(\bb,\bg_1,\eta_2)\}
\S_{\bg_1}(\o,\bb,\bg_1,\eta_2)f_\O(\o,\bb,\bg_1,\eta_2)d\o\\
&&+\int\delta\{\zeta(\bb,\bg_1,\eta_2)-r(\o,\bb)\}
\frac{\partial\zeta(\bb,\bg_1,\eta_2)}{\partial\bg_1}
f_\O(\o,\bb,\bg_1,\eta_2)d\o\\
&=&E\left[I\{ r(\O,\bb)<\zeta\}\S_{\bg_1}(\O,\bb,\bg_1,\eta_2)\right]
+\frac{\partial\zeta(\bb,\bg_1,\eta_2)}{\partial\bg_1}
\int\delta\{\zeta-r(\o,\bb)\}
f_\O(\o,\bb,\bg_1,\eta_2)d\o,
\ese
hence
\bse
\frac{\partial\zeta(\bb,\bg_1,\eta_2)}{\partial\bg_1}
=-
\frac{
E[I\{ r(\O,\bb)<\zeta\}\S_{\bg_1}(\O, \bb,\bg_1,\eta_2)]
}{E[\delta\{\zeta-r(\O,\bb)\}]}.
\ese
Finally, replacing $\eta_2$ with its parametric submodel, and 
taking derivative with respect to $\bg_2$, we get
\bse
\0&=&\int I\{r(\o,\bb)<\zeta(\bb,\eta_1,\bg_2)\}
\S_{\bg_2}(w,\z,\bg_2)f_\O(\o,\bb,\eta_1,\bg_2)d\o\\
&&+\int\delta\{\zeta(\bb,\eta_1,\bg_2)-r(\o,\bb)\}
\frac{\partial\zeta(\bb,\eta_1,\bg_2)}{\partial\bg_2}
f_\O(\o,\bb,\eta_1,\bg_2)d\o\\
&=&E\left[I\{ r(\O,\bb)<\zeta\}\S_{\bg_2}(W,\Z,\bg_2)\right]
+\frac{\partial\zeta(\bb,\eta_1,\bg_2)}{\partial\bg_2}
\int\delta\{\zeta-r(\o,\bb)\}
f_\O(\o,\bb,\eta_1,\bg_2)d\o,
\ese
hence
\bse
\frac{\partial\zeta(\bb,\eta_1,\bg_2)}{\partial\bg_2}
=\frac{-E[I\{ r(\O,\bb)<\zeta\}\S_{\bg_2}(W,\Z,\bg_2)]
}{E[\delta\{\zeta-r(\O,\bb)\}]}.
\ese

Let $\phi(\o,\bb,\zeta)$ be the efficient influence function for estimating
$\zeta$. Then it has the form
$\phi(\o,\bb,\zeta)=a_1(\o)+a_2(w,\z)+\epsilon\a\trans\m_\bb'(w,\z,\bb)/\var(\epsilon^2\mid
w,\z)$, where 
$E\{\epsilon a_1(\O)\mid w,\z\}=E\{a_1(\O)\mid w,\z\}=0$ and $E\{a_2(W,\Z)\}=0$.
In addition, at the true
parameter values, 
$\phi(\o,\bb,\zeta)$ satisfies
\bse
\frac{\partial\zeta(\bb,\eta_1,\eta_2)}{\partial\bb}&=&E\{\phi(\O,\bb,\zeta)\S_{\bb}(\O)\},\\
\frac{\partial\zeta(\bb,\bg_1,\eta_2)}{\partial\bg_1}&=&E\{\phi(\O,\bb,\zeta)\S_{\bg_1}(\O)\},\\
\frac{\partial\zeta(\bb,\eta_1,\bg_2)}{\partial\bg_2}&=&E\{\phi(\O,\bb,\zeta)\S_{\bg_2}(\W,\Z)\}.
\ese
This implies that $a_1(\o,\bb,\zeta),a_2(w,\z), \a$ must satisfy 
\bse
&&\frac{E[\delta\{\zeta-r(\O,\bb)\}
{\partial r(\O,\bb)}/{\partial\bb}]
-E[I\{
    r(\O,\bb)<\zeta\}\S_{\bb}(\O,\bb,\eta_1,\eta_2)]
}{E[\delta\{\zeta-r(\O,\bb)\}]}\\
&=&E\left(\{a_1(\O,\bb,\zeta)+a_2(W,\Z)+
\epsilon\a\trans\m_\bb'(W,\Z,\bb)/\var(\epsilon^2\mid
w,\z)\}\S_{\bb}(\O,\bb,\eta_1,\eta_2)\right)\\
&=&E\left[\{a_1(\O,\bb,\zeta)+\epsilon\a\trans\m_\bb'(W,\Z,\bb)/\var(\epsilon^2\mid
w,\z)\}\{-d\log f_{\epsilon\mid
  W,\Z}(\epsilon,W,\Z)/d\epsilon 
\m_\bb'(w,\z,\bb)\}\right]\\
&=&E\left\{a_1(\O,\bb,\zeta)
  \S_\bb(\O,\bb,\bg_1,\bg_2)\right\}+E\left\{\frac{\a\trans\m_\bb'(W,\Z,\bb)^{\otimes2}}{\var(\epsilon^2\mid
W,\Z)}
\right\},\\
&&
\frac{
-E[I\{ r(\O,\bb)<\zeta\}\S_{\bg_1}(\O, \bb,\bg_1,\eta_2)]
}{E[\delta\{\zeta-r(\O,\bb)\}]}\\
&=&E\left[\{a_1(\O,\bb,\zeta)+a_2(W,\Z)+\epsilon\a\trans\m_\bb'(W,\Z,\bb)/\var(\epsilon^2\mid
W,\Z)\}
\S_{\bg_1}(\O)\right]\\
&=&E\{a_1(\O,\bb,\zeta)\S_{\bg_1}(\O)\},\\
&&\frac{
-E[I\{ r(\O,\bb)<\zeta\}\S_{\bg_2}(W,\Z,\bg_2)]
}{E[\delta\{\zeta-r(\O,\bb)\}]}\\
&=&E\left[\{a_1(\O,\bb,\zeta)+a_2(W,\Z)+\epsilon\a\trans\m_\bb'(W,\Z,\bb)/\var(\epsilon^2\mid
W,\Z)\}
\S_{\bg_2}(W,\Z)\right]\\
&=&E\left\{a_2(W,\Z)
\S_{\bg_2}(W,\Z)\right\}.
\ese
Because $\S_{\bg_2}(W,\Z)$ can  be an arbitrary mean zero
  function of $W,\Z$, and $\S_{\bg_1}(\O)$ can be any function that
  satisfies $E\{\S_{\bg_1}(\O)\mid W,\Z\}=
  E\{\epsilon\S_{\bg_1}(\O)\mid W,\Z\}=\0$,
these requirements directly lead to
\bse
a_2(W,\Z)&=&\frac{(1-\alpha)
-E[I\{ r(\O,\bb)<\zeta\}\mid W,\Z]
}{E[\delta\{\zeta-r(\O,\bb)\}]},\n\\
a_1(\O)&=&\frac{
E[I\{
  r(\O,\bb)<\zeta\}\mid W,\Z]-
I\{
  r(\O,\bb)<\zeta\}}{E[\delta\{\zeta-r(\O,\bb)\}]}
+\epsilon \frac{E[\epsilon I\{ r(\O,\bb)<\zeta\}\mid W,\Z]
}{E[\delta\{\zeta-r(\O,\bb)\}]{E(\epsilon^2\mid W,\Z) }},\n\\
\a
&=&\left[E\left\{\frac{\m_\bb'(W,\Z,\bb)^{\otimes2}}{\var(\epsilon^2\mid
W,\Z)}\right\}\right]^{-1}\\
&&\times E\left(\frac{E[\delta\{\zeta-r(\O,\bb)\}
{\partial r(\O,\bb)}/{\partial\bb}]
-E[I\{
    r(\O,\bb)<\zeta\}\S_{\bb}(\O,\bb,\eta_1,\eta_2)]
}{E[\delta\{\zeta-r(\O,\bb)\}]}\right.\\
&&\left.-
E\left\{a_1(\O,\bb,\zeta)\S_\bb(\O,\bb,\eta_1,\eta_2)\right\}\right)\\
&=&\left[E\left\{\frac{\m_\bb'(W,\Z,\bb)^{\otimes2}}{\var(\epsilon^2\mid
W,\Z)}\right\}\right]^{-1}\\
&&\times E\left(\frac{E[\delta\{\zeta-r(\O,\bb)\}
{\partial r(\O,\bb)}/{\partial\bb}]
}{E[\delta\{\zeta-r(\O,\bb)\}]}+ \frac{E[\epsilon I\{ r(\O,\bb)<\zeta\}\mid W,\Z]\m_\bb'(W,\Z,\bb)
}{E[\delta\{\zeta-r(\O,\bb)\}]E(\epsilon^2\mid W,\Z) }
\right).
\ese

In summary, our efficient influence function for estimating $\zeta$ is
\bse
\phi_{\rm naive}(\O,\bb,\zeta)&=&
\frac{1-\alpha-
I\{
  r(\O,\bb)<\zeta\}}{E[\delta\{\zeta-r(\O,\bb)\}]}
+\epsilon \frac{E[\epsilon I\{ r(\O,\bb)<\zeta\}\mid W,\Z]
}{E[\delta\{\zeta-r(\O,\bb)\}]E(\epsilon^2\mid W,\Z) }\\
&&+\epsilon \m_\bb'(w,\z,\bb)\trans\left[E\left\{\frac{\m_\bb'(W,\Z,\bb)^{\otimes2}}{\var(\epsilon^2\mid
W,\Z)}\right\}\right]^{-1}\\
&&\times E\left(\frac{\delta\{\zeta-r(\O,\bb)\}
{\partial r(\O,\bb)}/{\partial\bb}
}{E[\delta\{\zeta-r(\O,\bb)\}]}+
\frac{\epsilon I\{ r(\O,\bb)<\zeta\}\m_\bb'(W,\Z,\bb)
}{E[\delta\{\zeta-r(\O,\bb)\}]E(\epsilon^2\mid W,\Z) }
\right).
\ese

%%%%%%%%%%%%%%%%%%

\subsection{Additional Simulation}\label{sec:moresimu}

\begin{figure}[!h]
    \centering
    \caption{Simulation A1:  Misspecified regression model. Histogram of the 90\% coverage probability (CP) and
      prediction interval length (PI) of the estimated prediction
      interval in the three models using the six methods. $n=100$.}\label{fig:simu31}
    \includegraphics[scale = 0.33]{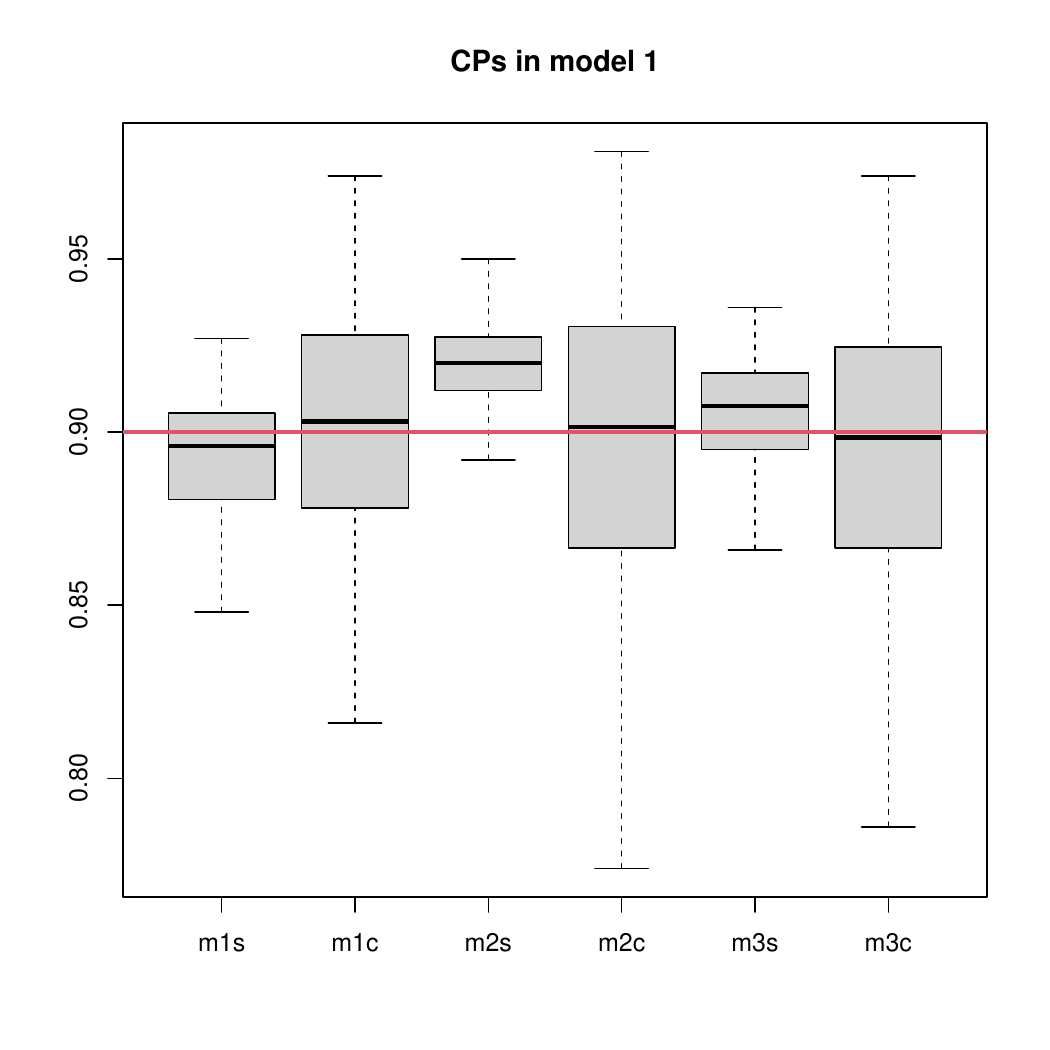}
    \includegraphics[scale = 0.33]{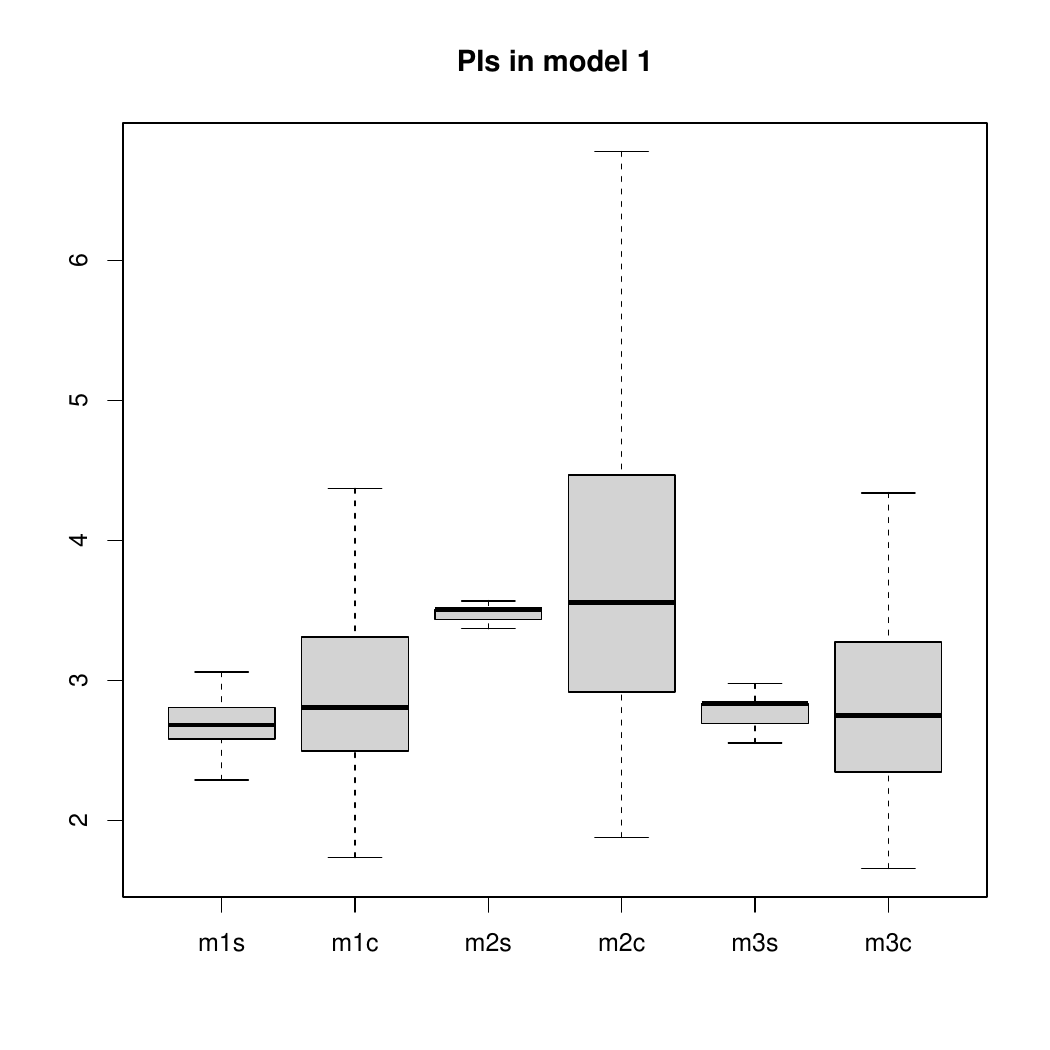}
        \includegraphics[scale = 0.33]{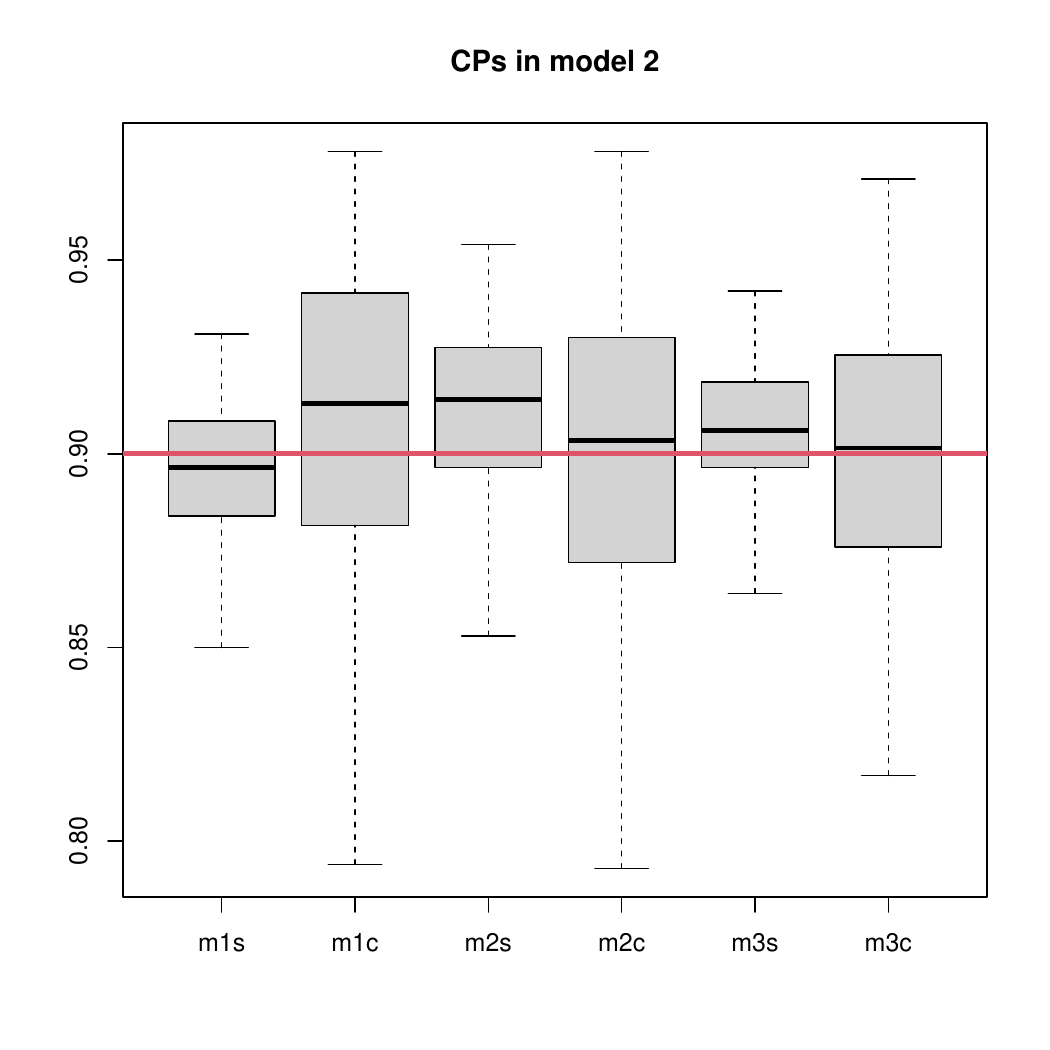}
        \includegraphics[scale = 0.33]{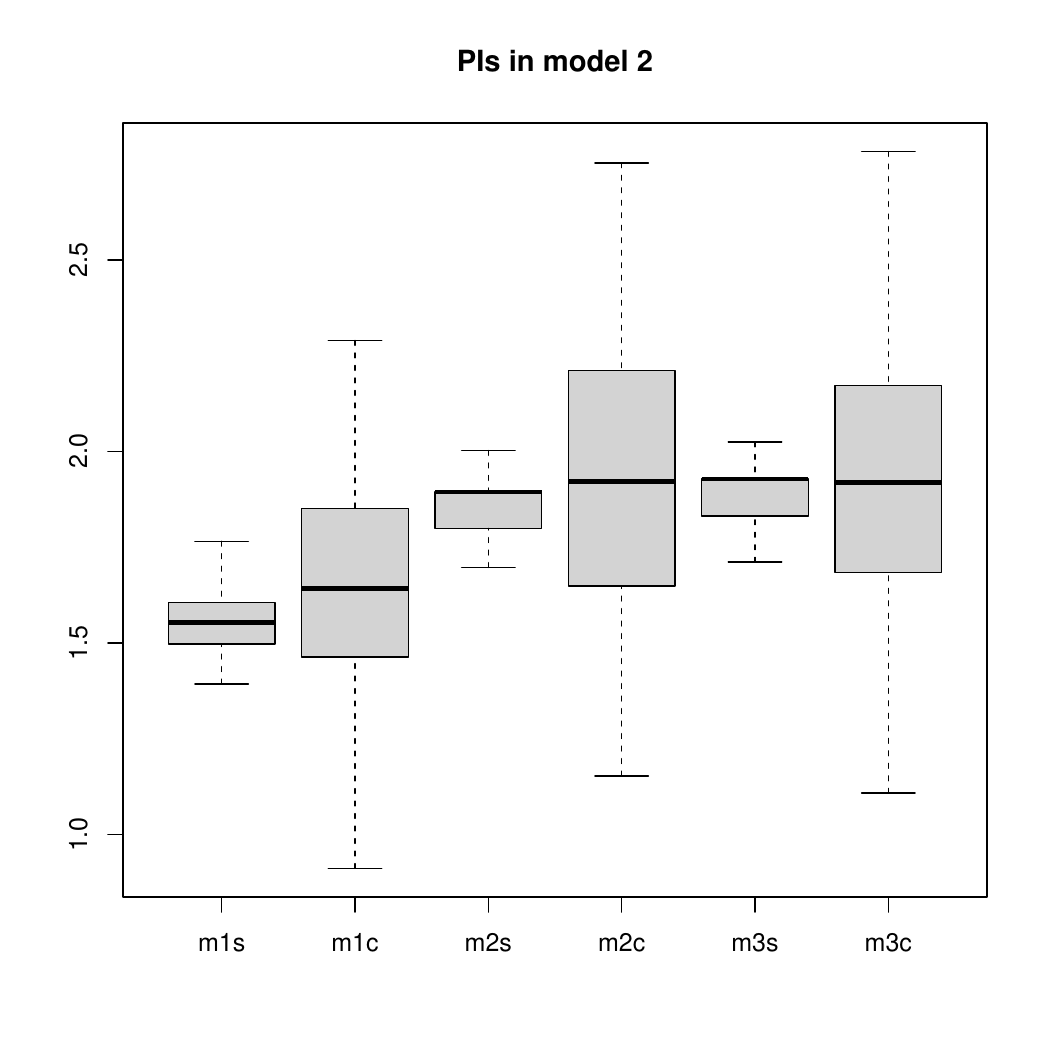}
            \includegraphics[scale = 0.33]{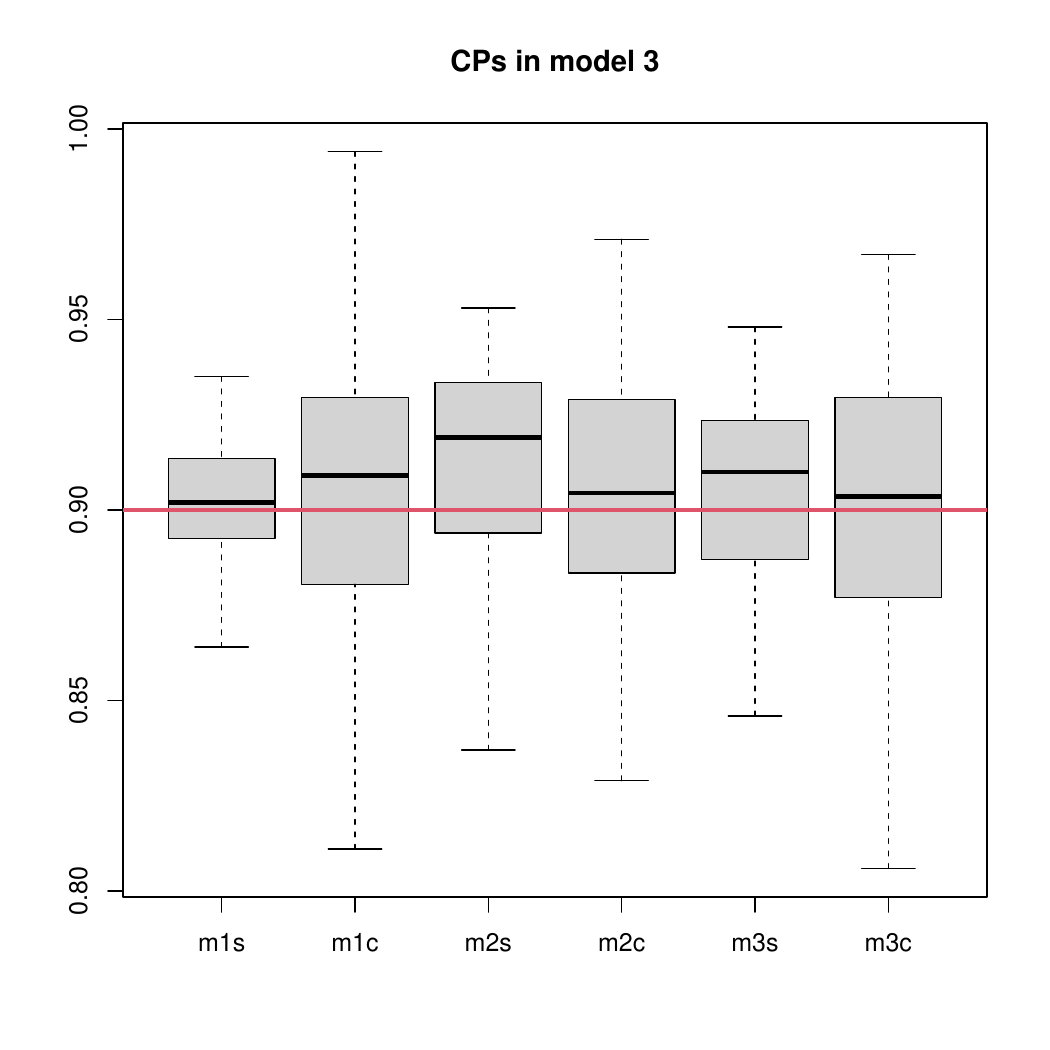}
     \includegraphics[scale = 0.33]{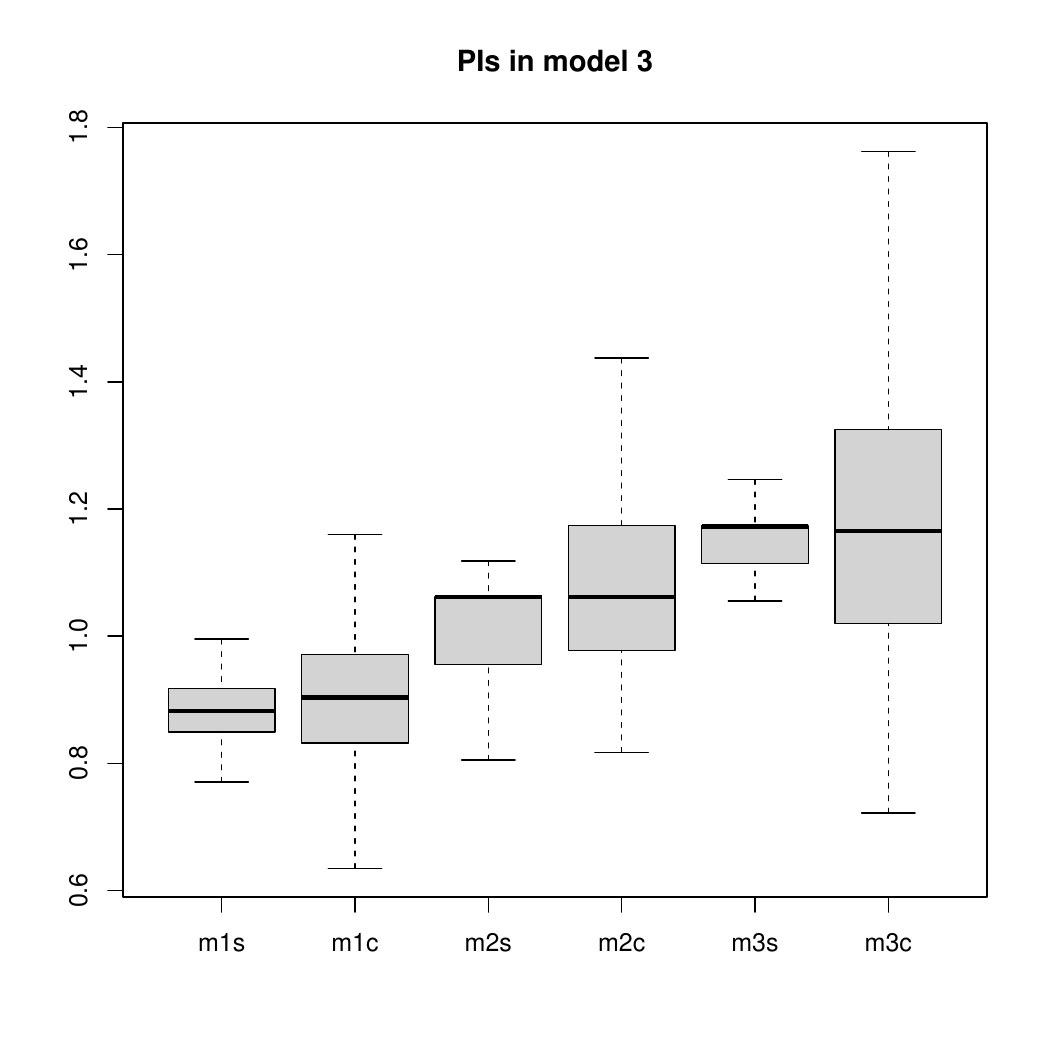}
   \end{figure}

\begin{figure}[!h]
    \centering
    \caption{Simulation A1: Misspecified regression model. Histogram of the 90\% coverage probability (CP) and
      prediction interval length (PI) of the estimated prediction
      interval in the three models using the six methods. $n=500$.}\label{fig:simu32}
    \includegraphics[scale = 0.33]{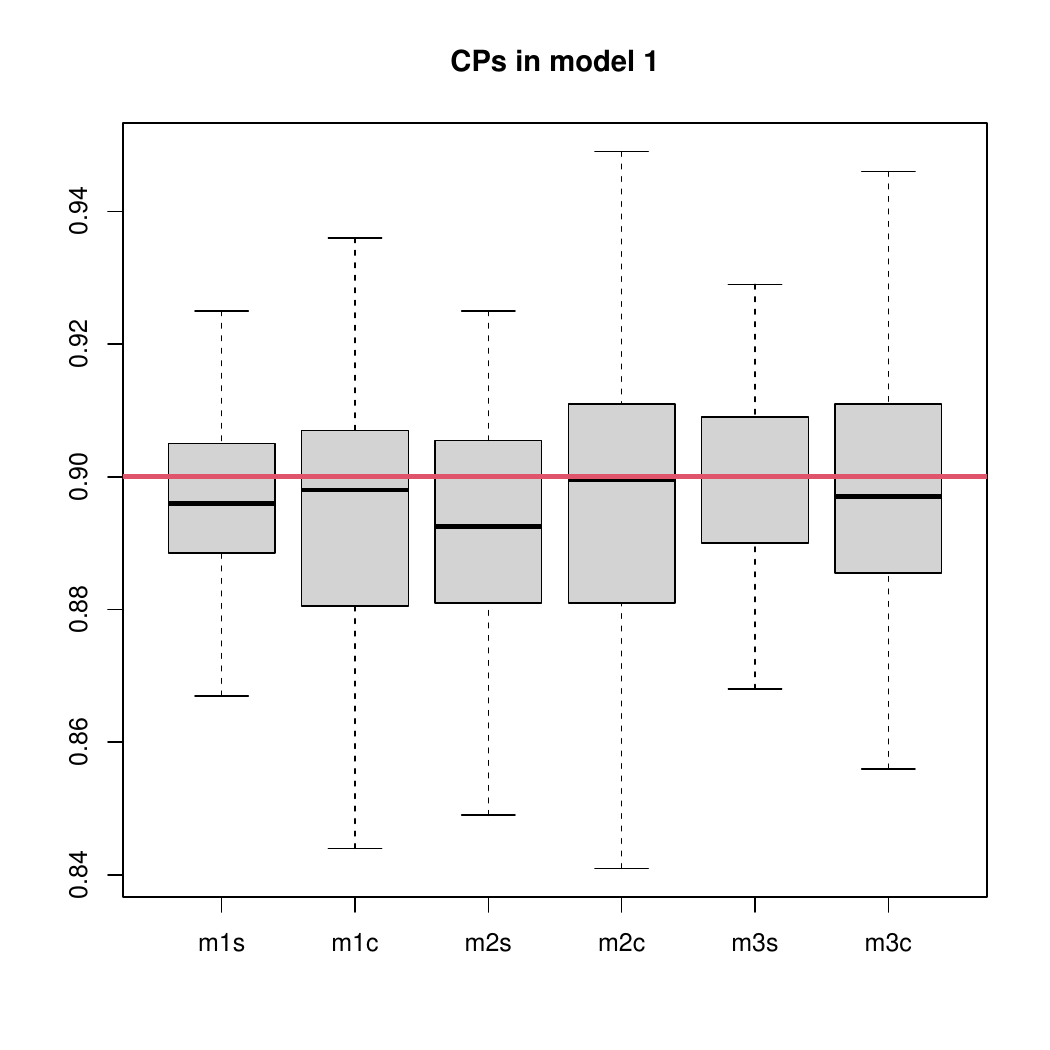}
    \includegraphics[scale = 0.33]{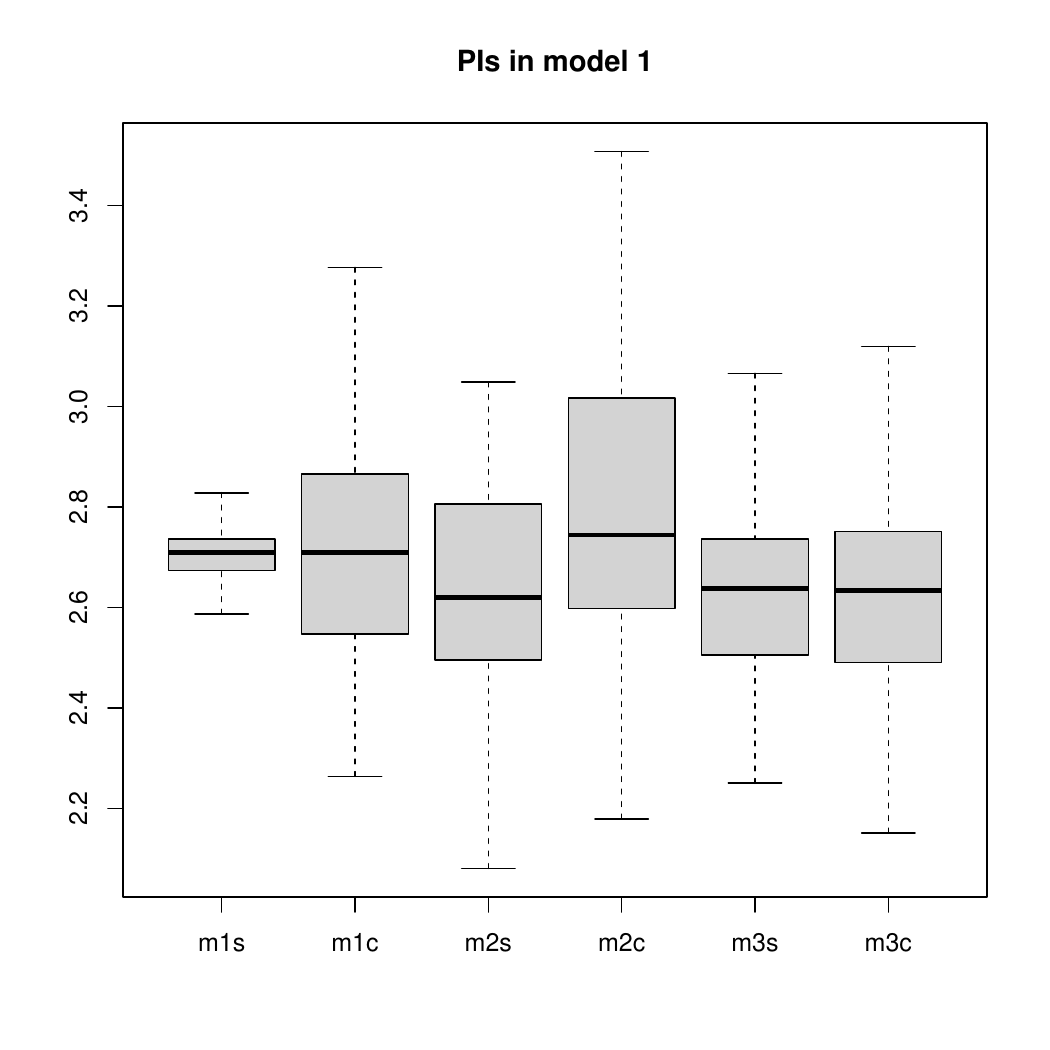}
        \includegraphics[scale = 0.33]{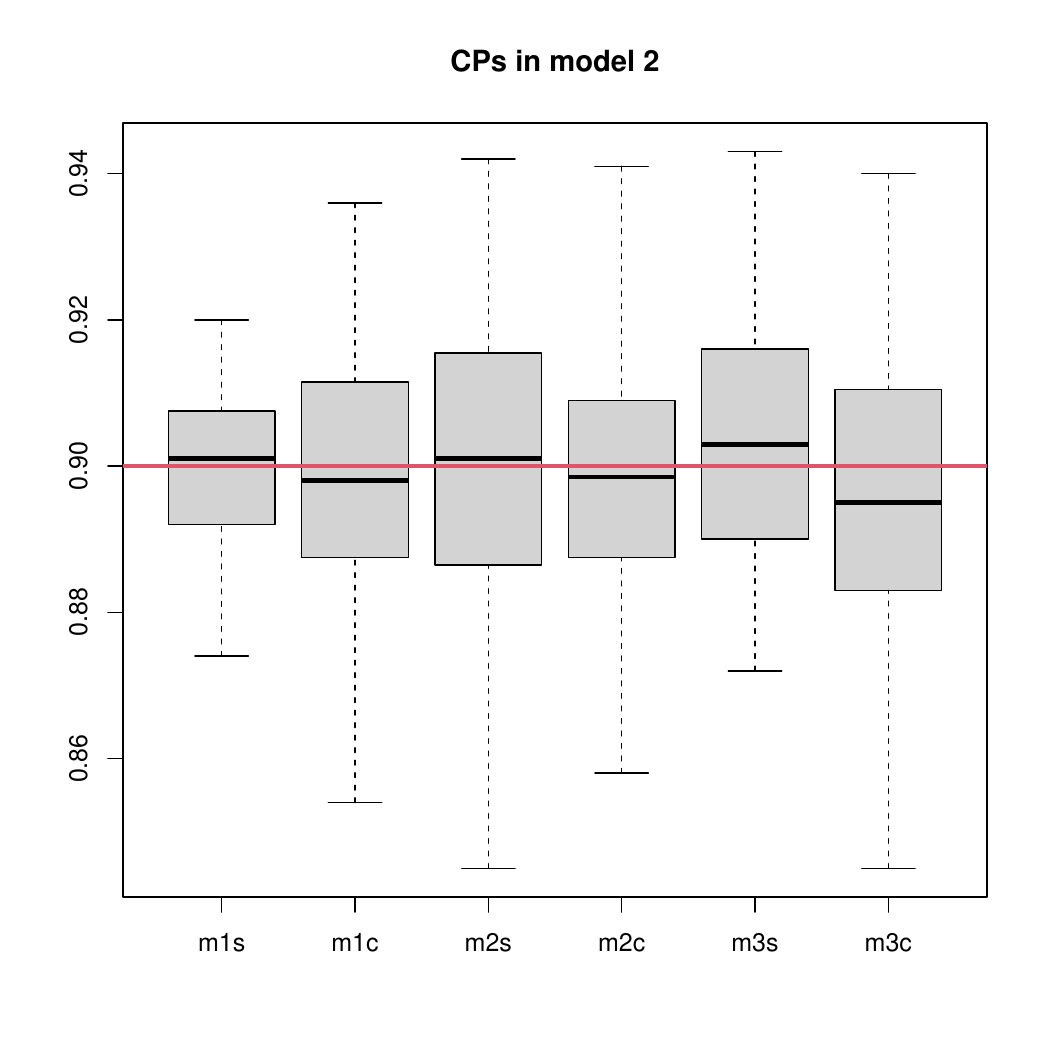}
        \includegraphics[scale = 0.33]{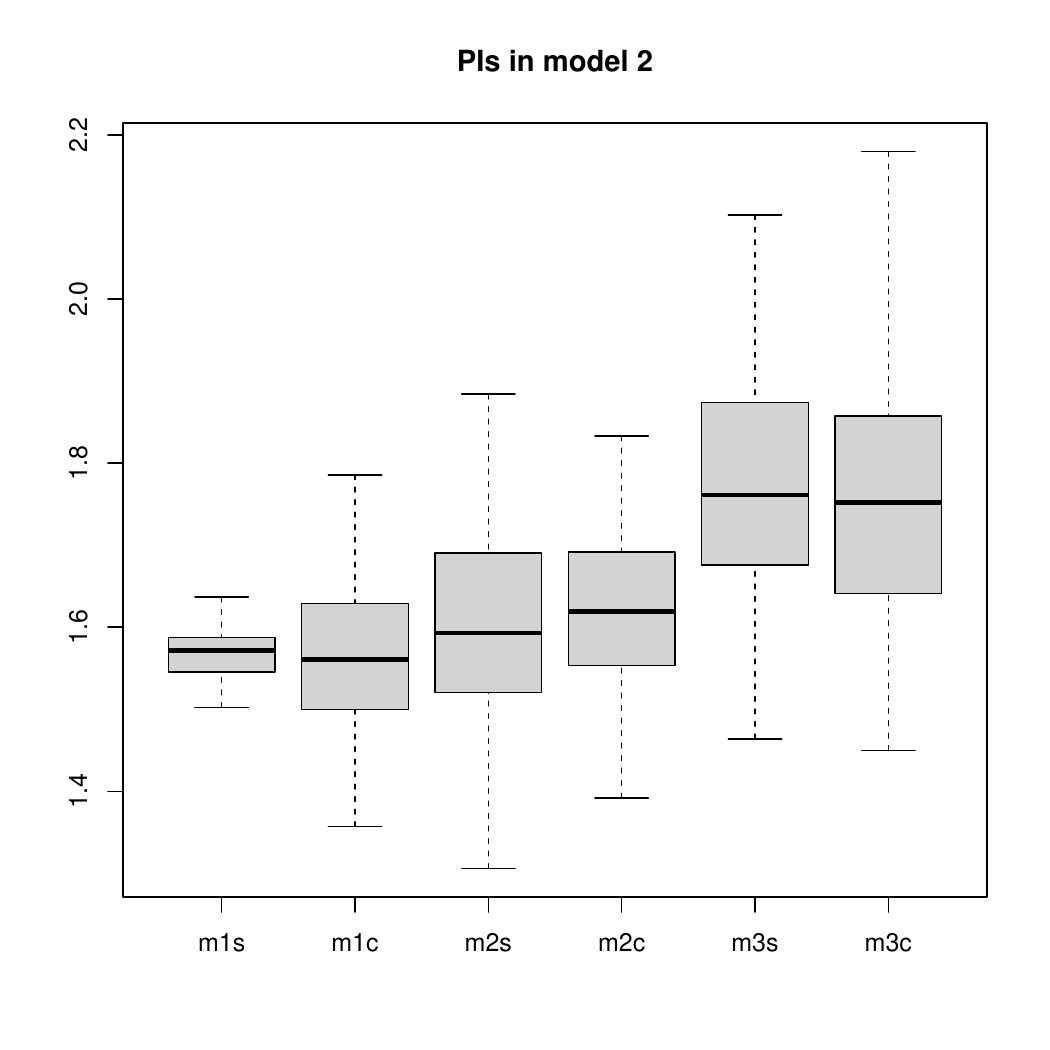}
            \includegraphics[scale = 0.33]{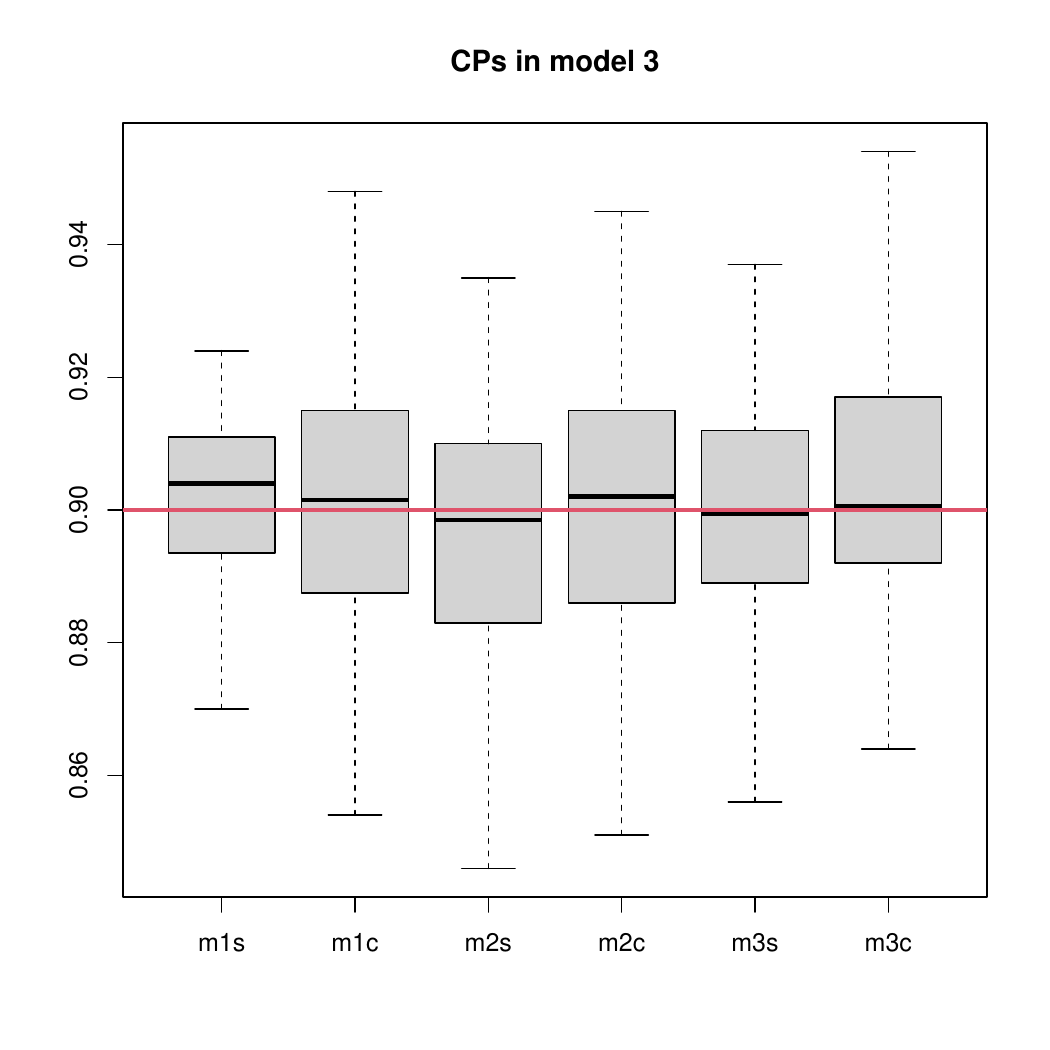}
     \includegraphics[scale = 0.33]{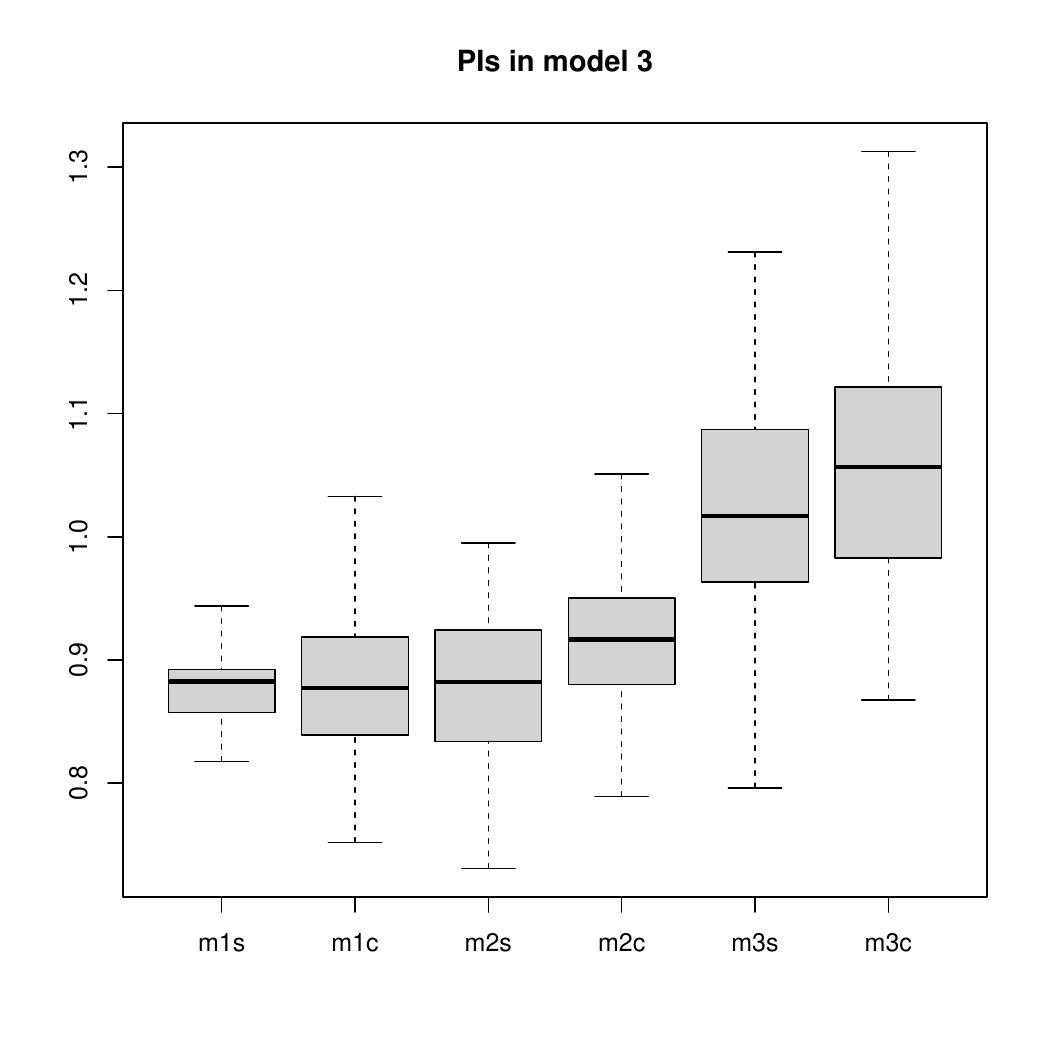}
   \end{figure}

To further investigate the sensitivity of the prediction
  methods to model misspecification, we conducted  three
  additional simulations. 
The additional simulation A1 is identical to Simulation 2 in Section
\ref{sec:simu}, except that when we generated the data, we generated
the regression errors $\epsilon_i$'s from $\epsilon \sim
t(3)\sigma_{\epsilon} /\sqrt{3}$.  Thus, the regression model is
misspecified in Simulation A1. The results presented in Table
\ref{tab:sim3}, Figures \ref{fig:simu31} and \ref{fig:simu32} show
that the coverage probabilities of 
all methods are still reasonably close to the nominal level and the
semiparametric methods 
m1s, m2s and m3s have shorter LPI, smaller variation in CP and LPI
compared to  its conformal counterparts.

\begin{figure}[!h]
    \centering
    \caption{Simulation A2: Misspecified measurement
        error model.
      Histogram of the 90\% coverage probability (CP) and
      prediction interval length (PI) of the estimated prediction
      interval in the three models using the six methods when the
      distribution of $U$ is misspecified. $n=100$.}\label{fig:simu51}
    \includegraphics[scale = 0.33]{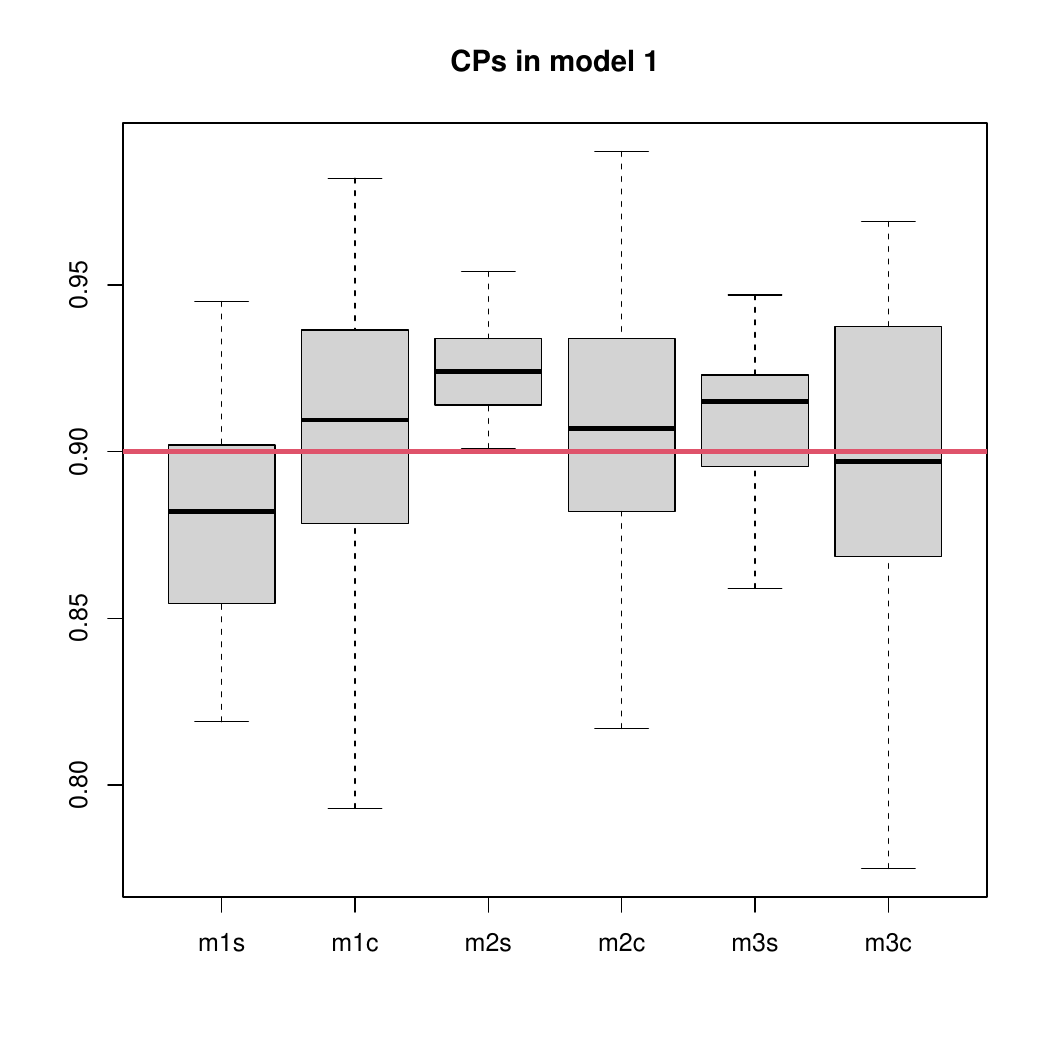}
    \includegraphics[scale = 0.33]{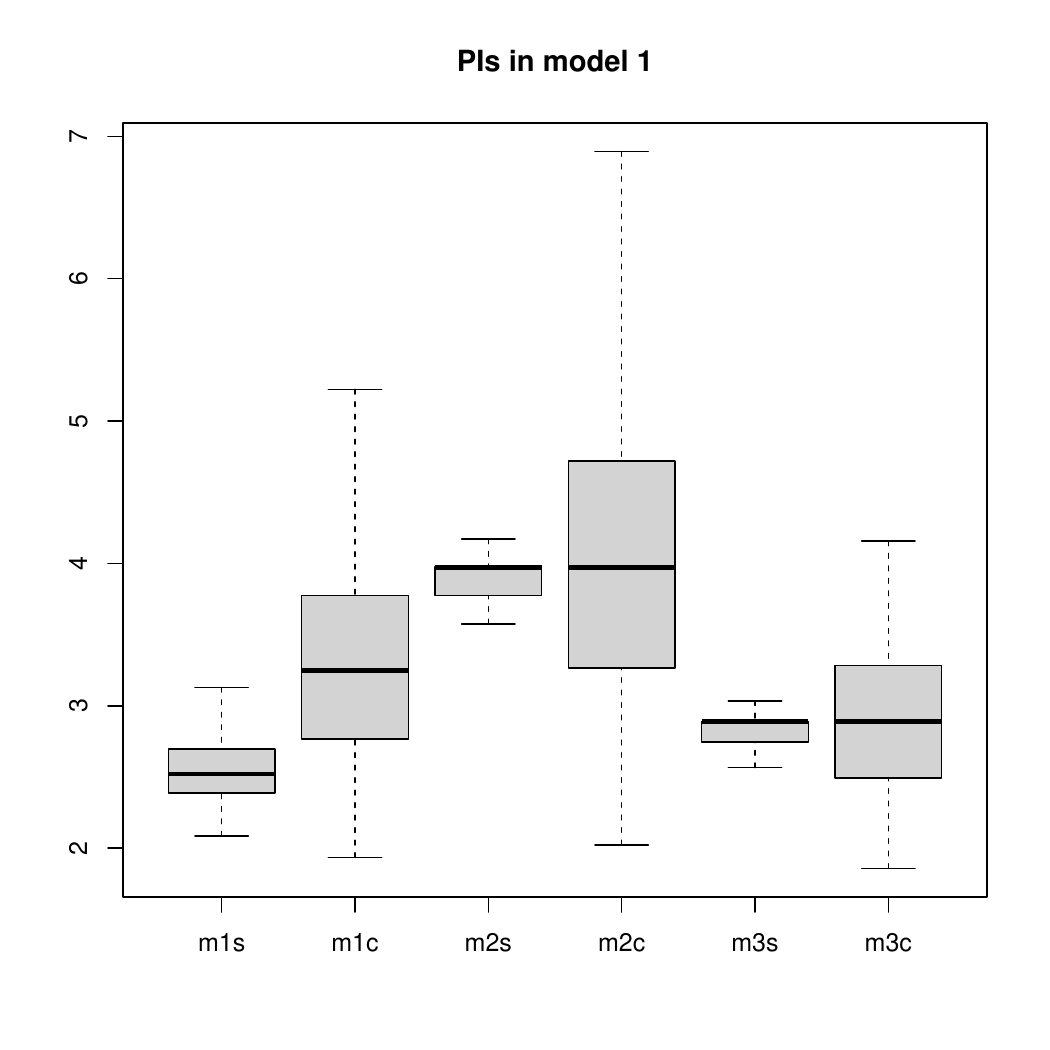}
        \includegraphics[scale = 0.33]{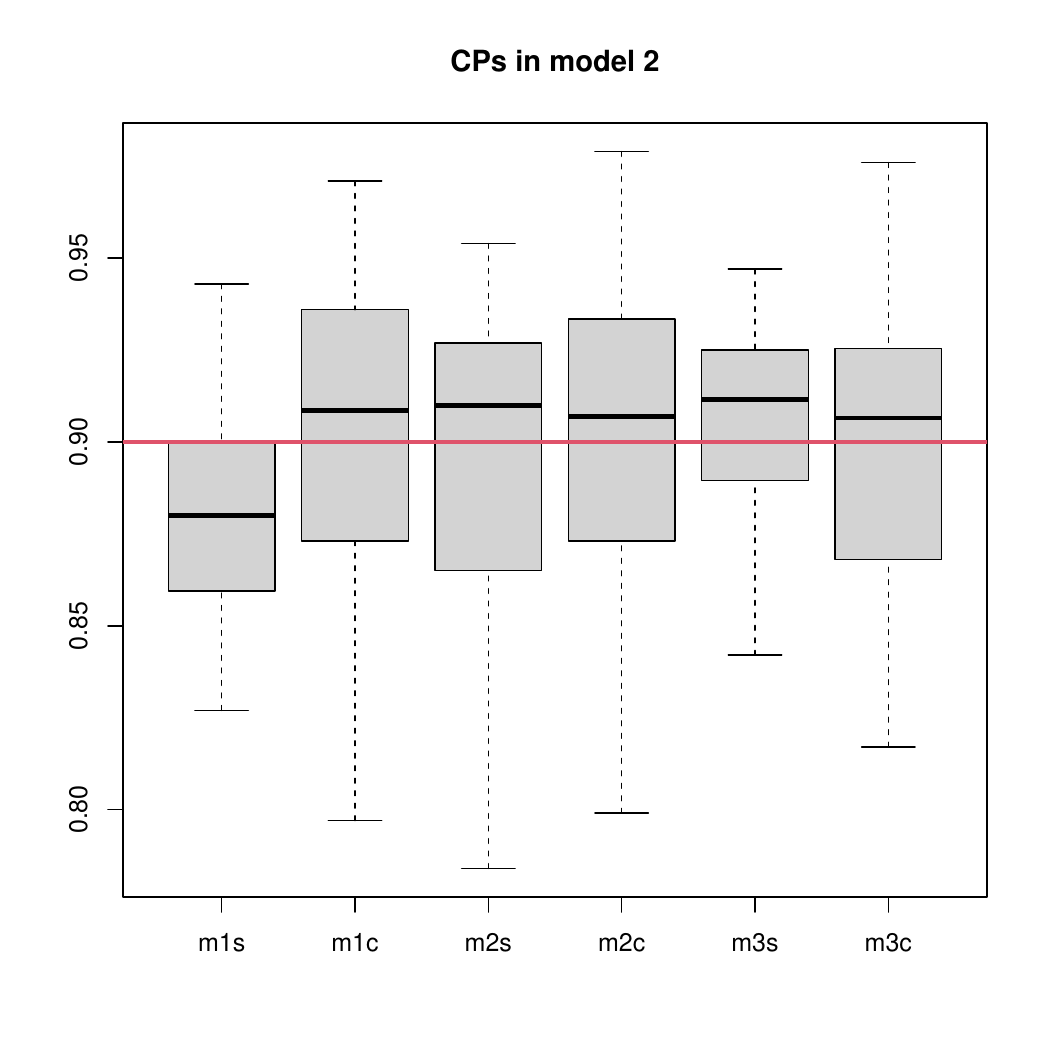}
        \includegraphics[scale = 0.33]{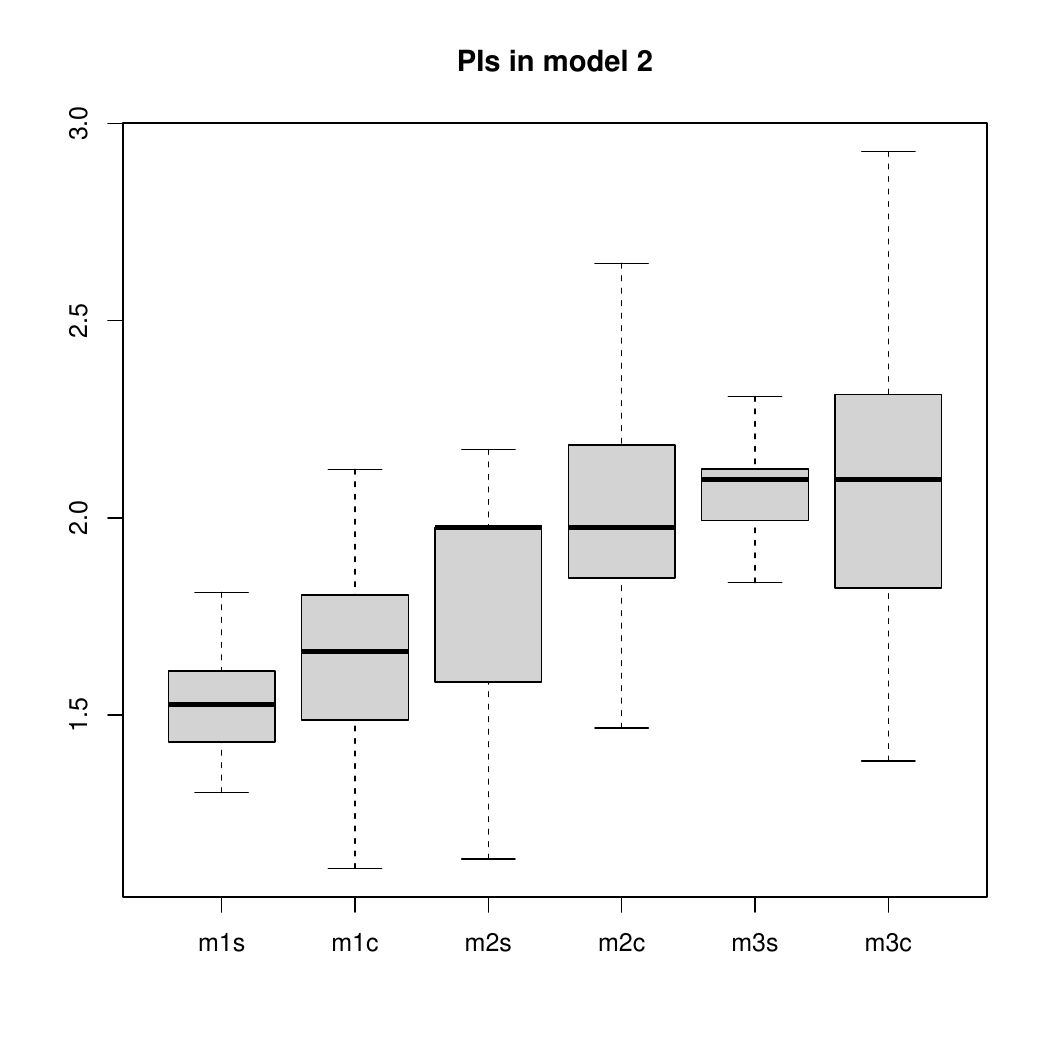}
            \includegraphics[scale = 0.33]{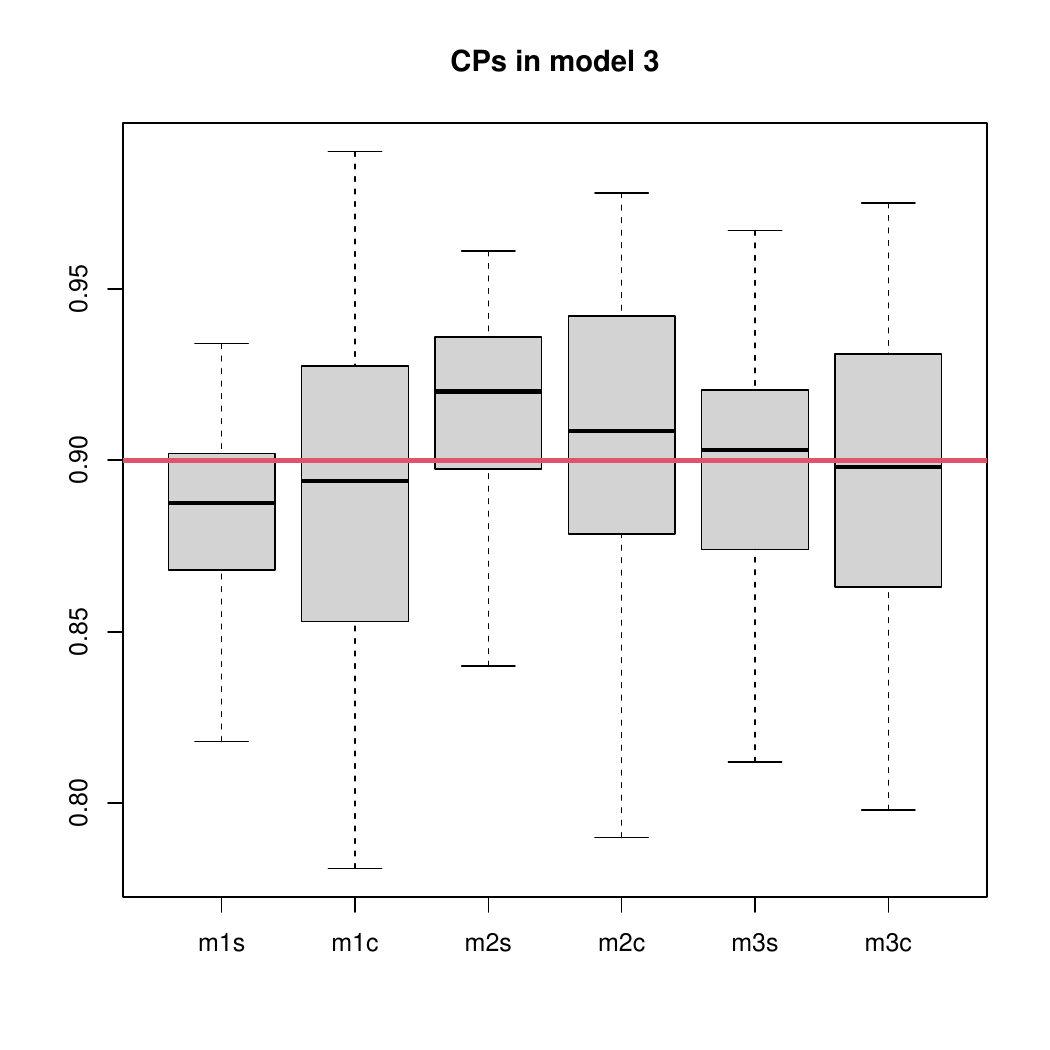}
     \includegraphics[scale = 0.33]{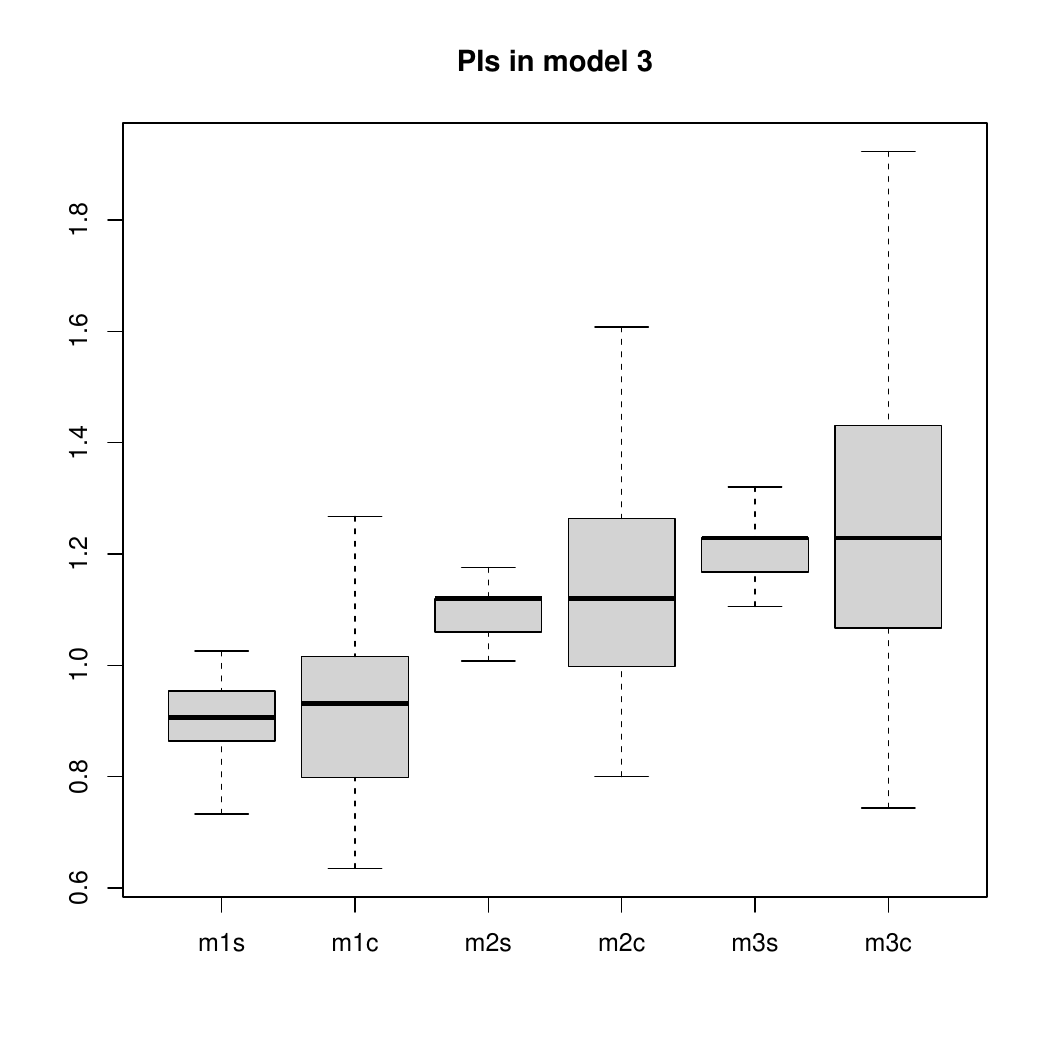}
   \end{figure}

\begin{figure}[!h]
    \centering
    \caption{Simulation A2: Misspecified measurement
        error model.
      Histogram of the 90\% coverage probability (CP) and
      prediction interval length (PI) of the estimated prediction
      interval in the three models using the six methods when the
      distribution of $U$ is misspecified. $n=500$.}\label{fig:simu52}
    \includegraphics[scale = 0.33]{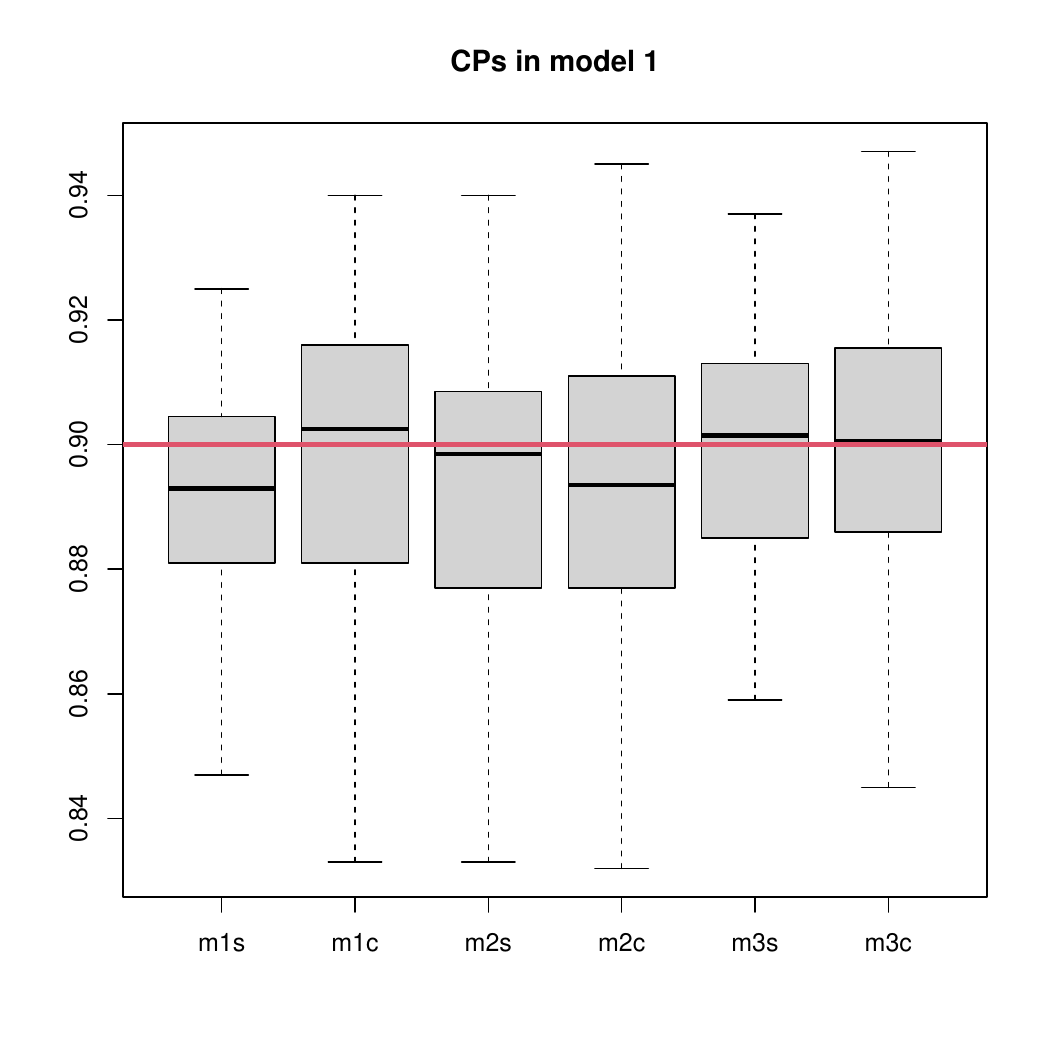}
    \includegraphics[scale = 0.33]{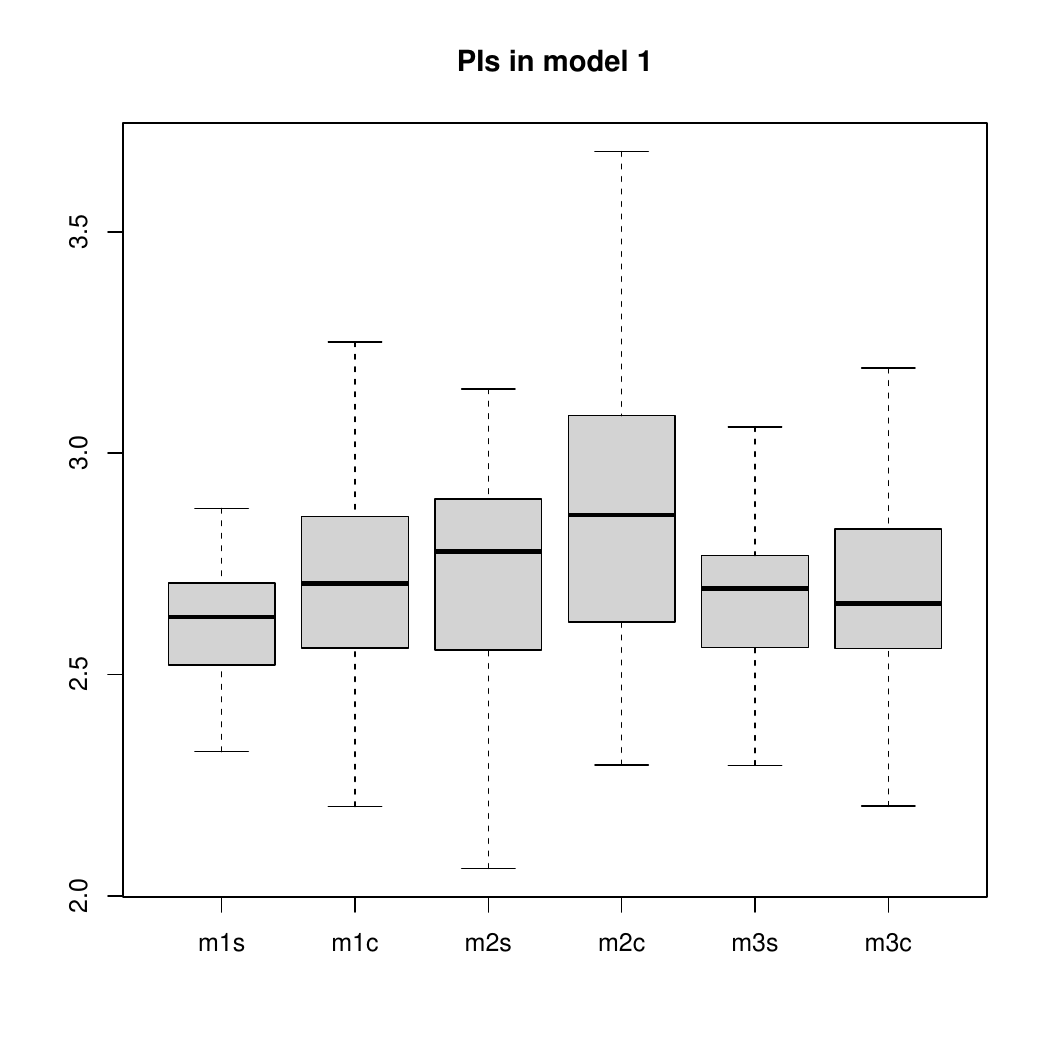}
        \includegraphics[scale = 0.33]{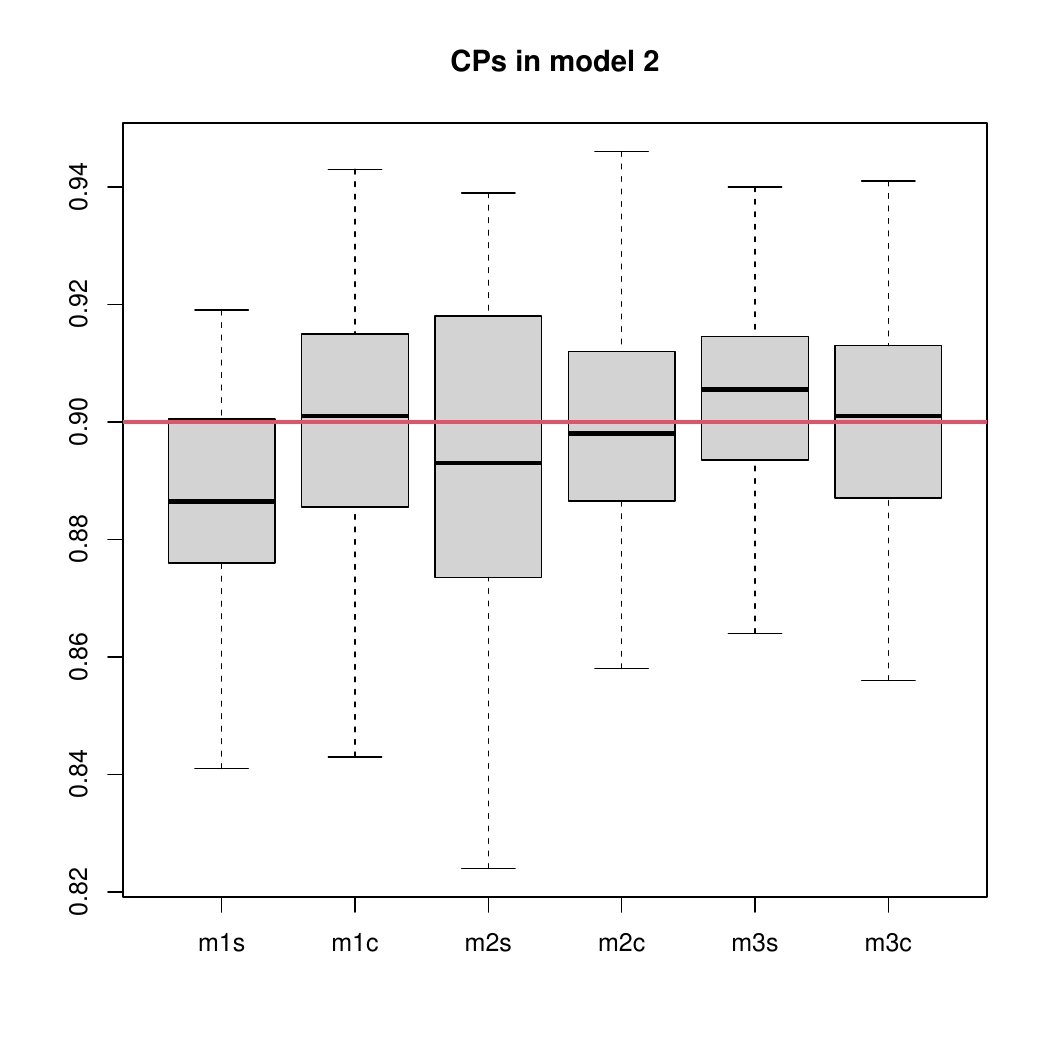}
        \includegraphics[scale = 0.33]{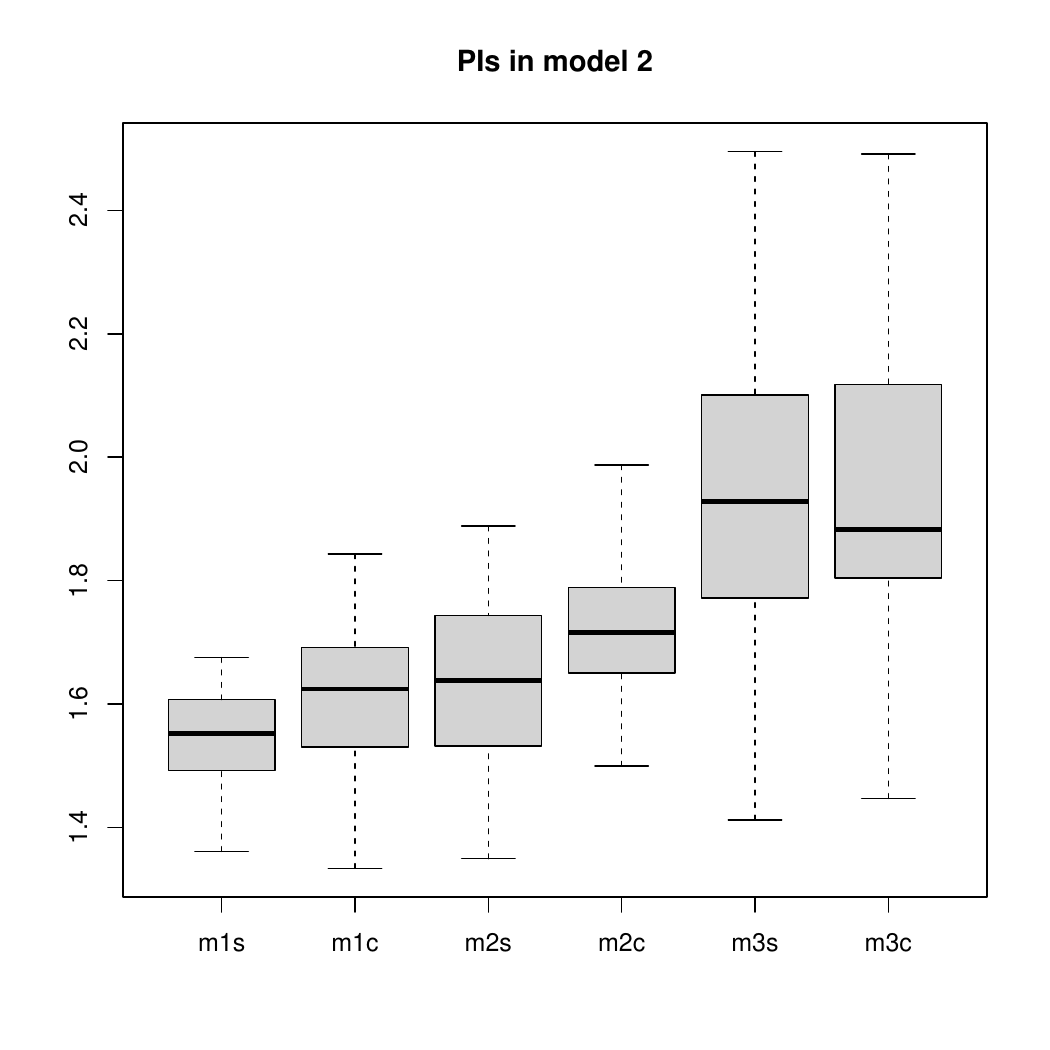}
            \includegraphics[scale = 0.33]{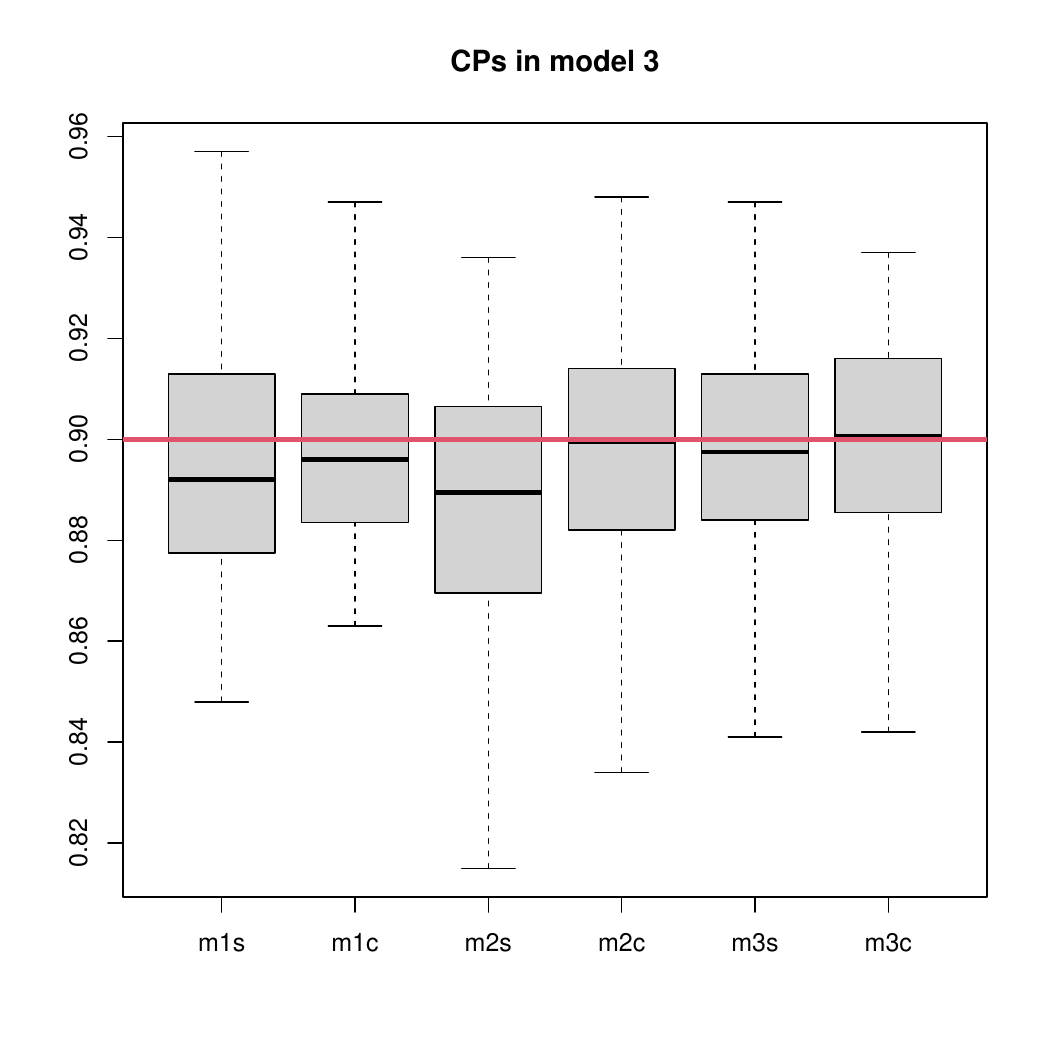}
     \includegraphics[scale = 0.33]{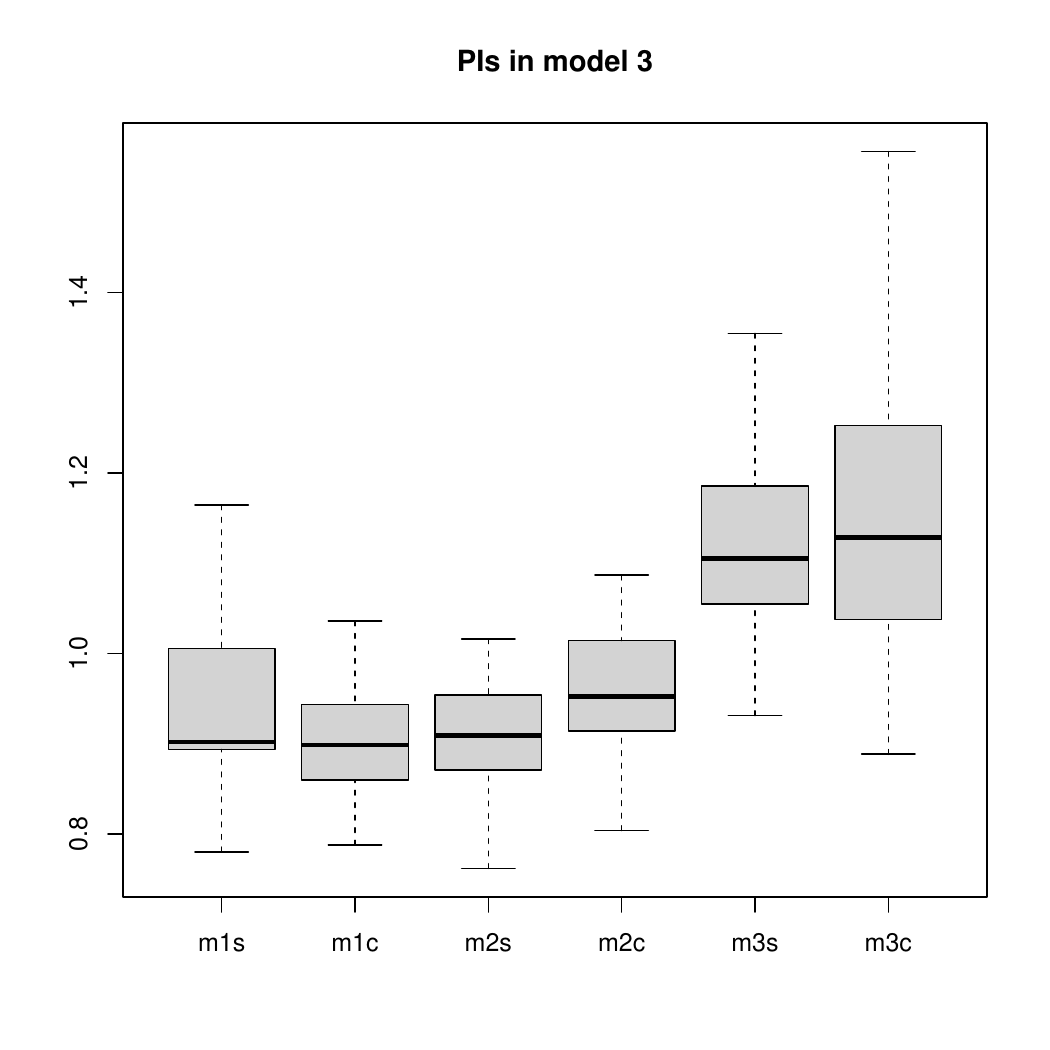}
   \end{figure}

We further conducted Simulation A2 to
  investigate the situation when the measurement error model is
  misspecified. Specifically, Simulation A2 is identical
  to Simulation 2 in Section 
\ref{sec:simu}, except that when we generated the data, we generated
the  measurement errors  $U_i$'s from 
$U\sim\sqrt{3} U(-1, 1)\sigma_U$. The results presented in Table
\ref{tab:sim5}, Figures \ref{fig:simu51} and Figure \ref{fig:simu52}
show that the coverage probabilities in the settings with sufficient
sample sizes ($n=500$)
are  close to the nominal level. 

\begin{figure}[!h]
    \centering
    \caption{Simulation A3: Misspecified regression model  
      and measurement error model. Histogram of the 90\% coverage probability (CP) and
      prediction interval length (PI) of the estimated prediction
      interval in the three models using the six methods when both the
      distributions of $U$ and $\epsilon$ are misspecified. $n=100$.}\label{fig:simu41}
    \includegraphics[scale = 0.33]{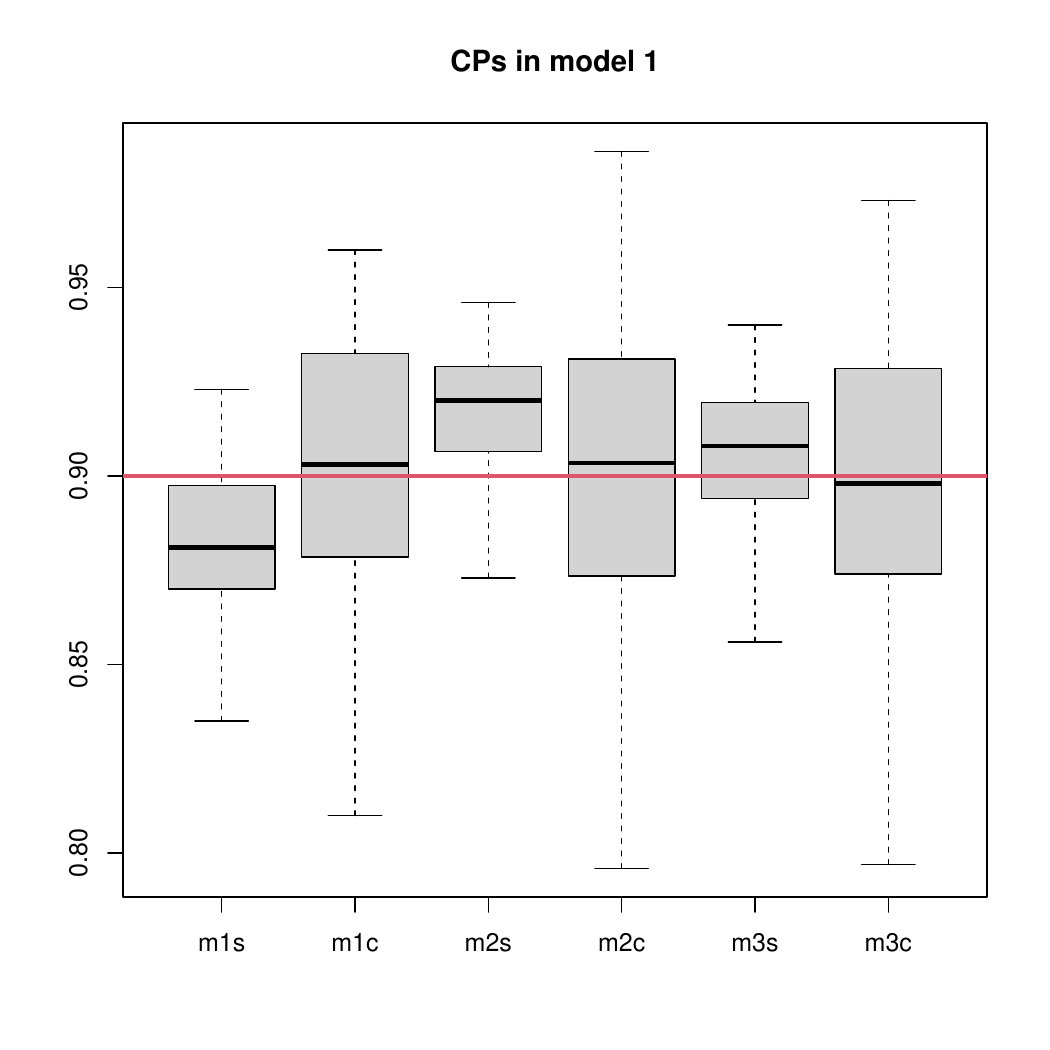}
    \includegraphics[scale = 0.33]{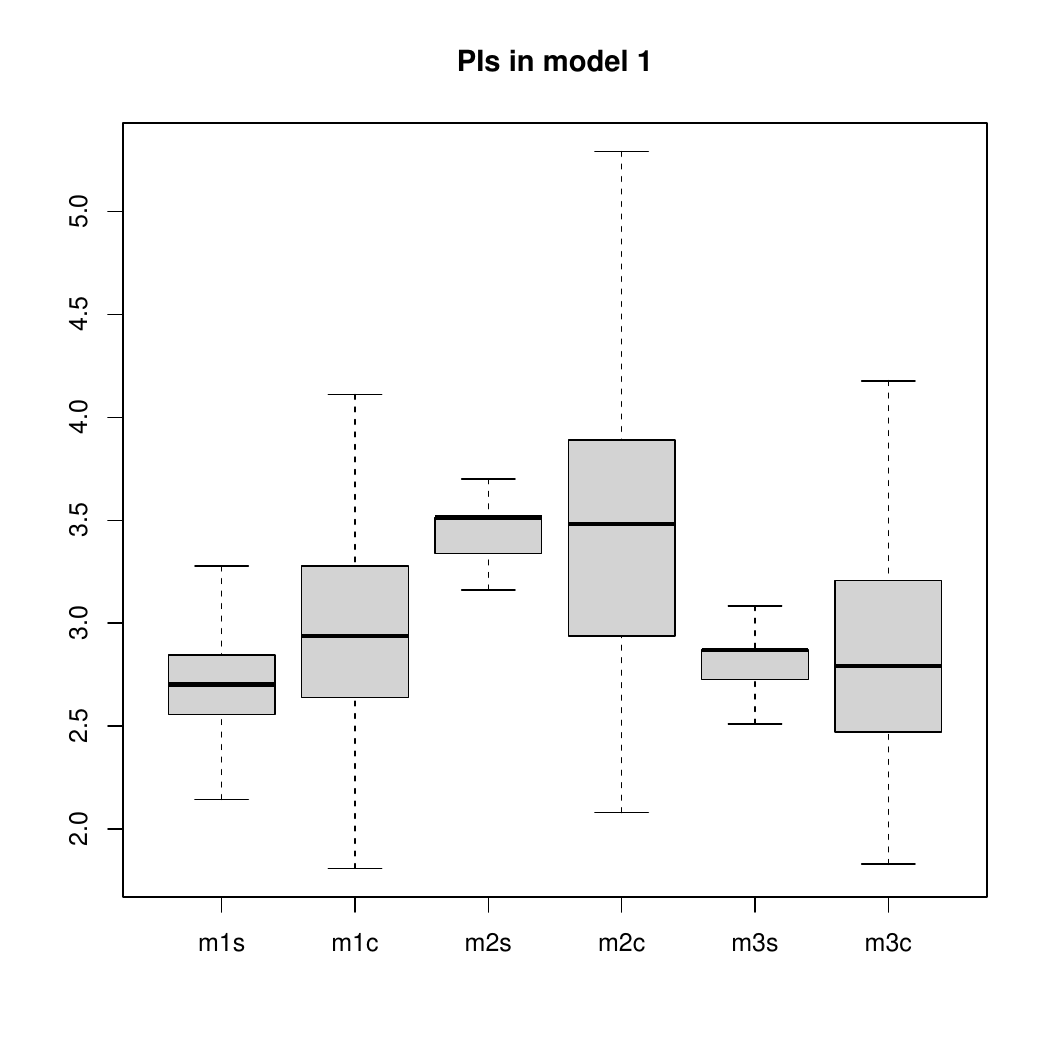}
        \includegraphics[scale = 0.33]{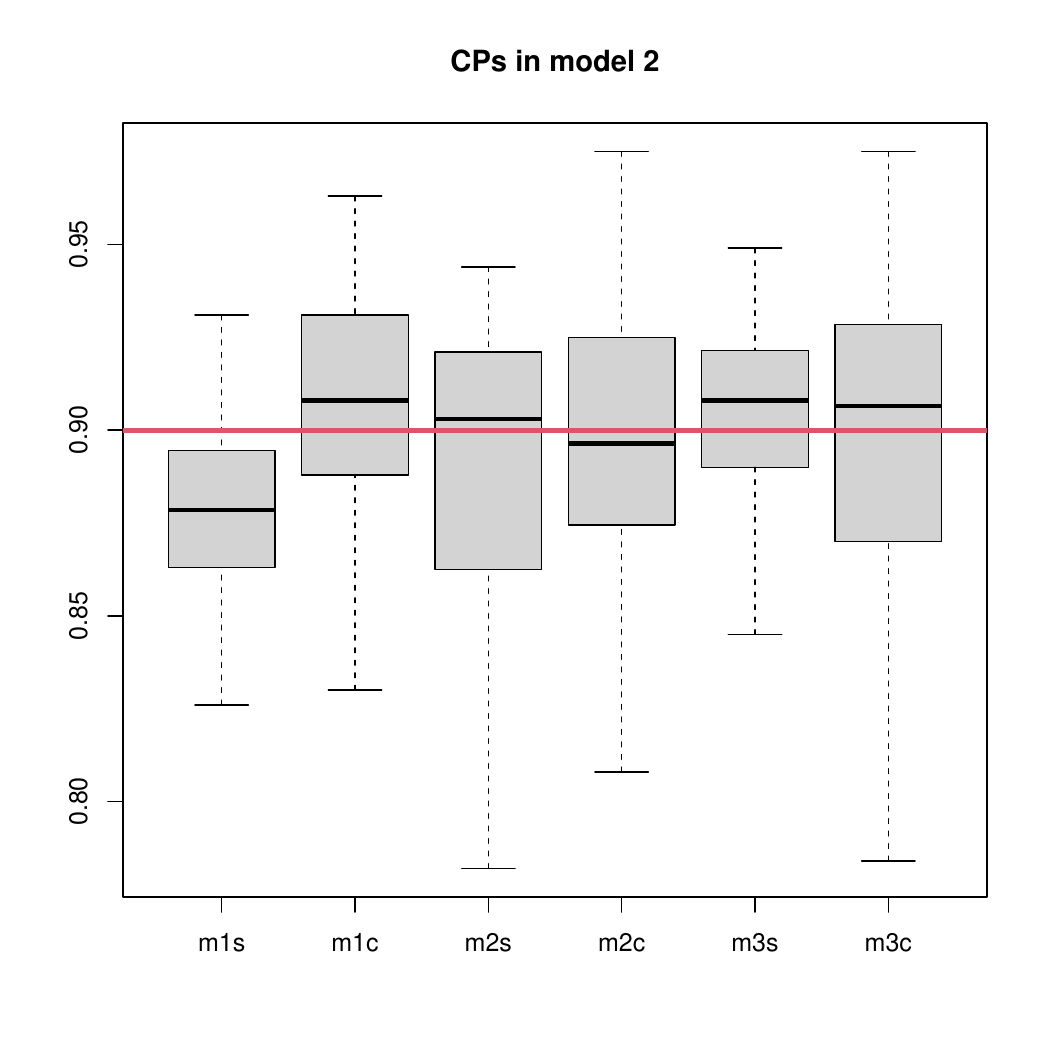}
        \includegraphics[scale = 0.33]{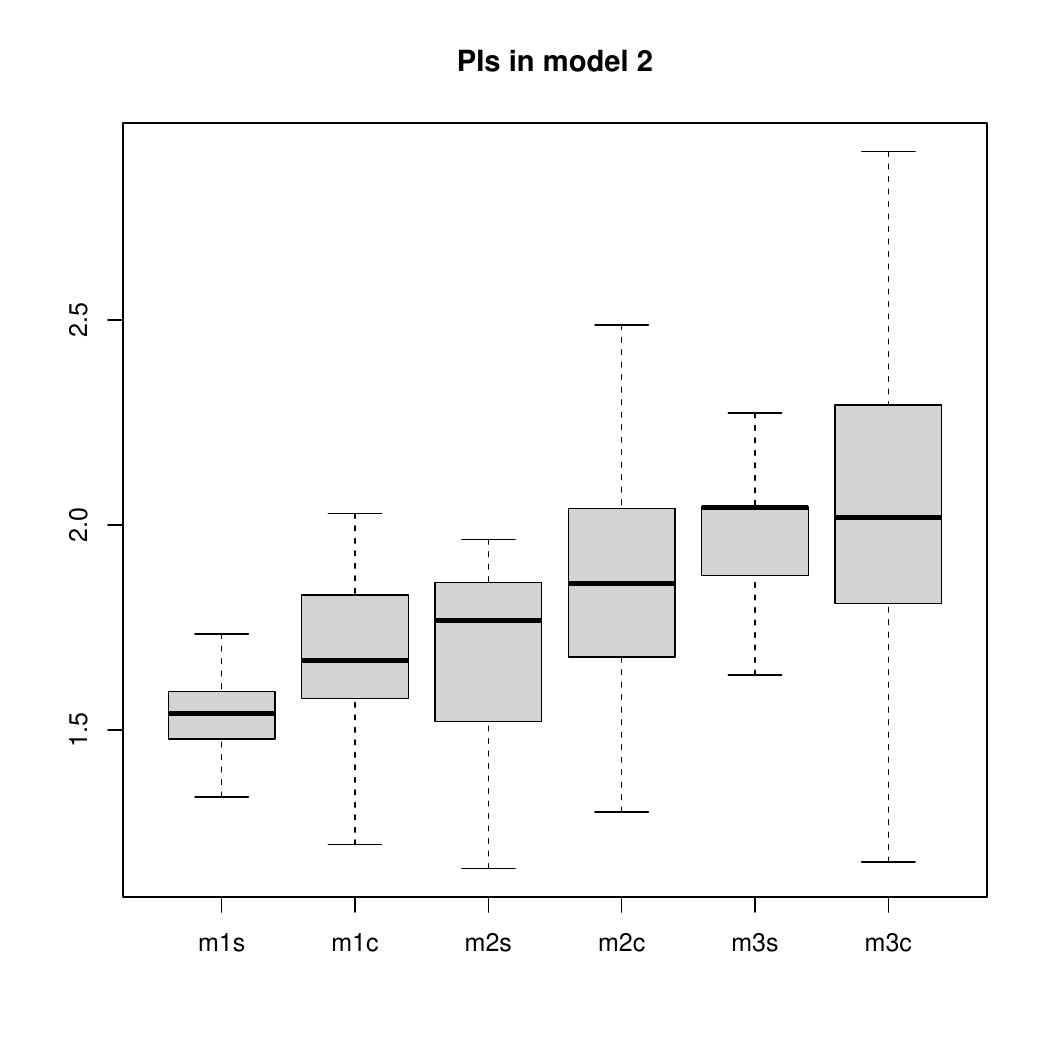}
            \includegraphics[scale = 0.33]{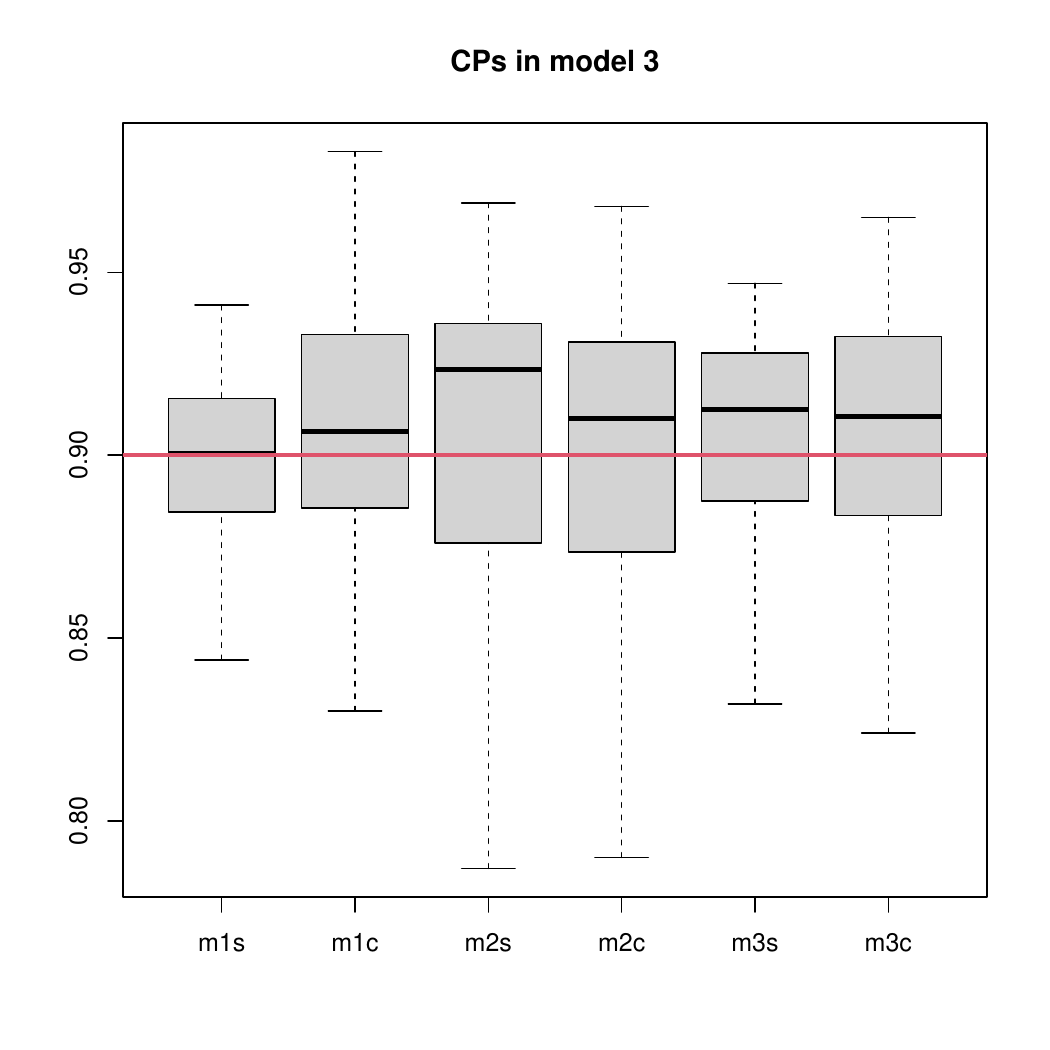}
     \includegraphics[scale = 0.33]{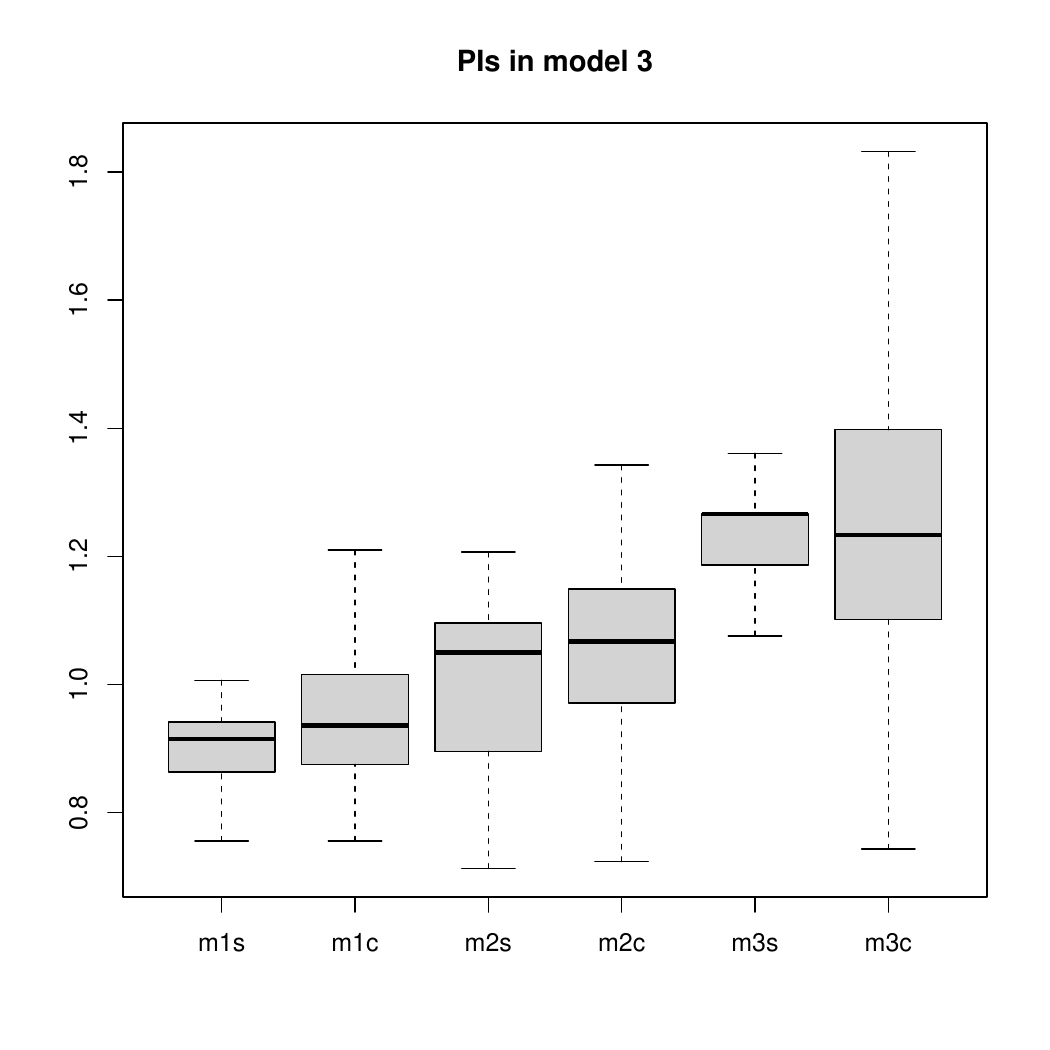}
   \end{figure}

\begin{figure}[!h]
    \centering
    \caption{Simulation A3: Misspecified regression model  
      and measurement error model. Histogram of the 90\% coverage probability (CP) and
      prediction interval length (PI) of the estimated prediction
      interval in the three models using the six methods when both the
      distributions of $U$ and $\epsilon$ are misspecified. $n=500$.}\label{fig:simu42}
    \includegraphics[scale = 0.33]{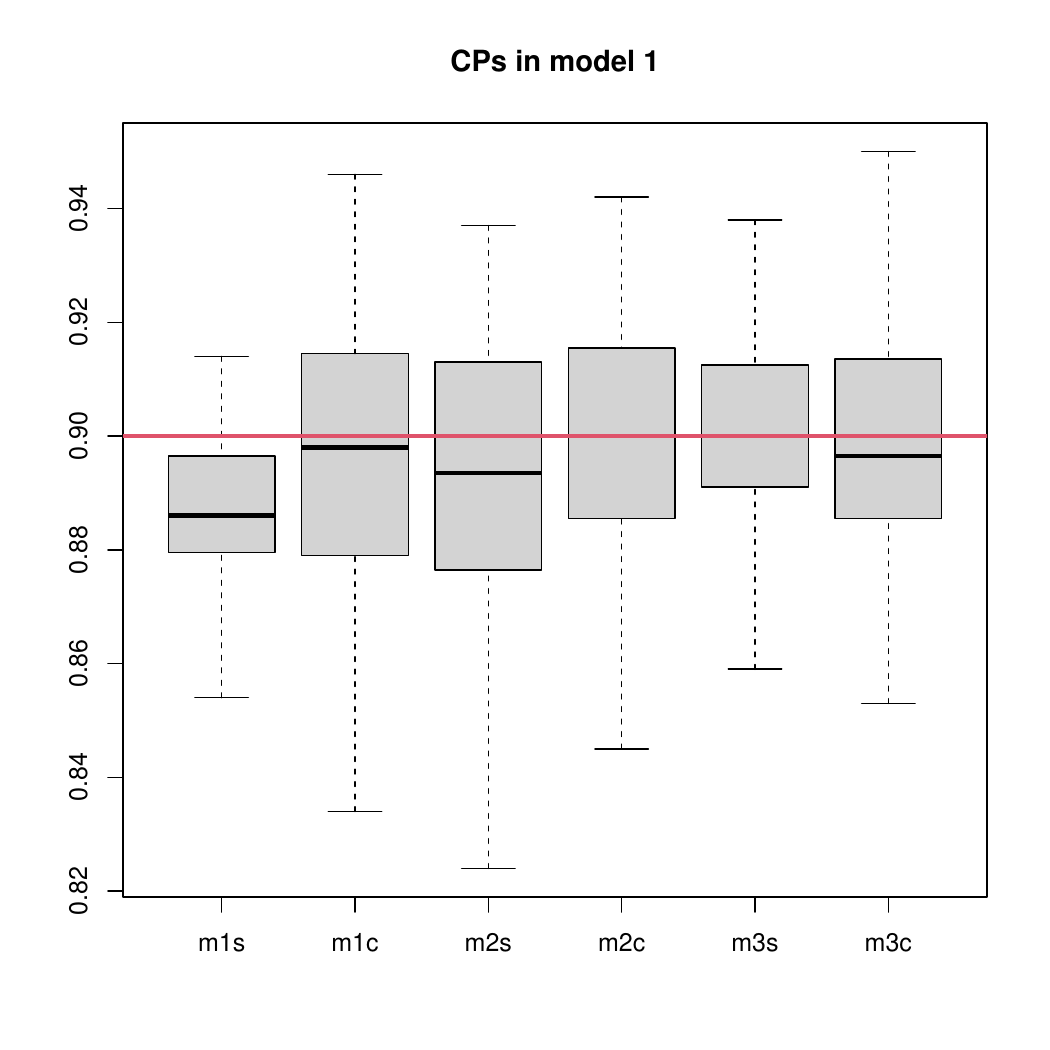}
    \includegraphics[scale = 0.33]{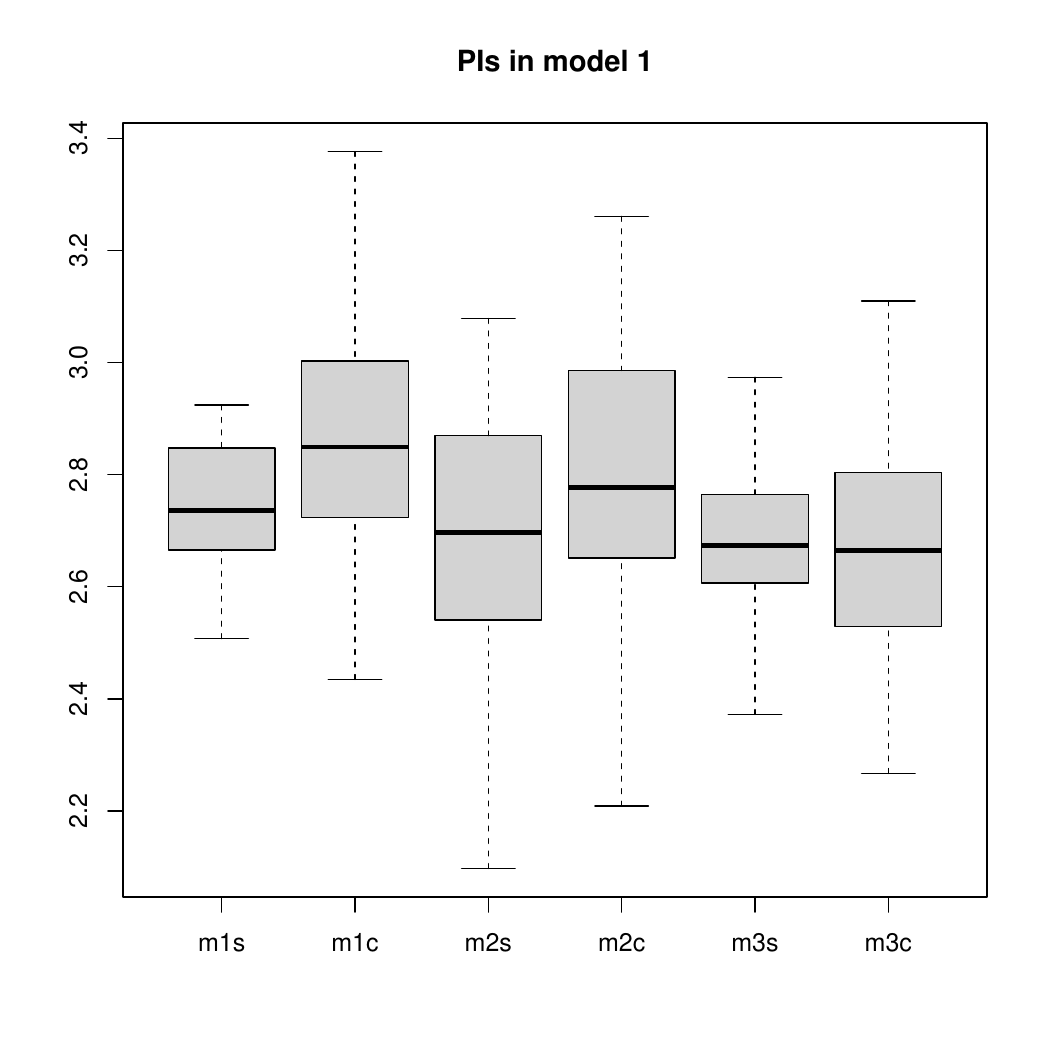}
        \includegraphics[scale = 0.33]{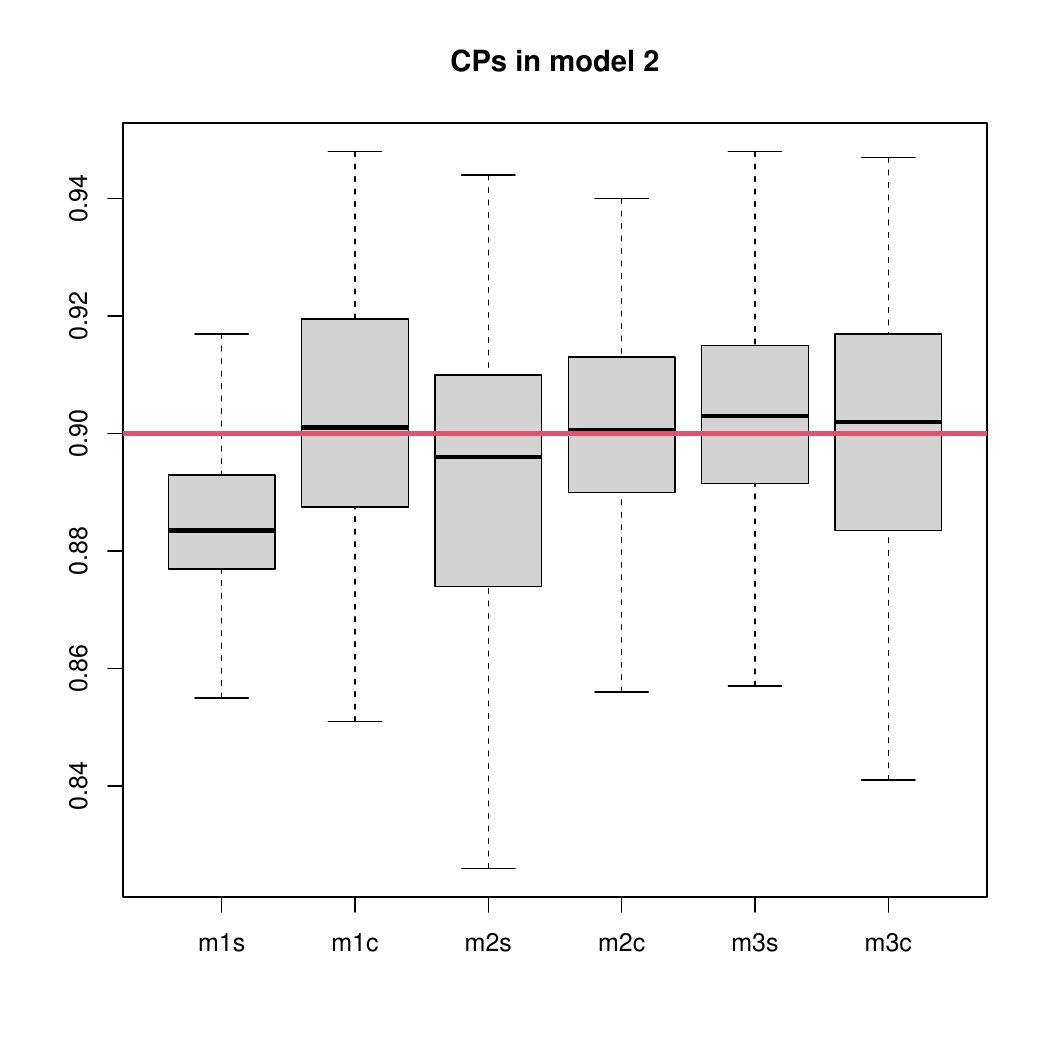}
        \includegraphics[scale = 0.33]{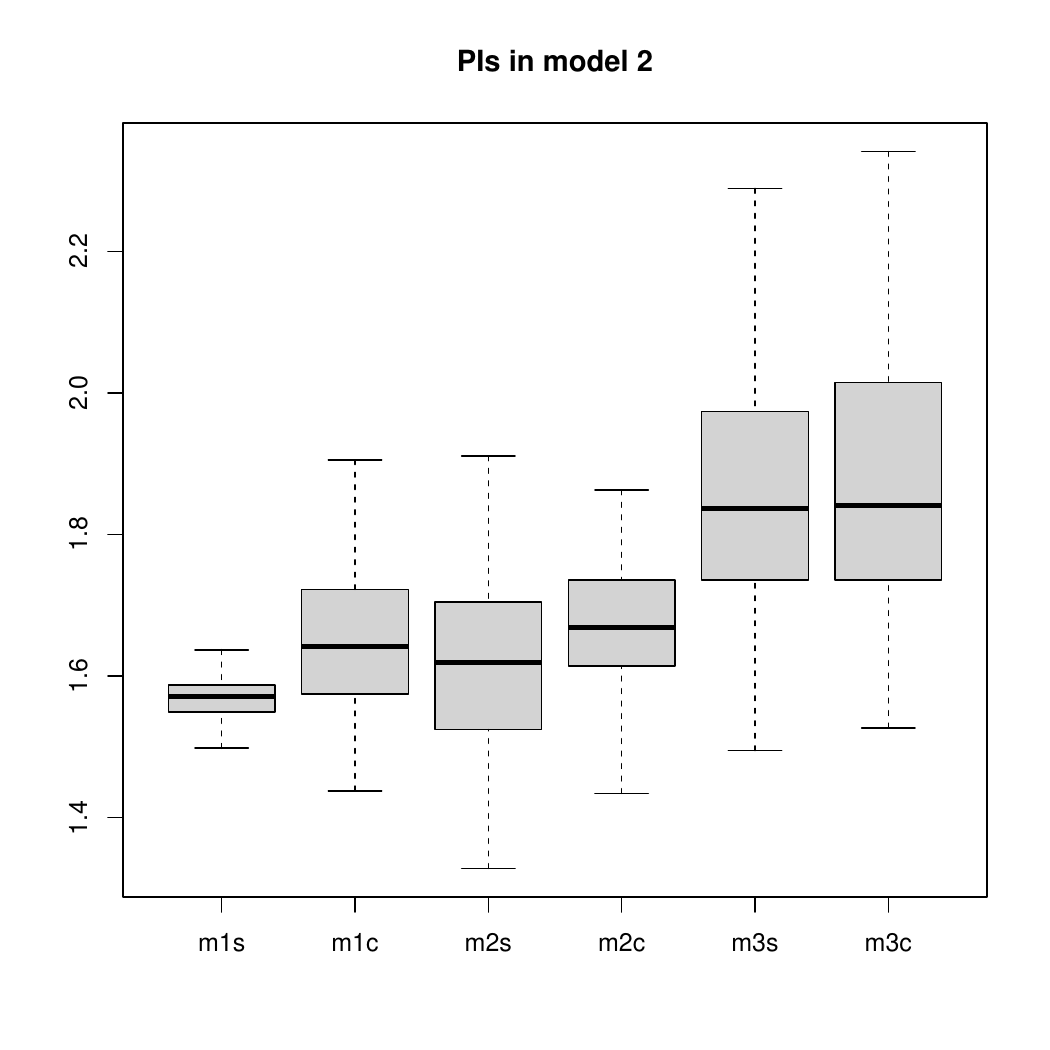}
            \includegraphics[scale = 0.33]{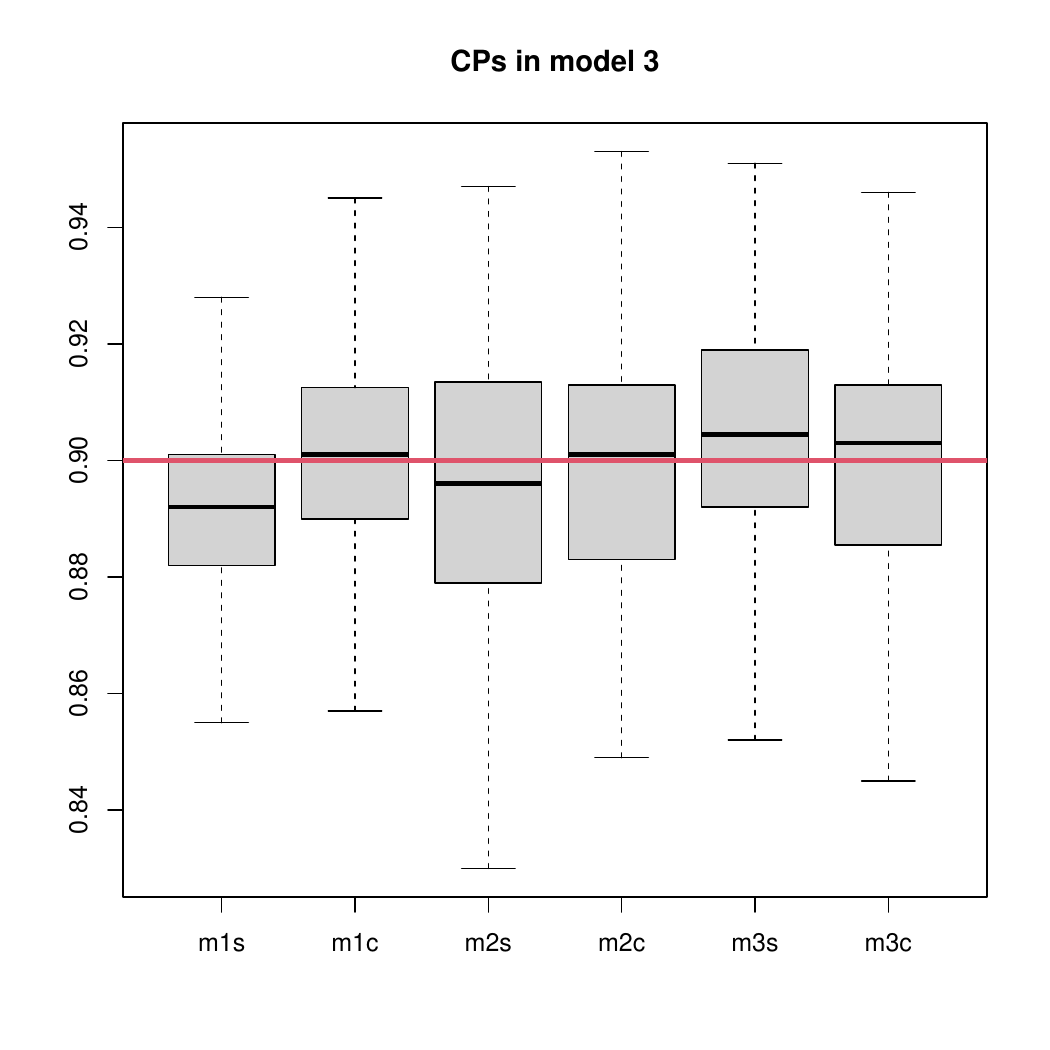}
     \includegraphics[scale = 0.33]{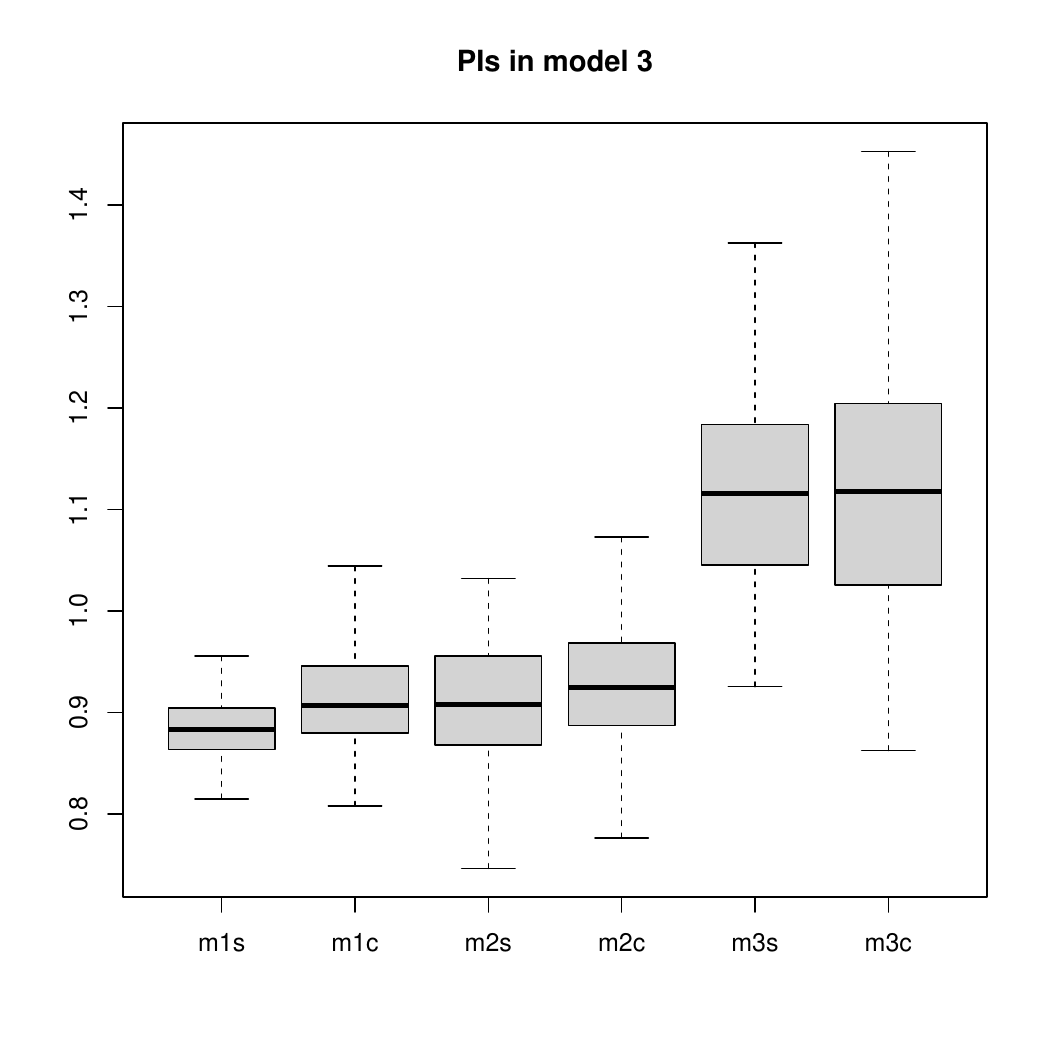}
   \end{figure}

We further conducted Simulation A3, where
  investigated the situation when both
  the regression and the measurement error models are
  misspecified. We combined the data generation procedure
    in Simulations A1 and A2 by generating
the regression errors $\epsilon_i$'s from $\epsilon \sim
t(3)\sigma_{\epsilon} /\sqrt{3}$ and generating the
  measurement errors $U_i$'s from 
$U\sim\sqrt{3} U(-1, 1)\sigma_U$. Other settings are  identical to
Simulation 2 in Section \ref{sec:simu}. 
  The results presented in Table
\ref{tab:sim4}, Figures \ref{fig:simu41} and Figure \ref{fig:simu42}
shows that when both the distributions of
$U$ and $\epsilon$ are misspecified, the coverage probabilities of the
estimated prediction intervals of our method tends to
slightly smaller than the nominal level.

  \begin{table}[!h]
    \caption{Simulation A1: Misspecified regression model.
      %The distribution of $X$ and
%  $\epsilon$ are  mis-specified.  
The average and
    standard deviation of the coverage
    probabilities (CP (SD)), and the average and standard deviation of 
    the  lengths  (LPI (SD)) of the estimated 90\% prediction
    intervals.  }{\label{tab:sim3}}
  \footnotesize
   \begin{tabular}{c|c|c|c|c|c|c}
  \hline
    &m1s& m1c&m2s&m2c&m3s&m3c\\

     \hline
    &\multicolumn{6}{c}{Model 1}\\
     \hline
  &\multicolumn{6}{c}{
     $n = 100$ }\\
     \hline
     &  \multicolumn{6}{c}{ model 1}\\
  \hline
    CP (SD)&0.892 (0.03)&0.902 (0.038)&0.916 (0.022)&0.896 (0.049)&0.904 (0.022)&0.894 (0.043)\\
         LPI (SD) &  2.698 (0.174)&2.913 (0.595)&3.428 (0.288)&3.765 (1.19)&2.762 (0.226)&2.835 (0.62)\\
     \hline
    & \multicolumn{6}{c}{ model 2}\\
      \hline
    CP (SD)&0.896 (0.018)&0.906 (0.046)&0.904 (0.035)&0.898 (0.045)&0.905 (0.019)&0.897 (0.038)\\
       LPI (SD) &    1.553 (0.077)&1.651 (0.271)&1.793 (0.191)&1.926 (0.366)&1.879 (0.144)&1.928 (0.37)\\
     \hline
   &  \multicolumn{6}{c}{model 3}\\
     \hline
    CP (SD)&0.902 (0.018)&0.904 (0.04)&0.907 (0.038)&0.903 (0.038)&0.902 (0.032)&0.903 (0.038)\\
       LPI (SD) &   0.882 (0.05)&0.908 (0.141)&1.002 (0.105)&1.072 (0.145)&1.139 (0.1)&1.173 (0.225)\\
     \hline
 &\multicolumn{6}{c}{
     $n = 500$. }\\
     \hline
    &    \multicolumn{6}{c}{ model 1}\\
  \hline
    CP (SD)&0.896 (0.012)&0.896 (0.02)&0.892 (0.018)&0.897 (0.022)&0.899 (0.015)&0.897 (0.02)\\
         LPI (SD) & 2.705 (0.076)&2.73 (0.234)&2.641 (0.203)&2.785 (0.263)&2.624 (0.177)&2.638 (0.225)\\
     \hline
    & \multicolumn{6}{c}{ model 2}\\
 
\hline
    CP (SD)& 0.9 (0.011)&0.899 (0.019)&0.898 (0.023)&0.896 (0.02)&0.904 (0.017)&0.896 (0.021)\\
       LPI (SD) &   1.565 (0.033)&1.563 (0.091)&1.591 (0.124)&1.619 (0.113)&1.779 (0.131)&1.761 (0.166)\\
     \hline
  &   \multicolumn{6}{c}{ model 3}\\
     \hline
    CP (SD)& 0.902 (0.012)&0.901 (0.021)&0.896 (0.021)&0.901 (0.02)&0.9 (0.018)&0.904 (0.02)\\
       LPI (SD) &   0.879 (0.026)&0.881 (0.059)&0.879 (0.06)&0.916 (0.057)&1.023 (0.088)&1.071 (0.127)\\
      \hline
\end{tabular}
  \end{table}

   \begin{table}[!h]
     \caption{Simulation A2:  Misspecified measurement error model.
              The average and
    standard deviation of the coverage
    probabilities (CP (SD)), and the average and standard deviation of 
    the  lengths  (LPI (SD)) of the estimated 90\% prediction
    intervals. }{\label{tab:sim5}}
\footnotesize
    \begin{tabular}{c|c|c|c|c|c|c}
  \hline
    &m1s& m1c&m2s&m2c&m3s&m3c\\
\hline
  &\multicolumn{6}{c}{
      $n = 100$ }\\
      \hline
      &  \multicolumn{6}{c}{
$m_1$}\\
  \hline
    CP (SD)&0.878 (0.036)&0.905 (0.04)&0.919 (0.028)&0.903 (0.042)&0.908 (0.022)&0.897 (0.045)\\
         LPI (SD) &   2.558 (0.246)&3.417 (0.974)&3.798 (0.415)&4.204 (1.388)&2.815 (0.198)&2.877 (0.519)\\
      \hline
       &  \multicolumn{6}{c}{
         $m_2$}\\
      \hline
    CP (SD)&0.883 (0.037)&0.901 (0.045)&0.919 (0.028)&0.903 (0.042)&0.908 (0.022)&0.897 (0.045)\\
       LPI (SD) &  2.676 (0.321)&2.885 (0.495)&3.798 (0.415)&4.204 (1.388)&2.815 (0.198)&2.877 (0.519)\\
      \hline
        &  \multicolumn{6}{c}{
          $m_3$}\\
      \hline
    CP (SD)&0.884 (0.025)&0.891 (0.05)&0.906 (0.048)&0.906 (0.045)&0.899 (0.035)&0.896 (0.044)\\
       LPI (SD) &   0.901 (0.065)&0.931 (0.16)&1.055 (0.123)&1.136 (0.166)&1.218 (0.173)&1.254 (0.259)\\
\hline
        &\multicolumn{6}{c}{
      $n = 500$ }\\
      \hline
        &  \multicolumn{6}{c}{
          $m_1$ }\\
  \hline
    CP (SD)& 0.89 (0.02)&0.898 (0.022)&0.891 (0.026)&0.893 (0.025)&0.899 (0.017)&0.\
9 (0.021)\\
         LPI (SD) & 2.616 (0.125)&2.74 (0.286)&2.697 (0.256)&2.861 (0.312)&2.675 (0.149)&2.688 (0.212)\\
      \hline
        &  \multicolumn{6}{c}{
          $m_2$ }\\
\hline
    CP (SD)& 0.885 (0.021)&0.898 (0.023)&0.891 (0.028)&0.898 (0.02)&0.903 (0.018)&0.899 (0.023)\\
       LPI (SD) &    1.546 (0.073)&1.617 (0.103)&1.63 (0.13)&1.72 (0.112)&1.948 (0.222)&1.944 (0.237)\\
      \hline
      &  \multicolumn{6}{c}{
        $m_3$}\\
       \hline
    CP (SD)&0.896 (0.027)&0.898 (0.02)&0.887 (0.027)&0.898 (0.023)&0.895 (0.03)&0.899 (0.022)\\
       LPI (SD) & 0.94 (0.076)&0.905 (0.057)&0.911 (0.059)&0.961 (0.066)&1.153 (0.16)&1.184 (0.217)\\
      \hline
\end{tabular}
  \end{table}

 \begin{table}[!h]
  \caption{Simulation A3:  Misspecified regression model 
      and measurement error model.
     The average and
    standard deviation of the coverage
    probabilities (CP (SD)), and the average and standard deviation of 
    the  lengths  (LPI (SD)) of the estimated 90\% prediction
    intervals. }{\label{tab:sim4}}
\footnotesize
    \begin{tabular}{c|c|c|c|c|c|c}
  \hline
    &m1s& m1c&m2s&m2c&m3s&m3c\\
\hline
  &\multicolumn{6}{c}{
      $n = 100$ }\\
      \hline
      &  \multicolumn{6}{c}{
$m_1$}\\
  \hline
    CP (SD)&0.882 (0.021) &0.9 (0.041)&0.910 (0.030) &0.9 (0.04)&0.873 (0.025)&0.897 (0.042)\\
         LPI (SD) &   2.719 (0.207) &2.969 (0.484)&3.262 (0.326) &3.741
                                                                                  (1.15)&2.484
                                                                (0.201)&2.87
                                                                         (0.507)\\
      \hline
       &  \multicolumn{6}{c}{
         $m_2$}\\
      \hline
    CP (SD)&0.879 (0.021) &0.905 (0.04)&0.87 (0.031) &0.898 (0.038)&0.861 (0.03)&0.898 (0.042)\\
       LPI (SD) &    1.541 (0.085) &1.681 (0.187)&1.671 (0.154) &1.905 (0.313)&1.684 (0.209)&2.043 (0.378)\\
      \hline
        &  \multicolumn{6}{c}{
          $m_3$}\\
      \hline
    CP (SD)&0.897 (0.024) &0.91 (0.035)&0.894 (0.047) &0.901 (0.041)&0.871 (0.03)&0.905 (0.04)\\
       LPI (SD) &   0.903 (0.054) &0.953 (0.114)&0.984 (0.127) &1.062 (0.151)&1.08 (0.177)&1.266 (0.234)\\
\hline
        &\multicolumn{6}{c}{
      $n = 500$ }\\
      \hline
        &  \multicolumn{6}{c}{
          $m_1$ }\\
  \hline
    CP (SD)& 0.886 (0.016) &0.897 (0.025)&0.880 (0.022) &0.899 (0.021)&0.892 (0.016)&0.898 (0.022)\\
         LPI (SD) & 2.739 (0.094) &2.86 (0.213)&2.551 (0.167) &2.802 (0.243)&2.6 (0.152)&2.673 (0.184)\\
      \hline
        &  \multicolumn{6}{c}{
          $m_2$ }\\
\hline
    CP (SD)& 0.884 (0.013) &0.901 (0.023)&0.875(0.022) &0.9 (0.019)&0.87 (0.019)&0.9 (0.022)\\
       LPI (SD) &    1.565 (0.035) &1.65 (0.1)&1.514 (0.092) &1.669 (0.095)&1.642 (0.174)&1.868 (0.192)\\
      \hline
      &  \multicolumn{6}{c}{
        $m_3$}\\
       \hline
    CP (SD)&0.891 (0.017) &0.9 (0.021)&0.883 (0.022) &0.9 (0.022)& 0.86 (0.028)&0.9 (0.022)\\
       LPI (SD) &   0.884 (0.03) &0.914 (0.056)&0.872 (0.048) &0.93 (0.059)&0.967 (0.104)&1.148 (0.179)\\
      \hline

\end{tabular}
  \end{table}

\end{document}